\documentclass{article}    
\usepackage{arxiv}
\usepackage{bookmark}

\usepackage[utf8]{inputenc} 
\usepackage[T1]{fontenc}    
\usepackage{hyperref}       
\usepackage{url}            
\usepackage{booktabs}       
\usepackage{amsfonts}       
\usepackage{nicefrac}       
\usepackage{microtype}      
\usepackage{lipsum}         
\usepackage{graphicx}
\usepackage{natbib}
\usepackage{doi}
\usepackage[noblocks]{authblk}

\usepackage{amsfonts}
\usepackage{amssymb}
\usepackage{amsthm}
\usepackage{amstext}
\usepackage{algorithmic}
\usepackage{array}
\usepackage{amsmath}
\usepackage{blkarray}
\usepackage{cleveref}       

\newcommand{\tablefont}[1]{\fontsize{#1}{#1}\selectfont}
\newtheorem{defi}{Definition}
\newtheorem{claim}{Claim}
\newtheorem{theorem}{Theorem}

\newtheorem{lemma}{Lemma}
\newtheorem{proposition}{Proposition}
\newtheorem{corollary}{Corollary}

\begin{document}

\title{Improving Teacher-Student Interactions in Online Educational Forums  using a Markov Chain based Stackelberg Game Model \mbox{\protect\footnotemark[2]}}
\footnotetext[2]{A preliminary version of this paper appeared in: \textbf{A Stackelberg game approach for incentivizing participation in online educational forums with heterogeneous student population}. \\Rohith~Dwarakanath Vallam, Priyanka Bhatt, Debmalya Mandal,  and Y.~Narahari. \newblock In \emph{ Twenty-Ninth AAAI Conference on Artificial Intelligence, {AAAI} 2015, Austin, Texas, USA. }}

\renewcommand{\shorttitle}{A Markov chain based Stackelberg game for online education forums}

\author[1]{Rohith Dwarakanath Vallam }
\author[2]{Priyanka Bhatt }
\author[3]{Debmalya Mandal}
\author[4]{Y Narahari}
 
\affil[1]{IBM India Research Labs, Bengaluru, India}
\affil[2]{Walmart Labs, Bengaluru, India}
\affil[3]{Harvard University, Cambridge,  USA}
\affil[4]{Indian Institute of Science, Bengaluru, India}
{
    \makeatletter
    \renewcommand\AB@affilsepx{: \protect\Affilfont}
    \makeatother

    \makeatletter
    \renewcommand\AB@affilsepx{, \protect\Affilfont}
    \makeatother

    \affil[1]{\texttt{rohithdv@gmail.com}}
    \affil[2]{\texttt{priyankabhatt91@gmail.com}}
    \affil[3]{\texttt{dmandal@g.harvard.edu}}
    \affil[4]{\texttt{hari@csa.iisc.ernet.in}}
}

\date{}

\maketitle

\begin{abstract}

With the rapid proliferation of the Internet, the area of education has undergone a massive transformation in terms of how students and instructors interact in a classroom. Online learning environments now constitute a very important part of any academic course. Online learning now takes more than one form, including the use of technology to enhance a face-to-face class, a hybrid class that combines both face-to-face meetings and online work, and fully online courses (popularly known as massive open online courses or MOOCs). Further, online classrooms are usually composed of an online education forum (OEF) where students and instructor discuss open-ended questions for gaining better understanding of the subject. However, empirical studies have repeatedly shown that the dropout rates in these online courses are very high partly due to the lack of motivation among the enrolled students. The objective of our work is to come up with an appropriate analytical model for OEFs and use this model to design an effective incentive-based game to enhance student-instructor participation in these OEFs considering for heterogeneity in the skills among the students as well as the limited budget of the instructor. 

We first undertake an empirical comparison of student behavior in OEFs associated with a graduate-level  course offered in the Indian Institute of Science during two terms. We identify key parameters dictating the dynamics of OEFs like effective incentive design, student heterogeneity, and  super-posters phenomenon. Motivated by empirical observations, we propose an analytical model based on continuous time Markov chains (CTMCs) to capture instructor-student interactions in an OEF. Using concepts from lumpability of CTMCs, we compute steady state and transient probabilities along with expected net-rewards for the instructor and the students. We formulate a mixed-integer linear program  which views an OEF  as a single-leader-multiple-followers Stackelberg game. Through simulations, we observe that students exhibit varied degree of non-monotonicity in their participation (with increasing instructor involvement). We also study the effect of instructor bias and budget on the student participation levels. Our model exhibits the empirically observed super-poster phenomenon under certain parameter configurations and recommends an optimal plan to the instructor for maximizing student participation in OEFs.
\end{abstract}

\section{Introduction}

With the rapid proliferation of the Internet, the area of education has undergone a massive transformation in terms of how students and instructors interact in a classroom. Online learning environments now constitute a very important part of any academic course. Online learning now takes more than one form, including the use of technology to enhance a face-to-face class, a hybrid class that combines both face-to-face meetings and online work (which is typical of courses in many universities today), and  online courses (popularly known as massive open online courses or MOOCs).  Broadly, any academic  course may thus be modelled in three ways: the traditional \textit{face-to-face} model, an \textit{online} model and the \textit{hybrid} model constituting both face-to-face interactions as well as online interactions between students and the instructor. Traditional face-to-face classrooms are identified by an instructor who is conducting the course and students enrolled to the course. Periodic  lectures, tests and assignments  require the physical presence of the students and the instructor in the classroom at some fixed timings. Hybrid  and online classrooms offer a great deal of flexibility in conducting the class which was not possible in the pre-Internet era.  One of the forms of online education that has garnered active interest of researchers due to their immense popularity are the massive open online courses (MOOCs) offered by MOOC platforms like Coursera, edX, Udacity, etc.  
As of January 19, 2017, Coursera \citep{coursera} had about 23 million enrollments from students representing 190 countries. Further, Coursera students (as of January 2014) voiced themselves in 590,000 discussion threads in the education forums for a total of 343,014,912 minutes of learning across 571 courses~\citep{butean2015approach}. 

A recent US Department of Education report \citep{MEANS09} made some key observations regarding effectiveness of online learning namely:
\begin{itemize}
\item {Students taking online courses performed better, on an average, than those taking the same course in a traditional face-to-face environment.}

\item {Hybrid courses having both an online instruction as well as face-to-face elements had a larger advantage relative to purely face-to-face instruction than did purely online instruction.}

\item {The efficacy of online classrooms varies across different courses as well as \textit{heterogeneous skills} of students.}
\end{itemize}

Many hybrid and online courses typically have an online discussion forum (also known as online education forum (OEF)\citep{GHOSH13}) wherein students with heterogeneous skills can interact with each other as well as the instructor and discuss a variety of questions related to the subject. Such environments offer a great deal of promise for education as well as pose a number of challenges.  Surveys of student perceptions about online threaded discussions have shown tremendous positive interest among the students to access these OEFs as an effective way for acquisition of knowledge (\citep{pendry2015individual},\citep{martyn2003hybrid}). Studies in this area have concluded that developing an actively participating learning community along with the instructor participating as an equal member is the vehicle through which online education is best delivered \citep{palloff2007building}.  An important part of such hybrid and online educational environments are OEFs which aim at filling the `physical' connect missing in these online platforms. Students and the instructor use these forums in order to interact and discuss various topics related to the course and otherwise, which helps in improving the understanding of the students \citep{ANDRESEN09}. 

It is  also noted that participation in OEFs provides opportunities for responsibility and active learning through the expectation of regular participation in online discussions \citep{hopperton1998computer}. This is not a scenario for a few prepared students to respond to a lecturer while the rest of the class sits back. Participation in the OEF demands that students become actively engaged with the course content and through the interaction with their peers, negotiate the meanings of the content. They construct knowledge through the shared experiences that each participant brings to the collaborative discussions \citep{markel2001technology}.  
Thus, the primary  objective of online classrooms is to enhance learning by stimulating active discussions on their OEFs and this is determined by the extent of student participations in these OEFs. In this work, we are addressing the problem of improving student participation in the OEFs . In particular, we quantify student participation in OEFs by measuring the number of answers given by enrolled students (with heterogeneous skills) to a number of open-ended questions posted by the instructor on these OEFs. Recent approaches to this problem have focused on an incentive-based approach for improving participation in OEFs \citep{GHOSH13} and Q\&A forums like Yahoo! Answers, Stackoverflow, WikiAnswers, etc (\citep{GHOSH12a},\citep{GHOSH12b},\citep{SAKURAI13}). 

In this work, we adopt a game-theoretic approach to address the issue of improving  participation levels among students (with heterogeneous skills) in an OEF. We incentivize students on a per question basis to keep up the momentum of participation in the class. Students are required to post answers to the specific open-ended/discussion-style questions that are posted on the OEF. These questions have the characteristic that they do not have a well-defined answer and hence necessitate multiple viewpoints/opinions. Students are thus incentivized to post answers to such questions even if other classmates have already posted their answers resulting in increased student participation and better understanding of the course content. In practice, these incentives could manifest in the form of book vouchers, food coupons or some extra grade points, as considered appropriate by the instructor.

\section{Online Education Forums - Key Issues and State of the Art}

Before we get into the details of our approach, we briefly discuss various properties of online classrooms as well as some important issues faced by online education forums (OEFs) in particular. We follow this with a review of some relevant game theoretic literature in this area.


\subsection{OEFs, Instructor Participation, and Open-Ended Questions} 

Numerous studies indicate that OEFs are instrumental in online classrooms for the construction of knowledge and for improving the understanding of the enrolled students. Andresen~\citep{ANDRESEN09} and Mazzolini~et.al.~\citep{MAZZOLINI03}
emphasize the critical role of discussion forums and optimal instructor participation in improving the effectiveness of online learning.  Breslow et. al~\citep{BRESLOW13} observe the data of participation patterns of students in a MOOC offered by edX and  their study indicates that $52\%$ of the certificate earners were active on the discussion forums. 
Williams et. al~\citep{WILLIAMS14} and Ballou~\citep{BALLOU08} emphasize that open-ended questions help building a rapport among the participants and also improve their understanding. Also, an experimental study by Hull et.al~\citep{HULL09} suggests that open-ended questions and controlled instructor participation on the OEFs  has positive effects on the learning of students.

\subsection{Heterogeneous Students}
Students participating in MOOCs have been classified in the literature based on various criteria. DeBoer~et.al.\citep{DEBOER14} identified that students enrolled for MOOCs differ based on geographical locations, educational backgrounds and age groups. 

Kizilcec~et.~al~\citep{KIZILCEC13} observe that there are  typically four types of students  in online courses based on their level of commitment to the course namely: `Completing', `Auditing', `Disengaging', and `Sampling'. Wilkowski et. al~\citep{WILKOWSKI14} note that there are about four main categories of students namely: `No-shows', `Observers', `Casual Learners', and `Completers',  classified based on their objective of joining the online classroom. 

 \begin{figure}[!hbtp]
 \begin{center}
  \includegraphics[scale = .36]{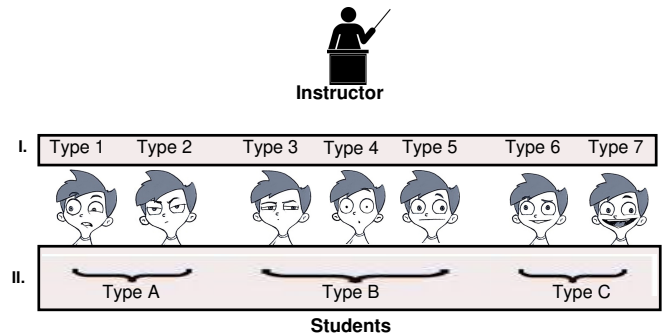}
 \end{center}
  \caption{This figure  shows a classroom and two sample ways in which the students in a class can be categorized: \textbf{I.} All students are different essentially and so each student belongs to a different type. \textbf{II.} Sets of students exhibiting properties more similar to each other but different from the rest are put together in one type. (Image Courtesy: BYU Animation) \label{student_types} }
 \end{figure}

One of the primary objectives of our work is to model such heterogeneity among the student population by categorizing the students into $L$ types according to some suitable criteria, for example, the expertise of a student in the related areas or the quality of answers given by the students, etc. 
The classification of students into different types (Figure~\ref{student_types}) can be as fine as each student belonging to a different type, or as coarse as all students belonging to a single type.  This can depend on the size of the class, for example, in a small class each student can be considered as belonging to a different type whereas in a larger class similar students can be grouped together into  one type to reduce the number of types. This grouping is essential in large classrooms (of the scale of online classrooms) as giving personal attention to each student is a humongous task for the instructor and it is easier for the instructor to focus based on grouping of the students.

\subsection{Low Student-Retention Rate}
Online classrooms witness massive enrollments but their retention rate is very dismal with around $90\%$ students not finishing the course~\citep{FOWLER13}. Halawa~et.~al~\citep{HALAWA14}, and Cheng~et.~al~\citep{CHENG13} claim that a timely intervention by the instructor might help decreasing the dropout rates in the online classrooms and so design predictors and tools which will assist the instructor in identifying the students at a high risk of dropping out.

\subsection{Incentives for Participation in OEFs}

The instructor of an online course might encounter distinct types of individuals in the OEFs and her objective is to maximize participation from each of them whereas the students' objective is to maximize rewards received (if any) and minimize the cost incurred towards answering questions on the forum. The proposed model thus introduces incentives in the form of some socio-psychological rewards or some physical rewards that are offered to the students on answering a question.
The socio-psychological rewards are in the form of satisfaction of the students on getting recognition from the instructor in the classroom and also that the instructor is reading their `good' answers (we assume that there are teaching assistants for the course who allow only a good quality answer to be posted). On the other hand, physical rewards may be in the form of some book vouchers or food coupons etc. that the instructor can provide to the students.
An important point to note is that both types of rewards can be awarded only on a limited scale as the instructor can read or reward only some limited number of answers to a question. This real-world  limitation drives us to consider a \textit{budgeted} rewarding scheme which ensures that the instructor has a limited budget $B$ per question, for giving out rewards to the students.

\subsection{Review of Relevant Literature}

Brafman~et.al.~\citep{brafman1996partially} analyze partially controlled multi-agent systems and in this context, the authors consider the scenario of  a two-agent system consisting of a teacher and a student. The work assumes the teacher is a knowledgeable agent, while the student is an agent that is learning how to behave in its domain. Their goal is to utilize the teacher (which is under our control) to improve the behavior of the student (which is not controlled by us).
Our work is similar in philosophy to the above work in that we are proposing an optimal plan to instructor (or teacher) to improve student behaviour in online educational forums. However, we consider solving the problem through designing of optimal incentive scheme in the presence of a budgeted instructor. Also, we customize the behaviour of the instructor taking into account the presence of students with heterogeneous skills.

Fave et.al~\citep{delle2014game} propose a Stackelberg game theoretic model to handle execution uncertainty in the security patrolling domain. They come up with equilibrium strategies for attacker and defenders and perform extensive experiments which implements the proposed model. Our work also models a Stackelberg game but in the domain of education. Our work models an educational forum involving  instructor and students in an online course and captures  heterogeneity in the skills of the students  through a Continuous time Markov Chain model for the online educational forum. Further, the parameters of the CTMC are optimized by viewing the classroom activities as a Stackelberg game model among the instructor and students. Equilibrium strategies are computed which provide an optimal plan to the instructor for maximizing student engagement in an online course.

Ghalme~et.al~\citep{GhalmeAAMAS2018} consider the problem of designing a robust credit score function in the context of online discussion forums. They design coalition resistant credit score functions for online discussion forums. They show that modularity is coalition identifying and provide theoretical guarantees on modularity based credit score function. Carbonara~et.al~\citep{Carbonara2015IncentivizingPG} model the problem of strategic auditing in peer grading modeling the student's choice of effort in response to a grader's audit levels as a Stackelberg game with multiple followers. In this theoretical analysis, a homogeneous student population is considered and provide a polynomial-time approximation scheme (PTAS) in order to compute an approximate solution to the problem of allocating audit levels.

Game theoretic literature that relates to online learning and discussion forums mostly concentrates on designing incentives for Q\&A forums like Yahoo! Answers, Stackoverflow, WikiAnswers etc.
Ghosh~et.al~\citep{GHOSH12a} study about two types of rewards, attention rewards and fixed-total rewards, and how they should be set in order to achieve good quality outcomes from the crowd in an online forum.
Sakurai~et.~al~\citep{SAKURAI13} make the agents indirectly report their efforts and allocate rewards such that agents have to be truthful and have to put in more efforts in order
to receive higher rewards.
Ghosh~et.al~\citep{GHOSH12b} investigate two settings, one in which a contribution's value depends upon an agent's expertise only, and the other in which a contribution's
value depends upon both the expertise of the agent and the effort put in by her. They show that optimal outcomes can be implemented in the first case, whereas in the
second case, optimal outcomes can be implemented only if the ranking of the contributions is noisy.

Ghosh~et.al~\citep{GHOSH13} focus on online forums for education which is closer to our work. They allow the students and the instructor to choose their rates of arrival
in order to maximize their utility. Each type of arrival is modeled as a Poisson process and then analysis on the relationship between the students' rate of arrival
and the instructor's rate of arrival is done for two types of questions: single-answer type and open-ended type. They are able to show that the students rate of 
participation varies non-monotonically with the instructor's rate of response to questions, as is generally observed. That is, the students' rate of participation initially 
increases as the rate of arrival of the instructor increases and then as the instructor's rate increases beyond a threshold, their rates start dropping until they stop 
participating. 

Some key assumptions in Ghosh~et.al~\citep{GHOSH13} are homogeneous student population and unlimited number of rewards given to the students, which are very unrealistic in practical settings. Our model generalizes their model in the following aspects: 
(a) Heterogeneity of student population: we consider a realistic model where students possess heterogeneous skill levels.  (b) Budgeted Rewards: the instructor has a limited budget to award students who are participating in the OEF.
The analysis in Ghosh and Kleinberg~\citeyear{GHOSH13}  assumes that all students are similar and this special case can be handled in our model by assuming that all the students belong to the same type. 
  
\section{Contributions and Outline}
We list the main contributions of this work below:
\begin{itemize}
\item We perform empirical analysis of a real-world online education forum by investigating the participation levels of students in the Game Theory course (E1 254) offered in the Indian Institute of Science, Bangalore during Jan-April, 2014. We design incentives to promote better participation in the online Piazza~\citep{piazza} forum associated with the course. We account for the heterogeneity among the students of the class through various parameters and study the improvement in participation levels in this course compared to the same course offered two years ago where no such incentives for participation were offered.

\item We devise a recipe for the instructor in an online course to increase student participation in OEFs. To achieve this goal, we propose a mixed integer linear program (MILP) framework to recommend an optimal plan to the instructor for maximizing student participation in OEFs. We approach this objective through the following intermediate steps:
 
\begin{itemize}
 \item[(a)] Motivated by the empirical study of the online classroom, we first analytically model the activities of an online education forum (OEF) in which there is a course instructor and students with heterogeneous skill levels. In order to promote a higher level of participation by the students in the forum, the instructor periodically posts open-ended (discussion type) questions pertaining to the course in the forum. We propose a continuous time Markov chain (CTMC) to model the non-strategic (i.e. when the rates of arrival of the students and the instructor to  the forum are known) student-instructor interactions in an OEF. We model the cost-benefit trade-offs of the students for participating in the forum  as well as the limited budget of the instructor by defining corresponding reward functions on the CTMC.
 
\item[(b)] Using ideas from lumpability of Markov chains, we simplify the CTMC into simpler, student-specific lumped CTMCs. We then compute the steady state and the transient probability distributions of the lumped CTMCs and evaluate the expected net rewards to the instructor and students.
 
\item[(c)]  We then formulate a strategic-OEF setting where both the instructor and the students have a set of possible rates of arrival to choose from and each of them will choose an arrival rate in order to maximize their respective (transient/steady state) utilities. Using a Stackelberg game formulation, we propose a bi-level optimization problem between the instructor and the students in the form of a mixed integer quadratic program (MIQP) which we convert to a mixed integer linear program (MILP) using an approach similar to the  analysis done for the Bayesian Stackelberg setting by Parachuri~et.al~\citeyear{PARACHURI08}.

\item[(d)] We undertake detailed simulations and develop new insights into the activities of OEFs.  We observe that the students exhibit varied degree of non-monotonicity (with increasing instructor rate) in their participation. We also study the effect of instructor bias and budget on the student participation levels. Our model exhibits  the empirically observed super-poster phenomenon under certain parameter configurations and recommends an optimal plan to the instructor for maximizing student participation in OEFs.

\end{itemize}
\end{itemize}

The rest of the paper is organized as follows. We discuss the empirical analysis of a real-world online education forum in Section~\ref{section_participation_case_study}. We analytically model the activities of an online education forum (OEF) using a CTMC in Section~\ref{Section_CTMC_model}. Lumpability of this CTMC is discussed in Section~\ref{Section_lumpaing}. A Stackelberg game formulation is provided in Section~\ref{Section_OEF_Game}. Simulation results are discussed in Section~\ref{Section_simulations}. We finally conclude in Section~\ref{Section_conclusions} and discuss possible avenues of future work in Section~\ref{futurework}. The proofs for all results are provided in the appendix.

\section{Participation in Online Education Forums - A Case Study}
~\label{section_participation_case_study}

\begin{figure*}[h]
\begin{tabular}{cc}
 \begin{minipage}{8cm}
	\footnotesize
	\centering
	\textbf{OEF Statistics}\\
	\begin{tabular}{|c|c|c|} 
	\hline
	E1 254 & \textbf{2012} & \textbf{2014}\\
	\hline
	\# \textbf{Enrolled}& 72 & 38 \\
	\text{ } \textbf{Students}  &    &    \\
	\hline
	\textbf{Forum} 	&$21^{\text{st}}$ Jan&$1^{\text{st}}$ Mar\\
	\textbf{Duration}	& to $1^{\text{st}}$ May& to $1^{\text{st}}$ May\\
	\hline		
	\end{tabular}

	\vspace{0.5cm}

	\textbf{Incentive System (2014)}
	\begin{tabular}{|c|c|c|} 
	\hline 
	\textbf{Action} & \textbf{Reward} \\
	& (\textbf{in Points})\\
	\hline 
	Registration & 20\\
	\hline
	Log in per day & 02\\
	\hline
	First Q/A & 20\\
	\hline
	Post per Q/A & 15\\
	\hline
	Endorsed by TAs & 20\\
	\hline
	Endorsed  by Instructor   & 20\\
	\hline
	Junk/Duplicate Post & -100\\
	\hline
	\end{tabular} 
	\vspace{0.5cm}
\end{minipage}
&
\begin{minipage}{8 cm}
	\centering
	\includegraphics[scale = .32]{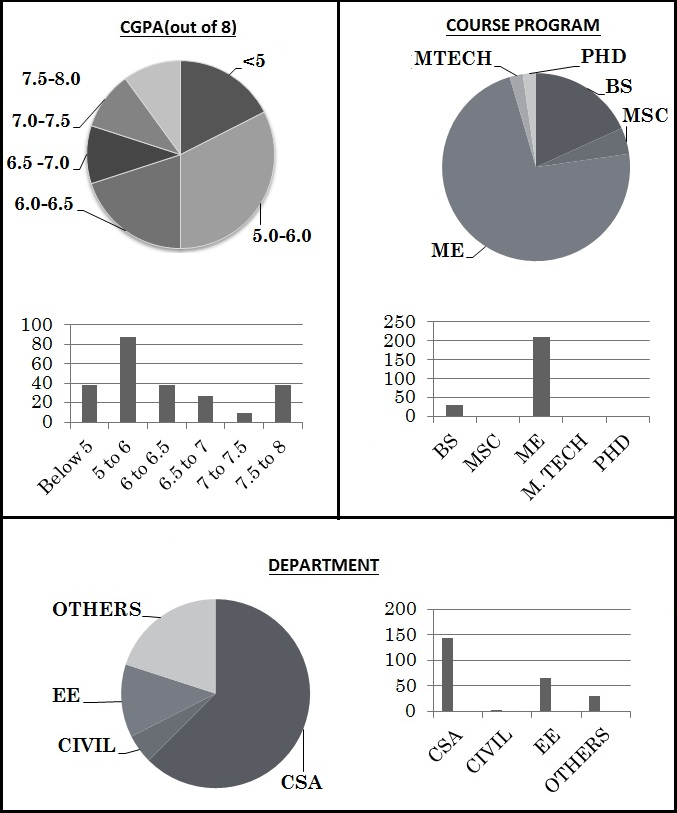}\\
\end{minipage}
\\
(a) & (b) 
\end{tabular}
\caption{Figures~\ref{Figure_OEF}(a) to (b) correspond to empirical observations in the Piazza forums of the Game Theory Course for two terms (2012 and 2014) held at the Department of CSA, IISc. Note that the y axis labels in the bar graphs in this figure correspond to ``Number of Answers posted on the Piazza forum". }\label{Figure_OEF}
\end{figure*}

We conducted an experiment using the Piazza~\citep{piazza} education forum to study the effect of incentives on  improving participation in the online classrooms. As part of this experiment, we created user accounts on Piazza for the students, instructors  and teaching assistants of the ongoing offline course on Game Theory (E1 254) in the Department of Computer Science and Automation, Indian Institute of Science, in Spring 2014. An explicit incentive system was formulated where participation on forum was rewarded with points. These points could in turn be encashed as an increase of upto $5\%$ in the marks awarded in the offline course of Game Theory. The details of the OEF statistics and  the incentives offered  are given in Figure~\ref{Figure_OEF}(a). In order to further incentivize participation, we also organized polls every few weeks on the 2014 OEF to choose the students who were actively posting questions and answers on the forum. The winners of these polls as well as the students who voted for the 
winners were given some additional reward points. We compared the participation levels we achieved in this forum with data obtained from the OEF for the same course in Spring 2012 at the Indian Institute of Science wherein there was no such incentive designed for improving participation. 

\begin{figure*}[h]
\begin{tabular}{cc}
\begin{minipage}{8 cm}
	\includegraphics[scale = .22]{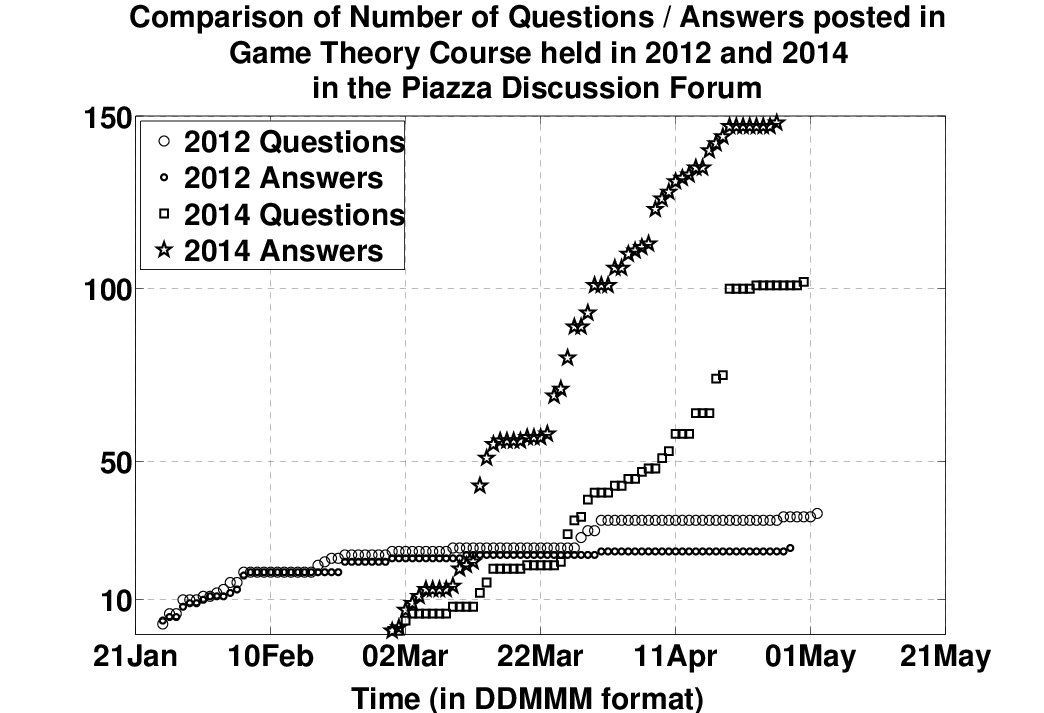}	
\end{minipage}
&
\begin{minipage}{8 cm} 
	\includegraphics[scale = .22]{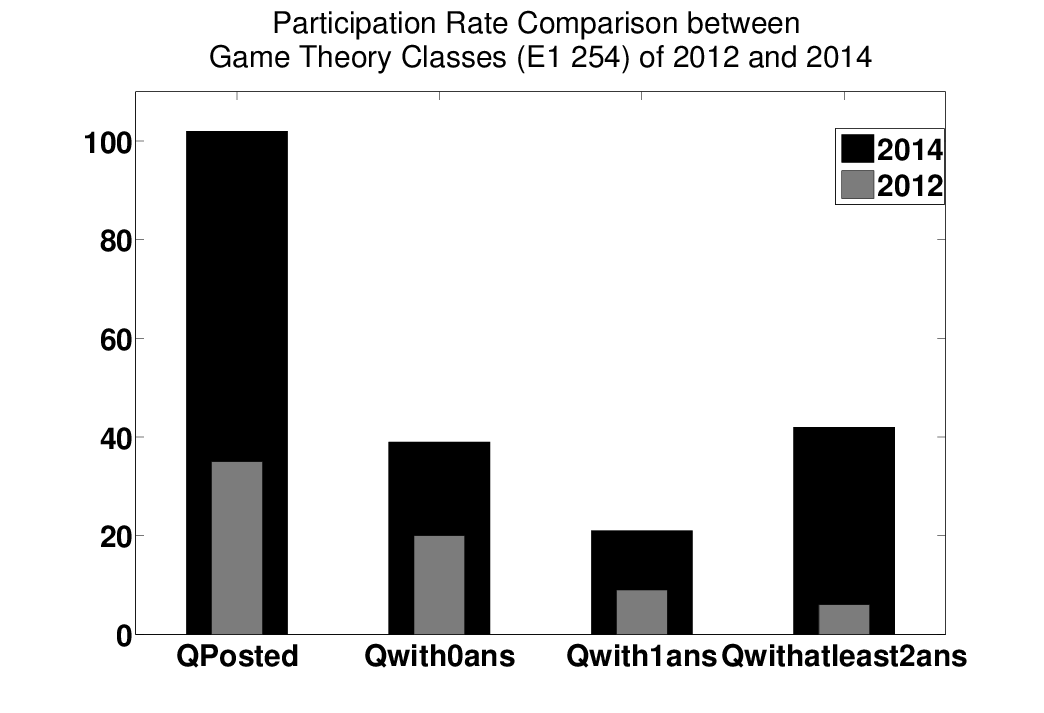} 
\end{minipage}
\\
\\
\begin{minipage}{8 cm} \centering (a) \end{minipage}& \begin{minipage}{8 cm} \centering (b) \end{minipage}
\\
\multicolumn{2}{c}{
\begin{tabular}{c}
\includegraphics[scale = .33]{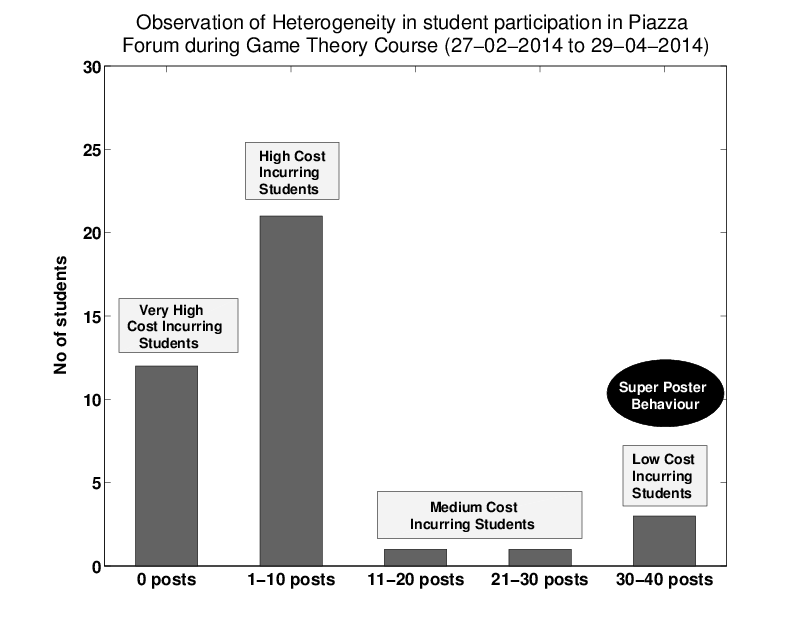}
\\(c)
\end{tabular}
}
\\
\end{tabular}
\caption{Figures~\ref{Figure_OEF2}(a) to (c) correspond to  evaluations of participation levels in the Piazza forums of the Game Theory Course for two terms (2012 and 2014) held at the Department of CSA, IISc.}\label{Figure_OEF2}
\end{figure*}

We put forth some of the key observations from our study of the OEFs from these two courses.
\begin{itemize}

\item The E1 254 (2014) OEF witnessed high diversity in terms of CGPA, course program and department of the different students that enrolled for the online forum. The different types of students also showed variation in their levels of participation on the forum (See Figure~\ref{Figure_OEF}(b)).
 
\item The participation levels on the 2014 OEF were higher as compared to the 2012 OEF in terms of the number of questions and answers posted by the enrolled students (See Figure~\ref{Figure_OEF2}(a)). This is despite the fact that the number of enrolled students and the forum duration were lower for the 2014 OEF. Figure~\ref{Figure_OEF2}(b) gives a fine-grained view of the difference in the participation levels in the two OEFs. We thus conclude that incentives that were offered in the 2014 OEF had a major impact on improvement of  participation levels in the 2014 class. 

\item We witnessed different levels of participation from the enrolled students with the total number of posts by a student varying from 0 to mid-30's (See Figure~\ref{Figure_OEF2}(c)). The students also vary in the amount of effort they are willing to put for participation in the online forums. This effort is subjective to a student based on various factors like her capabilities and commitments towards other courses etc. We thus roughly categorize the students as low, medium, high and very high cost (of effort) incurring students depending on their total participation levels through the forum duration (See Figure~\ref{Figure_OEF2}(c)). 

\item Low cost (of effort) incurring students posted more than 30 posts on the forum and were very active throughout the forum duration irrespective of participation levels of other students and the instructor. This behavior in online forums has also been observed by Huang~et.~al~\\citep{HUANG14} and they term such students as ``superposters'' (marked in a black oval shape in Figure~\ref{Figure_OEF2}(c)).
\end{itemize}

The incentives that were fixed for the OEF of the E1 254 course (offered in 2014) were decided by taking some basic guidelines as those offered in the Yahoo! Q\&A forums \\citep{JAIN09} and tweaking them a bit to suit our objective. So, the results we have observed are indicative of the general behavior of the students and the positive role that the incentives play in OEFs. The importance of OEFs towards improving the learning of the students and the issues these forums face along with the case study motivate us to do an extensive mathematical modeling of these OEFs.  We understand that we might not be able to incorporate all these real life observations in theory and so, in the mathematical modeling part of the report, we shall make suitable assumptions at appropriate junctures.
 
\begin{figure}[t]
\small
 \begin{algorithmic}[1]
 \STATE Instructor chooses an arrival rate to the forum and announces this to the class. 
 \STATE Students (of different types) observe the instructor announcement and decide their corresponding rate of arrival.
 \STATE Instructor and the student record the next arrival time based on their corresponding chosen arrival rates. 
 \WHILE{Course has not ended} 
    \IF{you are the instructor}
	\IF{there is time available for next arrival}  
	    \STATE Engage in other activities not related to the forum.
	\ELSE 
	    \STATE Close current (open-ended) question on the forum.
	    \STATE Reward points to students who have answered this question.
	    \STATE Post the next discussion-style question.
	    \STATE Record the next time for arrival to the forum.
	\ENDIF
    \ENDIF
    \IF{you are a student} 
	\IF {there is time available for next arrival}  
		\STATE Engage in any other activities not related to the forum.
	\ELSE  
	  \STATE Post a valid answer to the current open question in the OEF. 
	  \STATE Record the next time for arrival to the forum.
	\ENDIF 
    \ENDIF
 \ENDWHILE
\end{algorithmic}
\caption{ An instructor-driven interaction model in an OEF\label{instr-OEF}}
\end{figure}

\section{A CTMC Model for OEF (Non-Strategic Setting)}
\label{Section_CTMC_model}

In this section, we first discuss the setting of an online education forum (OEF) and then propose a CTMC to model the arrivals on this forum. We define a suitable reward structure on top of this CTMC and analyze the net-rewards received by the students as well as the instructor. We assume that the instructor and the students participating in the OEF are all non-strategic agents i.e., their rates of participation on the forum are known to the model. 
 
\begin{table}[h]
\centering
\tablefont{2.9mm}
\setlength{\extrarowheight}{4pt}
\begin{tabular}{|c|p{4.5in}|}
\hline
\textbf{Symbol} & \textbf{$\quad\quad\quad\quad\quad\quad\quad\quad\quad\quad\quad$Meaning} \\ 
\hline 
$m_l$ & Maximum number of answers considered by instructor per question to $Type_l$ students.\\ 
\hline 
$n_l$ & Number of $Type_l$ students in the course.\\
\hline 
$X(t)=(\mathcal{S}, Q)$ & The main OEF CTMC with finite state space $\mathcal{S}$ and generator matrix $Q$ (an example is shown in Figure~\ref{Figure_OEF2}(b)).\\
\hline 
$Q$ & Generator matrix of $X(t)$.\\ 
\hline 
$M$ & Maximum answers considered from a particular student per question posted on OEF.\\ 
\hline 
$\mu$ & Instructor arrival rate to the OEF.\\ 
\hline 
$\hat{q}$ & An upper bound on the entries in $Q$.\\ 
\hline 
$\Pi$ & Steady state probability vector of $X(t)$.\\ 
\hline 
$\pi^t$  & Transient probability vector of $X(t)$. \\ 
\hline 
$x = (x_1,\ldots,x_i,\ldots,x_n)$ & A state of $X(t)$.\\
\hline 
$\alpha_l$  & The cost per arrival of $Type_l$ student.\\ 
\hline 
$r^{l,i}(x)$ & Reward received by the student $i$ belonging to $Type_l$ in state $x$ of $X(t)$.\\ 
\hline 
$ R^{l,i}(x)$ & Net-reward received by a student $i$ belonging to $Type_l$ in a state $x$ of $X(t)$.\\ 
\hline 
$ c_i$ & Bias of the instructor towards answers from student $i$.\\ 
\hline 
$r^{I,i}(x)$ & Reward to the instructor if student $i$ gives $x_i$ answers in state $x: (x_1,\ldots,x_i,\ldots,x_n)$ of $X(t)$.\\ 
\hline 
$r^{I}(x)$ & Reward the instructor receives in some state $x: (x_1,\ldots,x_i,\ldots,x_n)$  of $X(t)$. \\ 
\hline 
$\delta^{\log \mu}$ & Discounting factor to the instructor. \\ 
\hline 
$\beta$ & Cost per arrival to the forum for the instructor. \\ 
\hline 
$R^{I}(x)$  & Net-reward received by the instructor on arriving once to the forum when $X(t)$ is in state $x$.\\ 
\hline 
$\pi^{0}$   & Initial distribution vector for the CTMC $X(t)$. \\ 
\hline 
$R_t^{l,i}$  & Expected transient net-reward received by a student $i$ of $Type_l$ at time $t$ in $X(t)$.\\ 
\hline 
$\pi^t(x)$  & Transient probability of being in state  x at time  t in $X(t)$.\\ 
\hline 
$R_t^{I}$  & Expected transient net-reward of the instructor at time $t$ in $X(t)$.\\ 
\hline 
$R_T^{l,i}$  & Expected transient aggregate net-reward of a student $i$ of $Type_l$ in $X(t)$.\\ 
\hline 
$R_T^{I} $  & Expected transient aggregate net-reward of the instructor $R_T^{I}$ over time $T$ in $X(t)$.\\ 
\hline 
$R_{\Pi}^{l,i}$  & Expected steady state net-reward for the student $i$ of $Type_l$ in $X(t)$.\\ 
\hline 
$R_{\Pi}^{I}$  & Expected steady state net-reward for instructor in $X(t)$.\\ 
\hline 
$q(x,z)$ & Entry in $Q$ whose row corresponds to state~$x$ and column corresponds to state~$z$.\\
\hline
\end{tabular}
\caption{Important notations related to OEF CTMC} \label{CTMCnotationtable1}
\end{table}  

We consider an online classroom setting with $n$ students and an instructor. We begin by proposing an instructor-driven approach to structure the activities in the OEFs (see Figure~\ref{instr-OEF}). Henceforth, we will assume the activities of the OEF follows as outlined in Figure~\ref{instr-OEF}. We assume that the arrivals of the $n$ students and the instructor to the OEF are independent Poisson processes with rate $\lambda_i$ for student $i$  ($\forall i \in \{1,\ldots,n\} $) and  rate $\mu$ for the instructor. Another modeling assumption is that only the instructor can post questions on to the OEF. Also, once the instructor arrives on the forum, the current question is closed, the students are rewarded based on the number of answers they have given to this question and a new question is posted. Hence, there is a single active question on the forum at any time which remains open until the next visit of the instructor. Closing a question here means not allowing to post further on that question and henceforth, not awarding for answering the closed question. Hence, the students will not find it beneficial to answer the closed questions and thus, it is assumed that they answer only the currently open question. An interesting 
future direction could be to model the scenario where questions may be posted by both the students and the instructor.

We capture the heterogeneity among students by allowing $L$ types of students in our OEF. Let $Type_l$ be a set containing the students of type $l$  ($\forall l \in \{1,\ldots,L\}$). 
Also, let $n_l$ 
be the number of students of type $l$ ($\forall l \in \{ 1,\ldots,L \}$) in the classroom  s.t. $\sum_{1 \leq l \leq L} n_l = n$. On arrival to the forum, a student answers the currently open question (if any) and hence, incurs a cost. Also, as the students belonging to the same type are similar, we assume that the cost of answering a question will be same for all students $i,j \in Type_l$ and this cost will be denoted by $\alpha_l$. The instructor has a budget $B$ per question and has to decide a suitable allocation of the budget among the  different students belonging to the $L$ types. Let $m_l$ denote the maximum number of answers \textit{ per question } for which the rewards will be given out, for each student $i \in Type_l$ s.t. $B=\sum_{ 1 \leq l \leq L} n_l m_l$. 
 
 \subsection{Description of the Model}\label{sectiondescriptionofthemodel}

We now model the activities of the OEF as a continuous time Markov chain (CTMC) $X(t) = (\mathcal{S}, Q)$,  where $\mathcal{S}$ is the set of states of the stochastic process $X(t)$ and $Q$ is the generator matrix\mbox{\protect\footnotemark[4]}. 
\footnotetext[4]{
For readers not familiar with CTMC, $q_{ij}$ i.e. the $(i,j)$-th entry of the generator matrix $Q$ governs the rate of transition from state~$i$ to state~$j$. To be precise, let $P_{ij}(t) = (X(t) = j | X(0) = i)$, then $d/dt (P_{ij}(t)) |_{t=0} = q_{ij}$}
We summarize important notations of this section in Table~\ref{CTMCnotationtable1}. We enumerate the students in the class as $1, 2, \ldots, n$ and define each state $x \in \mathcal{S}$ as: $x = (x_1,\ldots,x_n)$, where each $x_i$ corresponds to the number of answers received from the $i^{th}$ student for the current question. If $X(t)$ is in some state $x: (x_1,\ldots,x_i,\ldots,x_n)$ and student $i$ gives an answer on the forum then $X(t)$ goes into a state $y: (x_1,\ldots,x_i + 1,\ldots,x_n)$, thus capturing each arrival and reflecting the current forum status through its state space. When the instructor comes on the forum, the current question is closed and a new question is started, thus taking the CTMC to state $(0,\ldots,0)$. Hence, at any point of time, there is a single active question on the forum which when initially posted by the instructor had no answers from any student. Thus, the CTMC will be capturing arrivals of answers for the single question that is active in the forum at a particular instant of time. 

\begin{figure*}[t]
\begin{tabular}{cc}
\begin{minipage}{16 cm} 
\centering 	\includegraphics[scale = .3]{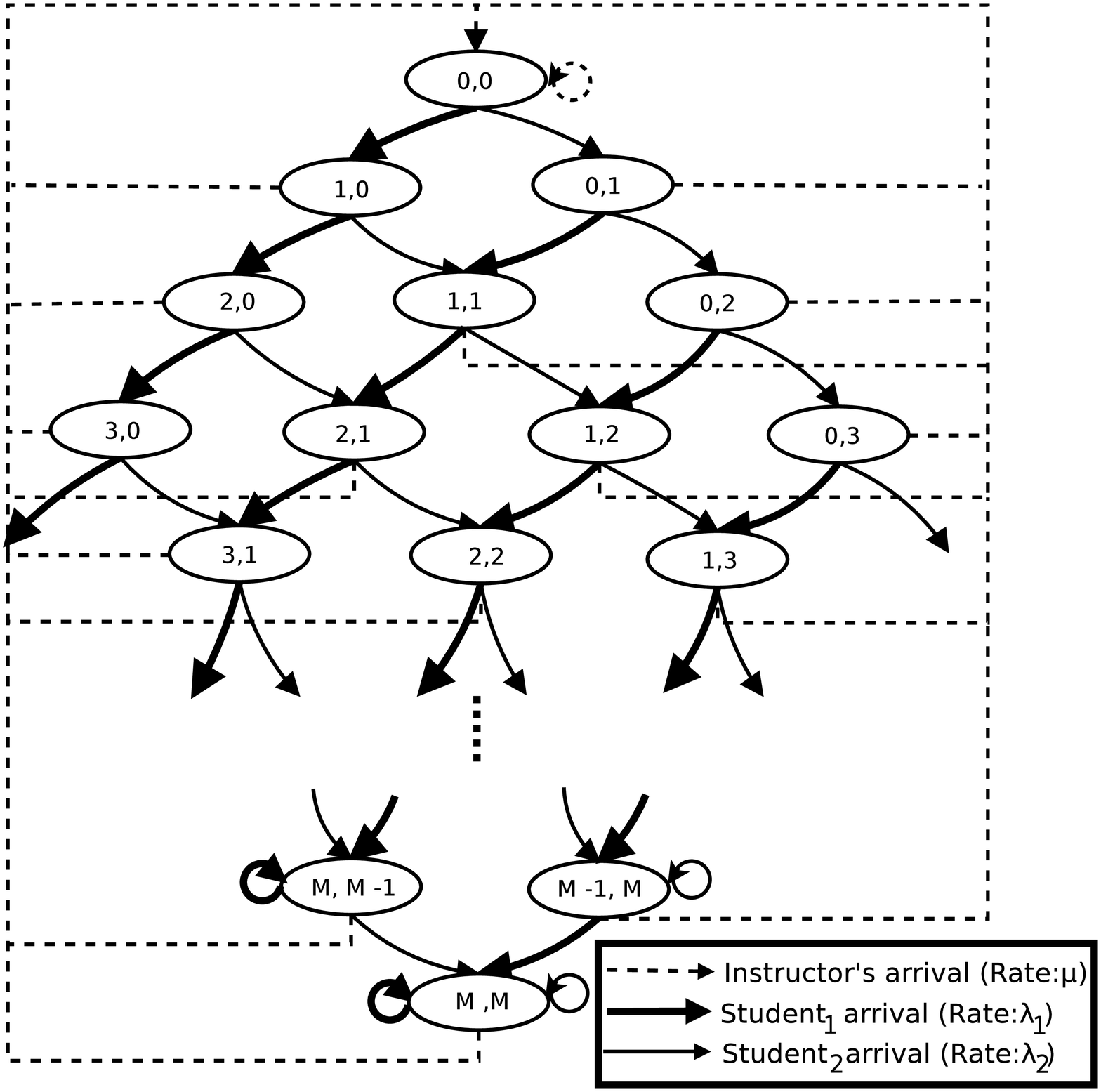}%
\end{minipage}
\end{tabular}
\caption{Representative CTMC in our model with 2 students: Each state is capturing the number of answers received from both the students, for a particular question.}\label{Figure_OEF3}
\end{figure*}

The CTMC being considered is actually an infinite state CTMC but since the course will be offered for a limited time and that the instructor and the students are arriving at some reasonable rates allow us to approximate a reasonably large upper bound $M$ on the number of answers that a single student can post on the forum  for each question.  This transforms our infinite state CTMC to a finite state CTMC where the `last' state will be $(M, \ldots, M)$, i.e. from this state, the arrival of only the instructor will be captured and the only transition possible from this state will be to the state $(0, \ldots, 0)$. So, the CTMC state space $\mathcal{S}$ will be defined as $\mathcal{S} = \{(x_1,\ldots,x_i,\ldots,x_n)| x_i \in \{0,1,\ldots,M\} \}$. We depict a sample CTMC in Figure~\ref{Figure_OEF3} where there are only two students in the OEF.  

Let $\lambda_i$ denote the rate of arrival of student~$i$ to the OEF. The instructor closes the active question on each arrival and starts a new one, so, because of the instructor's arrival, each state is transitioning to state $(0,\ldots,0)$ with rate $\mu$. Note that at any instant of time there can only be a single arrival and then the CTMC will evolve into the next state. Thus, for any two states $x:(x_1,\ldots, x_{i}, \ldots, x_{n})$,  $y:(y_1,\ldots, y_{i}, \ldots, y_{n})$ $\in \mathcal{S}$, the entry in the generator matrix $Q$ for the CTMC $X(t)$ will be defined as:
\begin{align}
\label{generator_matrix}
Q(x,y) &= \begin{cases}
\lambda_j & \text{if } \sum\limits_{i} |x_i - y_i| =  1  \text{ and } \exists j: y_j-x_j = 1 \text{ (Case i)}\\
 	  \mu       & \text{if } y = (0,\ldots,0) \neq x \text{ (Case ii)}\\
 	   \omega & \text{if } x = y 
 	   \text{ (Case iii)}\\
	  0 &o/w \text{ (Case iv)}.  
      \end{cases} \\
\nonumber \text{ where } \omega &= \sum \limits_{y' \in \mathcal{S} \setminus \{x \}} {-Q(x,y')}.        
\end{align}

Note that due to our finiteness assumption the number of entries in the $Q$ matrix and their values will be finite and
thus we can say that each entry in the $Q$ matrix is bounded by a large quantity $\widehat{q}< \infty$. 
Also, let $\Pi$ and $\pi^t$ respectively be the steady state (if it exists) and transient state probability vectors of $X(t)$.

\subsection{Reward Functions on the CTMC}
\label{section_rewards}
Each state $x \in X(t)$ indicates the number of arrivals on the OEF by $n$ students. The arrivals of  the students and the instructor have some costs and rewards associated with them.  In this section, we define a net-reward associated with each state in $\mathcal{S}$ for the students as well as the instructor.

The maximum number of answers for which the rewards will be given out, per question, is fixed for each $Type_l$ student as $m_l$. So, if a $Type_l$ student answers $k$ times, then the number of answers that are `rewardable' are $min(k,m_l)$. An additional factor which determines the quantity of rewards being given is that if the instructor might not be willing to give out much reward per answer if she is coming too often and answering the questions herself. So, we introduce $\delta \in \left(0,1 \right)$ as the willingness of the instructor to reward the students and $\delta^{h(\mu)}$ gives the actual discounting factor applied by the instructor for `rewardable' answer, where $h(\mu)$ is an increasing function of $\mu$. For our work, we consider $h(\mu) = \log \mu$.  So, if the instructor comes online when the Markov chain is in some state $x = (x_1,\ldots,x_i,\ldots,x_n)$, then the following will be the reward received by the student $i \in Type_l$:
 \begin{align} \label{reward_mainCTMC} r^{l,i}(x) = \begin{cases} 
 	  x_i \delta^{\log \mu}  & if\textbf{ } x_i \leq m_l, \\
	  m_l \delta^{\log\mu} &o/w. \\
      \end{cases}.
 \end{align}
      
The cost per arrival is assumed to be some constant $\alpha_l$ for each answer from a $Type_l$ student. So, if $X(t)$  is in some state $x: (x_1,\ldots,x_i,\ldots,x_n)$, then the cost incurred by a student $i$ belonging to $Type_l$ is $\alpha_l x_i$. Thus the net-reward received by a student $i$ belonging to $Type_l$ in a state $x$ will be given by
\begin{align} \label{net_reward_mainCTMC}
 R^{l,i}(x) = r^{l,i}(x) - \alpha_l x_i.
\end{align}
 
The instructor values each answer on the forum arriving from all students but can unequally value the contributions from different students. Instructor's total reward in a state $x$ from student arrivals will thus be a weighted sum over all students, of the reward she receives on arrival from a single student. We use $c_i$ to give the bias of the instructor towards answers from student $i$ such that  $\sum_{i=1}^{n} c_i =1$ and $ 0 \leq c_i\leq 1$ $ \forall i \in \{1,\ldots,n\} $.  As all students  belonging to the same type are assumed to be similar, so the instructor will value their arrivals equally $c_i =c_j (= c_l)$ $\forall i, j \in Type_l$. The instructor intends to maximize the student participation and hence the reward to the instructor if student $i$ gives $x_i$ answers will be given by $r^{I,i}(x) = c_i x_i \delta^{\log\mu}$, where $\delta^{\log \mu}$ is the discounting factor to the instructor. Thus, the reward the instructor receives in some state $x: (x_1,\ldots,x_i,\ldots,x_n)$ 
will be $r^{I}(x) = \sum_{i=1}^{n} r^{I,i}(x) $.

Instructor will also incur a cost per arrival to the forum, let us denote this by $\beta$. Then the net-reward received by the instructor on arriving once to the forum when $X(t)$ is in state $x$ will be: 
 \begin{align} 
  \nonumber R^{I}(x) &= \sum\limits_{i=1}^{n} \left( c_i x_i  \delta^{\log\mu} \right) - \beta = \sum\limits_{i=1}^{n} c_i ( x_i \delta^{\log\mu} - \beta )  (\text{as }\sum_{i=1}^{n} c_i = 1 ) 
\end{align}

The initial state of the CTMC will be the state $(0,\ldots,0)$ as the number of answers on the forum from each student will be zero. Thus, the initial distribution $\pi^{0}$ for the CTMC $X(t)$ will be defined as $\pi^0(0,\ldots,0) = 1$ and $\pi^0(x) = 0 \text{ } \forall x \in \mathcal{S}\setminus \{(0,\ldots,0) \}$. 
Now we shall analyze the expected net-rewards of the students and the instructor in $X(t)$. The expected transient net-reward $R_t^{l,i}$ received by a student $i$ of $Type_l$ at time $t$ will be given by :
\begin{align}
\nonumber & R_t^{l.i}= \sum \limits_{x \in \mathcal{S}} (R^{l,i}(x))\pi^t(x)\\
\nonumber \text{where, } R^{l,i}(x)  : & \text{ Net-reward to student } i \text{ of } Type_l \text{ if } X(t) \text{ is in} \text{ state } x,\\
\nonumber \pi^t(x)  :  & \text{ Transient probability of } X(t) \text{ being in state } x \text{ at} \text{ time } t.
\end{align}

We can define the expected transient net-reward of the instructor at time $t$ by:
\begin{align}
\nonumber & R_t^{I}= \sum \limits _{x \in \mathcal{S}} (R^{I}(x))\pi^t(x)\\
\nonumber \text{where, } R^{I}(x) \text{ : }  &   \text{Net reward to the instructor } \text{ if she arrives when}\\
			   \nonumber &			  X(t) \text{ is in state } x
\end{align}

The OEF for an online classroom will be functioning for a certain period of time $T$ (duration of the course) and hence, we need calculate the expected transient aggregate net-rewards over time $T$  and  the expected steady-state net-rewards (if the steady state probability vector of the CTMC exists) collected by the students and the instructor. The expected transient aggregate net-reward $R_T^{l,i}$ collected by a student $i$ of $Type_l$ and the expected transient aggregate net-reward collected by the instructor $R_T^{I}$ over time $T$ will be given by:
\begin{align}
\nonumber R_T^{l,i} = \int\limits_{t=0}^{T} R_t^{l,i} \text{ dt}, \quad R_T^{I} = \int\limits_{t=0}^{T} R_t^{I} \text{ dt}
\end{align}

Similarly, the expected steady state net-reward for the students $R_{\Pi}^{l,i}$ and the instructor  $R_{\Pi}^{I}$ will be given by:
\begin{align}
\nonumber   R_{\Pi}^{l,i} &= \sum \limits_{x \in \mathcal{S}} (R^{l,i}(x))\Pi(x), \quad
R_{\Pi}^{I} =  \sum \limits_{x \in \mathcal{S}} R^{I} (x)\Pi(x)\\
 \nonumber \text{ where, } & \Pi(x) \text{ : }   \text{Steady-state probability of } X(t) \text{ being in state } x.
\end{align}

As the state space of $X(t)$ is very large ($|\mathcal{S}|= (M+1)^n$), thus the calculation of the steady-state and transient probabilities of the CTMC $X(t)$ will be very difficult. Thus, computing the expected net-rewards will be very difficult for this complex CTMC and so we shall now use the approach of lumping a CTMC to simplify the analysis of the expected net-rewards for this CTMC. 

\newcommand{\ovs}{\overline{S}}
\section{Simplifying the OEF CTMC}

\label{Section_lumpaing}
 In this section, we introduce the concept of lumpability of Markov chains, in the context of discrete time as well as continuous time. We highlight some key results related to lumpability which we will use in reducing the CTMC for the OEF (shown in Figure~\ref{Figure_OEF2}(b)) into simpler student-specific lumped CTMCs. We summarize important notations of this section in Table~\ref{CTMCnotationtable2}.
 
 \begin{table}[!htb]
\centering
\tablefont{2.9mm}
\setlength{\extrarowheight}{4.5pt}
\begin{tabular}{|c|p{4.5in}|}
\hline
\textbf{Symbol} & \textbf{$\quad\quad\quad\quad\quad\quad\quad\quad\quad\quad\quad$Meaning} \\ 
\hline 
$X(t)=(\mathcal{S}, Q)$ & The main OEF CTMC with finite state space $\mathcal{S}$ and generator matrix $Q$ (an example is shown in Figure~\ref{Figure_OEF2}(b)).\\
\hline 
$Y_k=(\mathcal{S},P)$  & DTMC with finite state-space $\mathcal{S}$ and probability transition matrix $P$.\\ 
\hline 
$I$ & Identity matrix.\\ 
\hline 
$P^k(x)$  & Probability that the DTMC $(\mathcal{S},P)$ will be at state $x$ after $k$ steps.\\
\hline 
$\overline{S}=\{ {\overline{S}^j}|j \in \{ 1,\ldots, w\} \}$ & Partition of the state space $\mathcal{S}$ i.e. $\mathop{\cup}_{j=1}^w {\overline{S}^j} = \overline{S}$,  $\overline{S}^i \neq \emptyset$, and $\overline{S}^i \cap \overline{S}^j = \emptyset$ for $i\neq j$.\\ 
\hline 
$\overline{Y}_k = (\ovs,\overline{\pi})$ & DTMC obtained after applying lumping to DTMC $Y_k=(\mathcal{S},P)$.\\ 
\hline 
${\overline{S}_i}$ & Partition on the state space $\mathcal{S}$ of $X(t)$ w.r.t. a student $i$ in the OEF i.e., ${\overline{S}_i} = \{ {\overline{S}^a_i} | a \in \{ 0,1,\ldots,M \} \}$.\\ 
\hline 
${\overline{S}^a_i} $ & This is the set $\{ (x_i, x_{-i}) \in \mathcal{S} | x_i = a \}$.\\ 
\hline 
${\overline{X}_i}(t) = ({\overline{S}_i},{\overline{Q}_i})$ & The quotient (lumped) Markov chain we get on lumping the CTMC $X(t)$ w.r.t. partition ${\overline{S}_i}$  ($ i \in \{ 1,2,\ldots,n \}$) (an example is shown in Figure~\ref{ctmcLumped}). \\ 
\hline 
${\overline{\Pi}_i}$  & Steady state probability vector for the lumped CTMC ${\overline{X}_i}(t)$. \\ 
\hline 
${\overline{\pi}_i^t}$  & Transient state probability vector for the lumped CTMC ${\overline{X}_i}(t)$.\\
\hline 
${\overline{ r}^{l,i} }(\overline{x})$  & The reward received by a student $i$ of $Type_l$ in a state $\overline{x}$ of the ${\overline{X}_i}(t)$ (the lumped CTMC corresponding to student $i$).\\ 
\hline 
$\overline{R}^{l,i}(\overline{x})$ & Net-reward to a student $i$ of  $Type_l$  in a state $\overline{x}$ for the lumped CTMC ${\overline{X}_i}(t)$.\\ 
\hline 
$\overline{ R}_t^{l,i} $ & Expected net-reward at time $t$ to student $i$ of $Type_l$ in the lumped-CTMC ${\overline{X}_i}(t)$.\\ 
\hline 
$\overline{\pi}^t_i(\overline{x})$ & Transient probability of  being in state $\overline{x}$  at  time t in $\overline{X}_i(t)$.\\ 
\hline 
$\overline{R}_t^{I,i} $ & Expected transient net-reward to the instructor if she arrives on the forum  at time $t$, from the arrival of  student $i$ of $Type_l$ in the lumped-CTMC ${\overline{X}_i}(t)$.\\ 
\hline 
$\overline{ R}_T^{l,i} $ & Expected transient aggregate net-rewards over time $T$ for the student $i$ of $Type_l$ w.r.t. the CTMC ${\overline{X}_i}(t)$ . \\ 
\hline 
$\overline{ R}_T^{I,i} $ & Expected transient aggregate net-rewards over time $T$ for the instructor w.r.t. the CTMC ${\overline{X}_i}(t)$ . \\ 
\hline 
$\overline{ R}_{\overline{ \Pi}}^{l,i} $ & Expected steady state net-rewards over time $T$ for the student $i$ of $Type_l$ w.r.t. the CTMC ${\overline{X}_i}(t)$. \\ 
\hline 
$\overline{ R}_{\overline{ \Pi}}^{I,i}  $ & Expected steady state net-rewards over time $T$ for the instructor w.r.t. the CTMC ${\overline{X}_i}(t)$. \\ 
\hline 
$\overline{R}_T^{I}$   & Total transient aggregate net-reward received by the instructor from the arrival of all the students on the OEF. \\ 
\hline 
$\overline{R}_{\Pi}^{I}$  & Total steady state net-reward received by the instructor from the arrival of all the students on the OEF.\\ 
\hline 
${\overline{\pi}^0_i}$   & Initial state distribution for the CTMC ${\overline{X}_i}(t)$.\\ 
\hline 
${\overline{\Pi}_i}( \overline{x})$  & Probability that the CTMC ${\overline{X}_i}(t)$ (corresponding to student $i$) is in state $\overline{x}$, when ${\overline{X}_i}(t)$ has reached a steady state.\\ 
\hline 
${\overline{\pi}_i^t}(\overline{x})$ & Transient state probability of ${\overline{X}_i}(t)$ being in the state $\overline{x}$ at a time instant $t$.\\ 
\hline
\end{tabular}
\caption{Important notations related to lumped CTMC} \label{CTMCnotationtable2}
\end{table} 

\subsection{Lumpability of Markov Chains - Important Results from Literature}

\subsubsection{Lumpability of DTMC}
Consider a DTMC $Y_k$ with finite state-space $\mathcal{S}$ and probability transition matrix $P$. Let $\overline{S}=\{ {\overline{S}^j}|j \in \{ 1,\ldots, w\} \}$ be a partition of the state space $\mathcal{S}$ i.e. $\mathop{\cup}_{j=1}^w {\overline{S}^j} = \overline{S}$,  $\overline{S}^i \neq \emptyset$, and $\overline{S}^i \cap \overline{S}^j = \emptyset$ for $i\neq j$. Let us define the stochastic process $Z_k$ as follows
\begin{align*}
Z_k = j \quad \mathrm{if} \; Y_k \in \overline{S}^j \quad \mathrm{for}\;j \in \{1,\ldots,w\}
\end{align*}
Then $Z_k$ is called the lumped stochastic process of $Y_k$ with respect to the partition $\overline{S}$.

\begin{defi}\citep{sumita1989lumpability}
\label{DTMC_lumpable}
A finite DTMC $Y_k$ on $\mathcal{S}$ is said to be lumpable with respect to the partition $\ovs$ if the lumped process $Z_k$ is also a DTMC for any initial probability vector.
\end{defi}

It may be difficult to check Definition \ref{DTMC_lumpable} for all initial probability vectors. Kemeny~et.al~\citep{KEMSNELL60} provide the following necessary and sufficient condition for lumpability which is easy to verify. For notational ease, we use $p(x,A) = \sum_{z \in A} p(x,z)$ where $A \subseteq \ovs$.

\begin{theorem}\label{define_lumping_dtmc}\citep{KEMSNELL60}
\textbf{(i)} A finite DTMC $Y_k = (\mathcal{S},P)$ is lumpable with respect to the partition $\ovs$ iff for any two blocks $\ovs^a, \ovs^b \in \ovs$ and for all $x,y \in \ovs^a$ we have $p(x,\ovs^b) = p(y,\ovs^b)$. \textbf{(ii)} In such a case the lumped Markov chain $\overline{Y}_k = (\ovs,\overline{P})$ is determined by $\overline{P}(\ovs^a, \ovs^b) = p(x,\ovs^b)$ for some $x \in \ovs^a$.
\end{theorem}

\subsubsection{Lumpability of CTMC}

Consider a finite CTMC $X(t)$ with state space $\mathcal{S}$ and generator matrix $Q$. Analogous to DTMC, one can construct the lumped process of $X(t)$ with respect to a partition $\ovs$ of $\mathcal{S}$ and define the lumpability of
$X(t)$ with respect to $\ovs$. However, the lumpability of CTMC can be characterized in terms of a related DTMC as follows.
 
\begin{theorem}
\label{dtmc_underlying_and_lumping_other_paper_theorem} (\citep{sumita1989lumpability},\citep{buchholz:94})
Consider a CTMC X(t)= ($\mathcal{S},Q$). If the transition rates $Q(x,y)$ are bounded by some positive constant $\widehat{q} < \infty$, then there exists a DTMC $Y_k$ with transition matrix $P = Q/\widehat{q} + I$ such that
\begin{itemize}
 \item the distribution vector $\pi(t)$ of the CTMC at time $t$ starting with $\pi(0)$ at time $0$ is given by 
 \begin{equation}\label{CTMCtoDTMC}
 \pi(t) = \sum_{k=0}^{\infty} \exp(-\widehat{q} t)\left[ (\widehat{q} t)^k / k! \right] \pi^k
 \end{equation} 
 where $\pi^k$ is the distribution vector of $Y_k$ after $k$ jumps.
 \item the stationary probability vectors of $X(t)$ and $Y_k$ are identical.
 \item $X(t)$ is lumpable with respect to $\ovs$ iff $Y_k$ is lumpable with respect to $\ovs$.
 \end{itemize}
\end{theorem}

 Equation \ref{CTMCtoDTMC} states that the probability of being in a state $x \in \mathcal{S}$ at time $t$ for the CTMC $X(t)$ is equal to the sum over all possible number of jumps $k$ of the probability that exactly $k$ jumps occur in $t$ time \mbox{\protect\footnotemark[8]} and that after $k$ jumps the underlying DTMC $Y_k$ will be in state $x$.
For ease of notation, we use $q(x,A) = \sum_{z \in A} q(x,z)$ where $A \subseteq \ovs$. An immediate corollary of Theorem~\ref{dtmc_underlying_and_lumping_other_paper_theorem} is the following :
\footnotetext[8]{$exp(-\hat{q}t) ((\hat{q}t)^k / k!)$ is the probability that a Poisson process with rate $\hat{q}$ makes exactly $k$ jumps in time~$t$} 
\begin{corollary}(\citep{sumita1989lumpability},\citep{derisavi2003optimal},\citep{buchholz:94})
\label{define_lumping}
\textbf{(i)} A CTMC $X(t)$=($\mathcal{S},Q$)  is lumpable w.r.t. a partition $\overline{S}$ of $ \mathcal{S}$ iff for any two blocks ${\overline{S}^a}, {\overline{S}^b} \in \overline{S}$ and for every $x,y \in {\overline{S}^a}$ we have $q(x,{\overline{S}^b}) = q(y,{\overline{S}^b})$. 
\textbf{(ii)} In such a case, the lumped Markov chain $\overline{X}(t) = (\overline{S},\overline{Q})$ is determined by $\overline{Q}({\overline{S}^a},{\overline{S}^b}) = q(x,{\overline{S}^b})$  for some $x \in \overline{S}^a$.\end{corollary}

\subsubsection{Transient and Steady-State Probabilities of the Lumped Process}
For a DTMC $(S,P)$, let us define $P^k(x)$ \footnote{We omit the initial distribution $P^0(\cdot)$ in $P^k(\cdot)$, since this will be clear from the context.} to be the probability that the Markov chain will be at state $x$ after $k$ steps.
The following theorem relates the transient probability of the lumped Markov process with the transient probability of the original Markov chain. 

\begin{theorem}
\label{transient_state_other_paper_theorem} (\citep{sumita1989lumpability}, (Theorem~$5$ in \citep{buchholz:94})) Let the finite state DTMC $Y_k = (S,\pi)$ be lumpable w.r.t. the partition $\ovs$ of $S$ and let $\overline{Y}_k = (\ovs,\overline{\pi})$ be the lumped DTMC. Then  ${\overline{\pi}^{k}}(\overline{S}^a) = \sum_{y \in \overline{S}^a} \pi^{k}(y)$ $\forall \overline{S}^a \in \overline{S}$, for ${\overline{\pi}^0}(\overline{S}^a) = \sum_{y \in \overline{S}^a} \pi^0(y)$ $\forall \overline{S}^a \in \overline{S}$.
\end{theorem}

Therefore, the probability that the lumped DTMC is in a block after $k$ steps is equal to the sum of $k$-step transition probabilities of the states belonging to the same block in the original DTMC.
The next theorem provides a sufficient condition for the lumped CTMC to have a stationary distribution and connects this distribution with the stationary distribution of the original CTMC.

\begin{theorem}
\label{steady_state_other_paper_theorem} (\citep{Tian2006b}, $\S 2.12 $) Let X(t) be an {aperiodic}, irreducible CTMC with stationary distribution $\Pi$. If it is lumpable w.r.t. a partition of the state space, then the lumped chain also has a stationary distribution $\overline{\Pi}$ whose components can be obtained from $\Pi$ by adding corresponding components in the same block of the partition.
\end{theorem}
					
\subsection{Lumpability of the OEF CTMC}

 \begin{figure}[h]
 \begin{minipage}{16cm}
 \begin{center}
  \includegraphics[scale = .3]{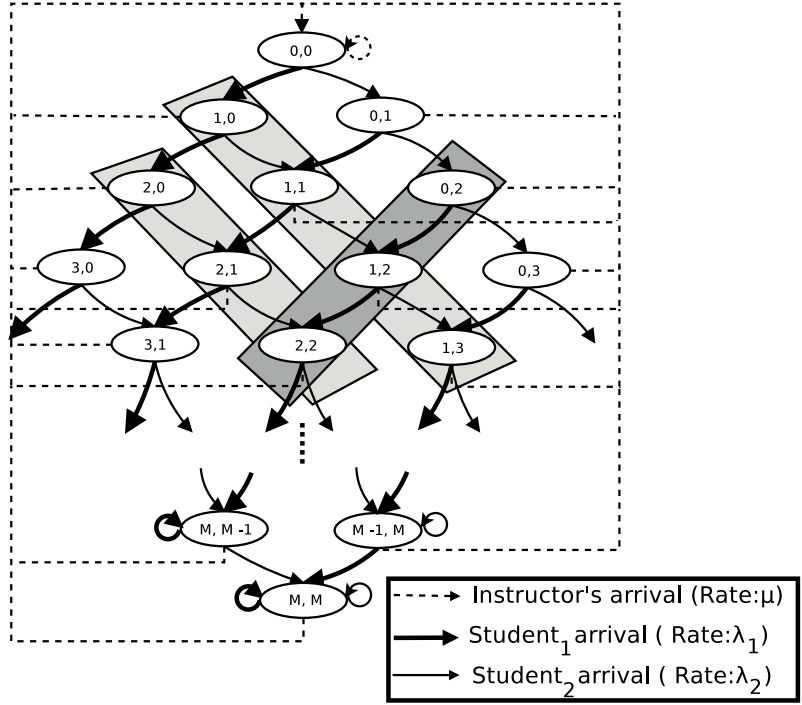}
 \end{center}
 \caption{ Light-gray and dark-gray quadrilaterals in the figure depict two partitions of the state space that can be defined on the example CTMC given in Figure~\ref{Figure_OEF3}. Each light-gray quadrilateral contains the states where the number of arrivals of student 1 are same.  Each dark-gray quadrilateral contains the states where the number of arrivals of student 2 are equal. \label{ctmcFull}}
 \end{minipage} 
 \end{figure}

Recall the CTMC $X(t) = (\mathcal{S}, Q)$ where $\mathcal{S} = \{(x_1,\ldots,x_i,\ldots,x_n)| x_i \in \{0,1,\ldots,M\} \}$ and $Q$ is defined as in  Equation~\ref{generator_matrix}. This CTMC represents the OEF but  is very complex to analyze because of its large state space ($|\mathcal{S}|= (M+1)^n$). So, we exploit the basic structure of the problem and come up with lumped-CTMCs with smaller state-space which will be easier to analyze. An important characteristic to note about our problem is that the rewards and costs incurred for each student in a particular state are independent of the other students and dependent only on their own strategy and the instructor's strategy. This is possible because of the property of open-ended questions that they have more than one possible answer and thus, even if an open-ended question has already been answered by a few students, still a new student can find it beneficial to give a new answer and potentially earn a good reward.

 Hence, instead of analyzing the complex CTMC, which keeps track of arrivals of all students (of all types), we show that we can analyze $n$ independent player-specific CTMCs (with $M+1$ states each) so that each of the player-specific CTMC keeps track of arrivals from only a particular student. This is possible by applying the lumping process on the original-CTMC which we describe next.

We first define a partition ${\overline{S}_i}$ on the state space $\mathcal{S}$ of $X(t)$ w.r.t. a student $i$ in the OEF as ${\overline{S}_i} = \{ {\overline{S}^a_i} | a \in \{ 0,1,\ldots,M \} \}$ where each block ${\overline{S}^a_i}$  of the partition ${\overline{S}_i}$ is  defined as ${\overline{S}^a_i} = \{ (x_i, x_{-i}) \in \mathcal{S} | x_i = a \} $. 

\begin{figure}[!htb]
\hspace{-1cm}
\begin{tabular}{|c||c|}
\hline
\begin{minipage}{8 cm}
 \vspace{0.1in}
  \includegraphics[scale = .32]{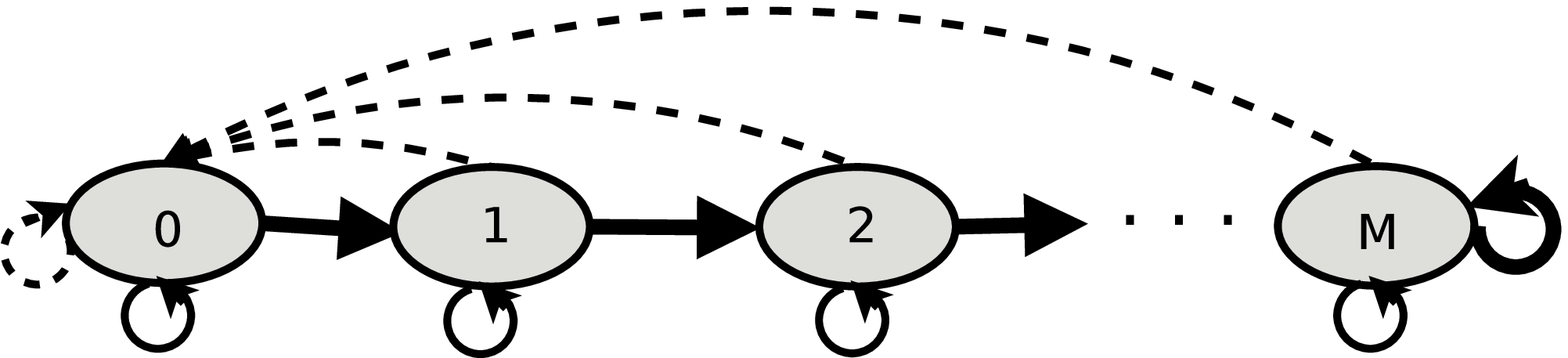}
\\
\centering(a) 
\end{minipage}&
\begin{minipage}{8 cm}
\vspace{0.1in}
  \includegraphics[scale = .32]{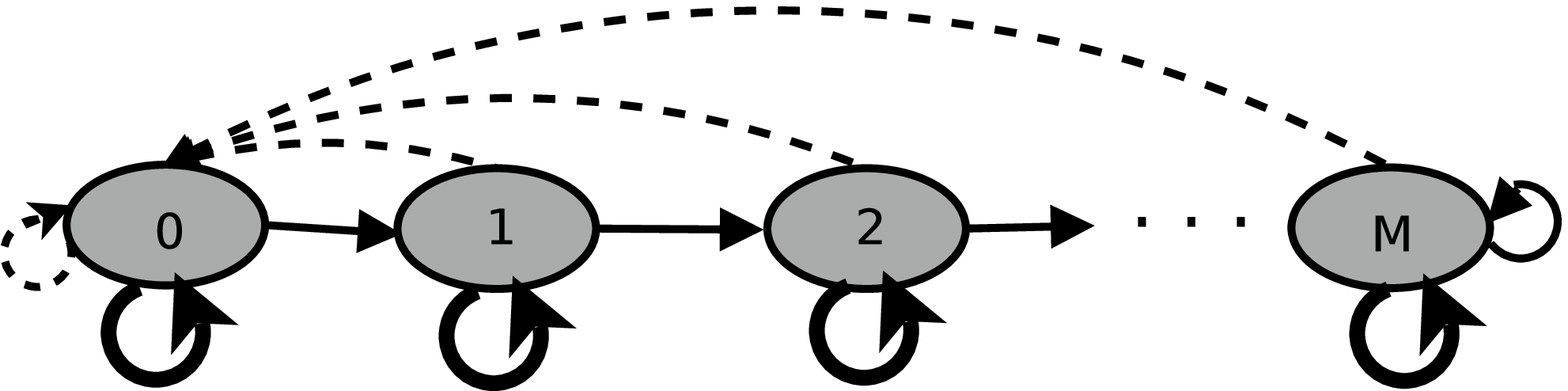}
\\
\centering(b) 
\end{minipage}\\
\hline
\end{tabular}
\caption{\textbf{a)} Lumped-$\text{Student}_1$ CTMC : Each light-gray state indicates the aggregation of all the states in the Figure~\ref{ctmcFull} CTMC which are enclosed by the light-gray quadrilateral. \textbf{b)} Lumped-$\text{Student}_2$ CTMC : Each dark-gray state indicates the aggregation of all the states in the Figure~\ref{ctmcFull} CTMC which are enclosed by the dark-gray quadrilateral.}
\label{ctmcLumped}
\end{figure}

 We lump together all the states in one light gray quadrilateral (having same number of answers from Student 1 in Figure~\ref{ctmcFull} as one state of the lumped chain in Figure~\ref{ctmcLumped}~(a). For example, the state $1$ in Figure~\ref{ctmcLumped}~(a) basically denotes the set of states $\{ (1,0), (1,1),\ldots, (1,M) \}$ from Figure~\ref{ctmcFull} and state $2$ in Figure~\ref{ctmcLumped}~(a) denotes the set of states $\{ (2,0), (2,1),\ldots, (2,M) \}$ from Figure~\ref{ctmcFull}, and so on. Similarly, all the states in the dark gray quadrilateral (having same number of answers from Student 2) are represented as one state of the lumped chain in Figure~\ref{ctmcLumped}~(b). 
We now present a result which states that, given student $i$, $X(t)$ is lumpable w.r.t. partition ${\overline{S}_i}$. Detailed proofs of the results presented in this section are provided in Appendix~A. 

\begin{theorem}\label{lumpable_theorem}\mbox{\protect\footnotemark[3]}
\textbf{(i)} $X(t) = (\mathcal{S},Q)$ is lumpable w.r.t. partition ${\overline{S}_i} = \{ {\overline{S}^a_i} | a \in \{ 0,1,\ldots,M \} \} $. 

\textbf{(ii)} The quotient (lumped) Markov chain ${\overline{X}_i}(t) = ({\overline{S}_i},{\overline{Q}_i})$ that we get on lumping the CTMC $X(t)$ w.r.t. partition ${\overline{S}_i}$  ($ i \in \{ 1,2,\ldots,n \}$) is given as :

\begin{align}
 \nonumber {\overline{Q}_i}({\overline{S}^a_i},{\overline{S}^b_i}) =&
			   \begin{cases} 
 	  				\lambda_i   & b=a+1, \\
					\mu   	    & b=0\neq a,\\
					\omega	    & b=a,\\
					0 	    & o/w.
			   \end{cases} \\
\nonumber  \text{where, } \omega = -\sum\limits_{c \in D \setminus \{ a \} } & {\overline{Q}_i}({\overline{S}^a_i},{\overline{S}^c_i}),D = \{0,1,\ldots,M \}.
\end{align}
\end{theorem}

\footnotetext[3]{
In fact a general form of Theorem~\ref{lumpable_theorem} is true . Consider a group of students G and define $\bar{S}_G = \{ \bar{S}_G^a : a \in \{0,1,...,M\}^{G} \}$, where $\bar{S}_G^a = \{(x_G,x_{-G}) : x_G = a\}, \forall a \in \{0,1,...,M\}^{G} $  i.e. we lump together all the states where the entries for the students in group $G$ is fixed at $a$. Then we can prove
that the CTMC $X(t) = (S,Q)$ is lumpable w.r.t. the partition $\bar{S}_G$. Moreover, the lumped Markov chain $\bar{X}_t = (\bar{S}_G, \bar{Q}_G)$ is given as
$\bar{Q}_G(\bar{S}_G^a, \bar{S}_G^b) = Q(a,b)$ where $Q(a,b)$ is defined in Equation~\ref{generator_matrix}.
We omit the proof of this general statement since the proof is quite similar to the proof of Theorem~\ref{lumpable_theorem}  and we will not be using the general lumpability result in our paper.}

So, from Theorem~\ref{lumpable_theorem}, we now have $n$ lumped-CTMCs ${\overline{X}_i}(t)=({\overline{S}_i},{\overline{Q}_i})$, $ 1 \leq i \leq n$ (See Figure~\ref{ctmcLumped}) with  finite state space ${\overline{S}_i} = \{\overline{x} | \overline{x} \in \{0,1,\ldots,M \} \}$ . Each block ${\overline{S}_i^{\overline{x}}} \in {\overline{S}_i}$ has been represented as a state $\overline{x}$ of the lumped-CTMC ${\overline{X}_i}(t)$ i.e. a state $\overline{x}$ of the lumped-CTMC ${\overline{X}_i}(t)$ is  representative of the block ${\overline{S}_i^{\overline{x}}}$ which contains all the states in $\mathcal{S}$ in which the student $i$ arrives $\overline{x}$ number of times. Each state $\overline{x} \in {\overline{S}_i} $ thus simply means how many answers have been received from student $i$. Now, using Theorem~\ref{lumpable_theorem}~(ii) the generator matrix ${\overline{Q}_i}$ for ${\overline{X}_i}(t)$  can be re-written as :
 \begin{align}
 \label{final_generator_matrix}
 {\overline{Q}_i}(\overline{x},\overline{y}) =
			   \begin{cases} 
 	  				\lambda_i   																		&\overline{y}=\overline{x}+1, \\
					\mu   																				&\overline{y}=0\neq \overline{x},\\
					-\sum\limits_{\overline{z} \in {\overline{S}_i} \setminus \{ \overline{x} \} } {\overline{Q}_i}(\overline{x},\overline{z}) 		& \overline{y}=\overline{x},\\
					0 																					& o/w.
      		   \end{cases}
 \end{align}
 
The initial state distribution ${\overline{\pi}^0_i}$ for the CTMC ${\overline{X}_i}(t)$ is defined  as  ${\overline{\pi}^0_i}(0) =1$ and  ${\overline{\pi}^0_i}(\overline{x}) =0 \forall \overline{x} \in {\overline{S}_i} \setminus \{ 0\}$ as, initially, no answer would be posted on the forum by any student. Also,  let ${\overline{\Pi}_i}$ and ${\overline{\pi}_i^t}$ respectively denote the steady state probability vector (if it exists) and the transient state probability vector for the lumped CTMC ${\overline{X}_i}(t)$.

We shall now define the notions of rewards, net-rewards, and expected net-rewards corresponding to the $i^{th}$ lumped-CTMC, for both the student $i$ and the instructor. The definitions will be very similar to the ones discussed for the original-CTMC (as given in Section ~\ref{section_rewards}).  The reward ${\overline{ r}^{l,i} }(\overline{x})$ received by a student $i$ of $Type_l$ when the instructor visits the forum and finds ${\overline{X}_i}(t)$ (the lumped CTMC corresponding to student $i$) in a state $\overline{x}$ will now be defined as: \begin{align} \label{reward_lumpedCTMC}
{\overline{ r}^{l,i} }(\overline{x}) = \begin{cases} 
 	  \overline{x} \delta^{\log\mu}  & if\textbf{ } \overline{x} \leq m_l, \\
	  m_l \delta^{\log\mu} &o/w. \\
      \end{cases}
      \end{align}
      
As discussed before (in Section ~\ref{section_rewards}), a $Type_l$ student will also incur a cost of $\alpha_l$ per arrival therefore  the net-reward to a student $i$ of  $Type_l$  in a state $\overline{x}$ will be: 
\begin{align}
 \label{net_reward_lumpedCTMC}
{\overline{ R}^{l,i} }(\overline{ x}) = {\overline{r}^{l,i} }(\overline{x}) - \alpha_l \overline{x} \end{align}

The expected net-reward at time $t$ to student $i$ of $Type_l$ using the lumped-CTMC ${\overline{X}_i}(t)$ will thus be given by:
\begin{align}
\nonumber  & \quad \text{ } {\overline{ R}_t^{l,i} } = \sum \limits_{\overline{ x} \in {\overline{ S}_i}} \left(   {\overline{R}^{l,i}}(\overline{ x})\right) {\overline{ \pi}^t_i}(\overline{ x})\\
\nonumber \text{where, }{ \overline{ R}^{l,i}}(\overline{ x}) \text{ : }  & \text{Net-reward to student } i \text{ of } Type_l \text{ in state } \overline{ x} \text{ of } {\overline{X}_i}(t),\\\nonumber
		{\overline{\pi}^t_i}(\overline{ x})\text{ : }  & \text{Transient probability of } {\overline{X}_i}(t) \text{ being in state } \overline{x} \text{ at} \text{ time } t.
\end{align}

As each state of the lumped-CTMC ${\overline{X}_i}(t)$ captures the number of arrivals from student $i$, thus we use ${\overline{X}_i}(t)$ to calculate the reward ${\overline{ r}^{I,i}}$ collected by the instructor by arrivals from student $i$. As only one student's arrivals are being tracked in ${\overline{X}_i}(t)$, if the instructor arrives on the forum when  $ {\overline{X}_i}(t)$ is in state $\overline{ x}$, the reward she receives will be given by:  ${\overline{ r}^{I,i}}(\overline{x}) = \overline{x} \delta^{\log\mu}$ (Refer Section ~\ref{section_rewards} for details.). Also,  if we take into account the cost per arrival of the instructor then the net-reward to the instructor will be: ${\overline{ R}^{I,i}}(\overline{x}) = {\overline{ r}^{I,i}}(\overline{x}) - \beta$. The expected transient net-reward to the instructor if she arrives on the forum  at time $t$, from the arrival of  student $i$ of $Type_l$ using the lumped-CTMC ${\overline{X}_i}(t)$ will thus be given by:
\begin{align}
\nonumber & {\overline{ R}_t^{I,i} }= \sum \limits_{\overline{ x} \in {\overline{ S}_i}} \left(  \nonumber {\overline{R}^{I,i}}(\overline{ x})\right) {\overline{ \pi}^t_i}(\overline{ x})\\
\nonumber \text{where, } {\overline{ R}^{I,i}}(\overline{ x}) \text{ : }  & \text{Net-reward to the instructor in state } \overline{x} \text{ of } {\overline{X}_i}(t) \\
	\nonumber 									 & \text{ from student } i\text{'s participation,}
		&
\end{align}

We shall now define the expected transient aggregate net-rewards over time $T$ and also the expected steady state net-rewards (if steady state probability vector of the lumped CTMC ${\overline{X}_i}(t)$ exists), to the student $i$ and the instructor w.r.t. the CTMC ${\overline{X}_i}(t)$.

The expected transient aggregate net-rewards over time $T$ for the student $i$ of $Type_l$ and the instructor w.r.t. the CTMC ${\overline{X}_i}(t)$ will be denoted by ${\overline{ R}_T^{l,i} }$ and ${\overline{ R}_T^{I,i} }$ respectively. These are the rewards that the student $i$ and the instructor accumulate over time $T$ from the arrivals captured using the lumped-CTMC ${\overline{X}_i}(t)$ and will be defined as:
\begin{align}
\nonumber {\overline{ R}_T^{l,i} }    = \int\limits_{t=0}^{T} {\overline{R}_t^{l,i}} dt, \nonumber \hspace{3mm}{ \overline{ R}_T^{I,i} }  = \int\limits_{t=0}^{T} {\overline{ R}_t^{I,i} }  dt.
\end{align}
    
The expected steady state net-rewards for the student $i$ of $Type_l$ and the instructor w.r.t. CTMC ${\overline{X}_i}(t)$ will be the expected net-rewards received by the student $i$ and the instructor when ${\overline{X}_i}(t)$ is in the steady state and are denoted by ${\overline{ R}_{\overline{ \Pi}}^{l,i} }$ and   ${\overline{ R}_{\overline{ \Pi}}^{I,i} } $ respectively. These will be defined as:
\begin{align}
 \nonumber{\overline{ R}_{\overline{ \Pi}}^{l,i} } =& \sum\limits_{\overline{x} \in {\overline{S}_i}} {\overline{R}^{l,i}} (\overline{x})\overline{\Pi}(\overline{x}), \hspace{3mm} 
 {\overline{ R}_{\overline{ \Pi}}^{I,i} }= \sum\limits_{\overline{x} \in {\overline{S}_i}}  {\overline{R}^{I,i}}( \overline{x})\overline{\Pi}(\overline{x})\\
\nonumber\text{ where, }\overline{\Pi}(\overline{x}) =& \text{Steady-state } \text{probability of } {\overline{X}_i}(t) \text{ being in state }\overline{x}.
\end{align}

As each lumped-CTMC ${\overline{X}_i}(t)$ captures the arrival of only student $i$ on the OEF, thus the net-rewards received by the instructor in ${\overline{X}_i}(t)$ is from student $i$'s arrivals only. So lets define ${\overline{R}_T^{I}}$, ${\overline{R}_{\Pi}^{I}}$ as the total aggregate reward over time $T$ and the total steady state reward that the instructor receives from all the $n$ lumped-CTMCs ${\overline{X}_i}(t)$  $i \in \{1,2,\ldots,n\}$. The instructor has a bias $c_i$ for arrivals from student $i$ (as explained in Section~\ref{section_rewards}). So, we take a weighted sum over the expected net-rewards received by the instructor in all the $n$ lumped-CTMCs ${\overline{X}_i}(t)$ $i \in \{1,2,\ldots,n\}$ to get the total net-reward received by the instructor from the arrival of all the students on the OEF.
\begin{align}
 \nonumber {\overline{R}_T^{I}} = \sum\limits_{ 1 \leq i \leq n}    c_i { \overline{ R}_T^{I,i} }, \hspace{3mm}
 {\overline{R}_{\Pi}^{I}} = \sum\limits_{ 1 \leq i \leq n}    c_i  {\overline{ R}_{\Pi}^{I,i} }.
\end{align}

 The next lemma proves that the transient probability   ${\overline{\pi}^t_i}(\overline{ x})$  of  ${\overline{X}_i}(t)$   being in state $\overline{x}$ at time $t$ is obtained by summing the transient probabilities $\pi^t(x)$   of  $X(t)$ being in state $x$ at time $t$ over all $x$ belonging to the block ${\overline{S}_i^{\overline{x}}}$.
   
\begin{lemma} \label{lemma_state_probabilities_equivalence}
${\overline{\pi}^t_i}(\overline{ x}) =  \sum\limits_{x \in {\overline{S}_i^{\overline{x}}}} \pi^t(x)$
\end{lemma}

The following lemma proves that the expected transient net-reward to the student $i$ of $Type_l$ is same when calculated using the CTMCs $X(t)$ and ${\overline{X}_i}(t)$.
Thus instead of calculating the net-reward to a student $i$ using $X(t)$ , we can calculate the net-reward to the student $i$ using the $i^{th}$ lumped-CTMC ${\overline{X}_i}(t)$.
\begin{lemma}
\label{Lemma_Student_Utilities_equal_in_original_And_Lumped}
$ R_t^{l,i} = {\overline{ R}_t^{l,i} }$
\end{lemma}

 The following lemma proves that the  expected transient net-reward to the instructor is the same when calculated using the CTMC $X(t)$ and when calculated as a weighted sum of the instructor's expected transient net-rewards w.r.t. CTMCs ${\overline{X}_i}(t)$ for student $i :1 \leq i \leq n$. 
 Therefore we can utilize the lumped CTMCs ${\overline{X}_i}(t)$ instead of the original CTMC $X(t)$ to calculate the instructor's expected transient net-reward.
\begin{lemma}  
\label{Lemma_Instructor_Utility_can_be_calculated_in_original_And_Lumped} 
$ R_t^{I} = \sum\limits_{ 1 \leq i \leq n}    c_i  {\overline{ R}_t^{I,i} } $  
\end{lemma}

Next we prove that the original CTMC $X(t)$ and the lumped CTMCs ${\overline{X}_i}(t)$ have steady state vectors and provide a relationship between them.
\begin{lemma}
\label{Lemma_CTMC_irreducible_and_steady_state_exists}
\label{CTMC_irreducible_and_steady_state_exists}
The CTMC $X(t)=(\mathcal{S},Q)$ is irreducible and positive recurrent and a steady state vector $\Pi$ exists for this CTMC.
\end{lemma}

\begin{lemma}
\label{Lemma_CTMC_lumped_steady_state_exists}
The lumped-CTMC ${\overline{X}_i}(t)=({\overline{S}_i},{\overline{Q}_i})$ is irreducible and positive recurrent and a steady state vector ${\overline{\Pi}_i}$ exists for this CTMC. Also each component of ${\overline{\Pi}_i}$ will be obtained as: ${\overline{\Pi}_i}(\overline{x}) = \sum\limits_{x \in {\overline{S}^{\overline{x}}_i} } \Pi(x). $
\end{lemma}

\begin{theorem}\label{Theorem_final_all_rewards}
The expected transient aggregate net-rewards over time $T$ and the expected steady-state net-rewards received by the students and the instructor  is the same when calculated using the original CTMC $X(t)$ or the $n$ lumped CTMCs ${\overline{X}_i}(t)$ $i \in \{1,2,\ldots,n\}$. 
\begin{itemize}
\item Expected transient aggregate net-rewards over time $T$.\\
a) $  R_T^{l,i}  = {\overline{ R}_T^{l,i} }$ \hspace{3mm} 
b) $  R_T^{I} =  {\overline{ R}_T^{I} } $.

\item Expected steady-state net-rewards.	\\
a) $  R_{\Pi}^{l,i} = {\overline{ R}_{\Pi}^{l,i} }$ \hspace{3mm}
b) $ R_{\Pi}^{I} = {\overline{ R}_{\Pi}^{I} } $.	
\end{itemize}
\end{theorem}

Thus, the expected transient net-rewards to the students and the instructor at time $t$ can be calculated using the $n$ lumped CTMCs $ {\overline{X}_i}(t)$  ($ 1 \leq i \leq n$) instead of using the original complex CTMC $X(t)$. So, for our analysis, we shall now focus on the lumped-CTMCs for each student instead of the original-CTMC.

\subsection{Analysis of the Lumped CTMCs}
\label{section_analysis_of_the_lumped_ctmcs}

Recall that each lumped-CTMC ${\overline{X}_i}(t) = ({\overline{S}_i},{\overline{Q}_i})$ for student $i$ of $Type_l$ will have the following entities associated with it: max-arrival $M$,  max-reward $m_l$, course duration time $T$, rate of arrival $ \lambda_i$ for the student $i$ in the OEF and the rate of instructor arrival $\mu$ fixed by the instructor. Also recall that the state space ${\overline{S}_i}$ of ${\overline{X}_i}(t)$ is given by ${\overline{S}_i}= \{0,1,\ldots,M \}$ and the generator matrix ${\overline{Q}_i}$ is defined as given in Equation ~\ref{final_generator_matrix}. Each lumped-CTMC will thus be similar in structure to the CTMC given in Figure~\ref{ctmcLumped}~(a).

We give the   matrix representation of the generator matrix ${\overline{Q}_i}$  below for better clarity. Each entry of this matrix is bounded by a large value $\widehat{q} < \infty$(Discussed in Section~\ref{sectiondescriptionofthemodel}). Also recall that the initial state distribution ${\overline{\pi}^0_i}$ for the CTMC ${\overline{X}_i}(t)$ is defined  as  ${\overline{\pi}^0_i}(0) =1$ and  ${\overline{\pi}^0_i}(\overline{x}) =0 \forall \overline{x} \in {\overline{S}_i} \setminus \{ 0\}$.

\begin{align} \label{lumped_generator_matrix}
 {\overline{Q}_i} =\hspace{-0.2in}
\begin{blockarray}{cccccc}
States		&0			 & 1			 & 2			& \ldots                   & M\\
\begin{block}{c(ccccc)}
0		&-\lambda_i	         & \lambda_i  		 & 0			& 0\ldots 		   & 0\\
1		&\mu 	  		 & -(\mu + \lambda_i) 	 & \lambda_i		& 0\ldots	           & 0\\
2		&\mu 	  		 & 0			 & \ddots		& \ddots 		   & \vdots\\
\vdots		&\vdots		 	 & \vdots                & \vdots               & -(\mu + \lambda_i)	   & \lambda_i\\
M		&\mu     	 	 & 0                     & \ldots     		& 0     		   & -\mu\\
\end{block}
\end{blockarray}
\end{align}

Using this generator matrix, we shall now solve for the steady state and transient probabilities for the lumped-CTMC and use these to calculate the expected steady state and transient aggregate net-rewards received by the students and the instructor (given by Theorem~\ref{Theorem_final_all_rewards}). Detailed proofs of the results presented in this section are provided in Appendix~B.

\subsubsection{Steady State Analysis}
Steady state probability ${\overline{\Pi}_i}( \overline{x})$ is the probability that the CTMC ${\overline{X}_i}(t)$ (corresponding to student $i$) is in state $\overline{x}$, when ${\overline{X}_i}(t)$ has reached a steady state. Lemma~\ref{Lemma_CTMC_lumped_steady_state_exists} states that a steady state exists for this CTMC and so using the generator matrix ${\overline{Q}_i}$, we come up with the following equations for the steady state probability vector ${\overline{\Pi}_i}$ of ${\overline{X}_i}(t)$:
\begin{align}
\nonumber (\text{Inflow in state } \overline{x} 
					= &\text{ Outflow from state } \overline{x})\   \forall 0 \leq \overline{x} \leq M\\
\label{eqnSteady1} \sum\limits_{\overline{y} \geq 1} { \mu  {\overline{\Pi}_i}( \overline{y}) } 
					= & \lambda_i {\overline{\Pi}_i}( 0) \text{, for } \overline{x}  = 0 \\ 
\label{eqnSteady2} \lambda_i {\overline{\Pi}_i}( \overline{x}-1) =& \lambda_i {\overline{\Pi}_i}(\overline{x})  + \mu {\overline{\Pi}_i}( \overline{x})   \text{, } \forall \overline{x}: 0 < \overline{x}  < M \\
\label{eqnSteady3} \lambda_i {\overline{\Pi}_i}( M -1)  =& \mu {\overline{\Pi}_i}( M)   \text{, for } \overline{x}  = M.
\end{align}

\begin{lemma}\label{Lemma_Solving_for_steady_state_Probabilities}
The solution to equations ~\ref{eqnSteady1}, ~\ref{eqnSteady2}, and ~\ref{eqnSteady3} is:  
\begin{align} \nonumber {\overline{\Pi}_i}( \overline{x}) =& \frac{\rho^{ \overline{x}}_i}{(1+\rho_i)^{ \overline{x}+1}} \text{, } \forall \overline{x}: 0 \leq  \overline{x} < M, \text{ and}\\
\nonumber  {\overline{\Pi}_i}( M) =& \frac{\rho_i^{M}}{(1+\rho_i)^{M}} .\text{ where, } \rho_i = \frac{\lambda_i}{\mu}
\end{align}
\end{lemma}

Using these steady state probabilities, we calculate the expected steady state net-rewards of the students and the instructor (from student $i$'s participation) in ${\overline{X}_i}(t)$.

\begin{theorem}
\label{Thorem_SteadyState_reward_student_and_instructor}
 The  expected steady state net-rewards of the student $i$ and the instructor in ${\overline{X}_i}(t)$ are given by:					
\begin{align}
 \nonumber \text{(a) } {\overline{ R}_\Pi^{l,i} } =&  \delta^{\log\mu} {\rho_i}  \left(  1 - \left(  \frac{{\rho_i}}{1 + {\rho_i}}  \right)^{m_l}   \right) -   \alpha_l   {\rho_i} \left(  1  - \left(  \frac{{\rho_i}}{1 + {\rho_i}}  \right)^{M}  \right)\\
\nonumber \text{(b) }  {\overline{ R}_{\Pi}^{I,i} }	=&  \delta^{\log\mu}  {\rho_i} \left(  1  - \left(  \frac{{\rho_i}}{1 + \rho_i}  \right)^{M}  \right) - \beta  
\end{align}
\end{theorem}

\subsubsection{Transient Analysis}

Transient state probability ${\overline{\pi}_i^t}(\overline{x})$ is the probability of ${\overline{X}_i}(t)$ being in the state $\overline{x}$ at a time instant $t$. We need to solve the following differential equations for the transient probability vectors ${\overline{\pi}_i^t}$'s of the lumped-CTMC ${\overline{X}_i}(t)$:
\begin{align}
 \label{eq1_transient} \frac{d {\overline{\pi}_i^t}(0)}{dt} &= -\lambda_i {\overline{\pi}_i^t}(0) + \sum\limits_{\overline{y}=1}^{M} \mu  {\overline{\pi}_i^t}(\overline{y}) \\
 \label{eq2_transient} \frac{d {\overline{\pi}_i^t}(\overline{x})}{dt} &= -(\lambda_i+\mu) {\overline{\pi}_i^t}(\overline{x}) + \lambda_i {\overline{\pi}_{i}^t}(\overline{x} - 1)\text{, }\forall \overline{x}:  0 < \overline{x} <M \\ 
 \label{eq3_transient} \frac{d {\overline{\pi}_{i}^t}(M)}{dt} &= - \mu {\overline{\pi}_{i}^t}(M) +\lambda_i {\overline{\pi}_{i}^t}(M-1).
\end{align}
We now present some important results regarding transient probabilities. Detailed proofs of the results presented in this section are provided in Appendix~C.

\begin{lemma}
\label{Lemma_Solving_for_transient_Probailities}
Given the initial distribution ${\overline{\pi}_{i}^0}$ for $\overline{X}_i(t)$ (${\overline{\pi}_{i}^0}(0) =1$ and ${\overline{\pi}_{i}^0}(\overline{x}) =0$ $\forall \overline{x} \in \{1,2,\ldots,M \}$). The solution to the equations ~\ref{eq1_transient}, ~\ref{eq2_transient}, and ~\ref{eq3_transient} is given by:   
\begin{align}
\nonumber {\overline{\pi}_i^t}(0) =& \frac{\mu}{{K}_i} + \frac{\lambda_i}{{K}_i} e^{-{K}_it}\\
\nonumber {\overline{\pi}_i^t}(\overline{x}) =& \left( -  \left(\frac{\lambda_i}{{K}_i}\right)^{\overline{x}} \frac{\mu}{{K}_i}    - \sum\limits_{\overline{y}=1}^{\overline{x}-1} \frac{\lambda_i^ {\overline{x}} t^{\overline{y}} \mu}{\overline{y}! {K}_i^{\overline{x}-\overline{y}+1}} + \frac{{\lambda_i}^{\overline{x}+1} t^{\overline{x}}}{\overline{x}! {K}_i} \right)e^{-{K}_{i} t} 
+ \left(\frac{\lambda_i}{{K}_i}\right)^{\overline{x}} \frac{\mu}{{K}_i}  \hspace{2mm} \forall \overline{x}: 0 < \overline{x} < M\\
\nonumber {\overline{\pi}_{i}^t}(M) =& 1 - \sum\limits_{\overline{x}=0}^{M-1} {\overline{\pi}_i^t}(\overline{x}) \text {, where } {K}_i = (\lambda_i + \mu)
\end{align}

\end{lemma}
We now calculate the  expected  transient aggregate net-rewards over time $T$ of a  student $i$ and the instructor (from student $i$'s participation) using the lumped-CTMC ${\overline{X}_i} (t)$.

\begin{theorem}
\label{Thorem_Transient_reward_student_and_instructor}
 The  expected  transient aggregate net-rewards of the student $i$ and the instructor in ${\overline{X}_i}(t)$ are given by:		
\begin{align}
\nonumber \text{(a) } {\overline{ R}_T^{l,i}} =&  \delta^{\log\mu} \sum \limits_{\overline{ x} = 0}^{m_l}  ( \overline{x} - m_l)     \int\limits_{0}^{T}  {\overline{ \pi}_i^t}(\overline{ x}) dt  
+ m_l \delta^{\log\mu} T-  \alpha_l \sum \limits_{\overline{ x} = 0}^{M}   \overline{x}   \int\limits_{0}^{T} {\overline{ \pi}_i^t}(\overline{ x}) dt \\
\text{(b) } {\overline{ R}_{T}^{I,i} } =& \delta^{\log\mu}    \sum \limits_{\overline{ x} = 0}^{M} \overline{x}  \int\limits_{0}^{T}  {\overline{ \pi}_i^t}(\overline{ x}) dt -  \beta T.
\end{align}
 \end{theorem}

By Theorem ~\ref{Theorem_final_all_rewards}, we  can calculate the expected net-rewards received by the students in an  OEF from the net-reward obtained by the student $i$ in the lumped-CTMC ${\overline{ X}_i}(t)$. Thus we shall use the expected steady state reward ${\overline{ R}_{\Pi}^{l,i} }$ and the expected transient aggregate reward ${\overline{ R}_{T}^{l,i} }$ for student $i$ of the lumped-CTMC ${\overline{X}_i}(t)$ to derive the expected steady-state $R_{\Pi}^{l,i}$ and transient aggregate  net-rewards $R_{T}^{l,i}$ respectively received by the students in the OEF. 

 By Theorem ~\ref{Theorem_final_all_rewards}, we can calculate the net-reward received by the instructor in an OEF by taking a weighted sum over the expected net-rewards to the instructor received through the lumped-CTMCs ${\overline{X}_i}(t)$s $\left( 1 \leq i \leq n\right)$. Thus we shall now compute the instructor's expected steady state net-reward $ R_{\Pi}^{I}$ and the expected transient aggregate net-reward $ R_{T}^{I}$ in the OEF by  respectively taking weighted sums  over the expected steady state net-rewards $ \sum_{ 1 \leq i \leq n}    c_i  {\overline{ R}_{\Pi}^{I,i} }$ and  over the expected transient aggregate  net-rewards $\sum_{ 1 \leq i \leq n}    c_i { \overline{ R}_T^{I,i} }$   the instructor receives in the lumped CTMCs ${\overline{X}_i}(t)$s.

\section{OEF  as a Stackelberg Game (Strategic Setting)}
\label{Section_OEF_Game}

So far, we have assumed that the arrival rates of the instructor and the enrolled students are known to the model but this is unrealistic in the real world as the students and the instructor are strategic agents who will have a range of arrival rates to choose from and wish to maximize their utilities by choosing an optimal strategy (rate of arrival). We summarize important notations of this section in Table~\ref{CTMCnotationtable3}.

\begin{table}[!htb]
\centering
\tablefont{2.9mm}
\setlength{\extrarowheight}{5pt}
\begin{tabular}{|c|p{4.5in}|}
\hline
\textbf{Symbol} & \textbf{$\quad\quad\quad\quad\quad\quad\quad\quad\quad\quad\quad$Meaning} \\ 
\hline 
$n$   &  Number of students in the course.\\
\hline 
$L$   &  Number of types of students. \\
\hline 
$T$   &  Course Duration. \\
\hline 
$G^{ins}$& Finite strategy set of the instructor (of size $v$).\\ 
\hline 
$G^{st}$ & Finite strategy set of the students (of size $w$).\\ 
\hline 
$\Delta_a$ & A pure strategy strategy of the instructor.\\ 
\hline 
$\Lambda_b$ & A pure strategy strategy of a student .\\ 
\hline 
$\phi_a$ &  Proportion of times that pure strategy ${\Delta}_a$ is used.\\ 
\hline 
$\psi^i$ &  Pure/mixed strategy over $G^{st}$ of the student~$i$.\\ 
\hline 
$D^{\Pi,l,i}_{a,b}$ & Expected steady state net-reward received by the student $i$ of $Type_l$ if she chooses pure strategy $\Lambda_b$ and instructor chooses the pure strategy $\Delta_a$.\\ 
\hline 
$D^{T,l,i}_{a,b}$ & Expected transient aggregate net-reward (over time T) received by student $i$ of $Type_l$ when she chooses pure strategy $\Lambda_b$ and instructor chooses the pure strategy $\Delta_a$.\\ 
\hline 
$U_{\Pi}^{l,i}$ &  Expected steady state utility to a student~$i$ of $Type_l$ .\\
\hline 
$U_{T}^{I}$  &  Expected aggregate transient utility (over time $T$) to the instructor.\\
\hline 
$U_{\Pi}^{I}$  &  Expected steady state utility to the instructor.\\
\hline 
$B^{\Pi,I,i}_{a,b}$  &  Expected steady state utility received by the instructor if the student $i$ chooses pure strategy $\Lambda_b$ and the instructor chooses the pure strategy $\Delta_a$.\\
\hline 
$B^{T,I,i}_{a,b}$   &  Expected transient utility received by the instructor if the student $i$ chooses pure strategy $\Lambda_b$ and the instructor chooses the pure strategy $\Delta_a$.\\
\hline
\end{tabular}
\caption{Important notations in the Stackelberg game formulation } \label{CTMCnotationtable3}
\end{table} 

We assume that the instructor is a social-welfare maximizing agent whose goal is to maximize the students' understanding of the subject by enabling them to discuss frequently among themselves in the OEF. 
The students typically will have commitments towards other courses and hence, their objective will be to maximize the rewards they get from answering questions in the OEF while minimizing their cost (of visiting the forum and answering a question). We assume that the instructor's strategy  will be announced first  following which the students will decide their strategy. 

We thus formulate an OEF as a Stackelberg game where the players are: an instructor offering the online course and each of the $n$ students enrolled for the course. We make the following assumptions related to the formulation of the Stackelberg game.
\begin{itemize}
\item Instructor acts as a leader who decides her   strategy (rate of arrival) first and the students act as the followers who after observing the instructor's strategy will finalize their own strategy (rate of arrival) in order to maximize their utility.
\item We assume the strategy space of the players is finite.
\end{itemize}

The key idea behind the formulation of the Stackelberg game is to link the expected net-rewards of the CTMC (when the strategies are known to the model) as derived in Theorems~\ref{Thorem_SteadyState_reward_student_and_instructor} and~\ref{Thorem_Transient_reward_student_and_instructor}  to the strategic scenario where the players are optimizing their corresponding utility functions. We make this intuition clear in the following subsections by defining the utility functions of the students and instructor and putting these two together in a bi-level optimization problem which solves for the optimal instructor and students' strategies.

 We establish some basic notations before we formulate the game. We denote the finite strategy set of the instructor and the students by $G^{ins}$ (of size $v$) and $G^{st}$ (of size $w$) respectively. These are the set of arrival rates from which the instructor and the students can choose their respective rates of arrival. The rates in the sets $G^{ins}$ and $G^{st}$ are bounded and are in the range $[0, \widehat{q} ]$.
\begin{align}
\nonumber    G^{ins} &= \{ \Delta_a | a \in \{1,\ldots,v \}, {\Delta}_a \in [0, \widehat{q}], \widehat{q} < \infty  \}\\
\nonumber    G^{st} &= \{ \Lambda_b | b \in \{1,\ldots,w \},\Lambda_b \in [0, \widehat{q} ], \widehat{q} < \infty  \}
\end{align}

The instructor can choose a pure/mixed strategy over $G^{ins}$ and we denote the instructor's strategy by $\phi$ which is a point in probability simplex, $\Omega(G^{ins})$, where  
\begin{align}
\nonumber \Omega(G^{ins})&=\{\phi =(\phi_1, \cdots, \phi_v)| 0\leq \phi_a \leq 1, \sum_{a=1}^{v}\phi_a = 1, 1\leq a\leq v\}
\end{align}
The value $\phi_a$ is the proportion of times that pure strategy ${\Delta}_a$ is used. Similarly, $\psi^i$ denotes the pure/mixed strategy over $G^{st}$ of the student~$i$ which is a point in probability simplex $\Omega(G^{st})$, where
\begin{align}
\nonumber \Omega(G^{st}) &=\{\psi^{i} =(\psi^{i}_1, \cdots, \psi^{i}_w)| 0\leq \psi^{i}_b \leq 1, \sum_{b=1}^{w}\psi^{i}_b = 1, 1\leq b\leq w\}
\end{align}
The value $\psi^{i}_b$ basically represents the proportion of times in which pure strategy $\Lambda_b$ is used. 

\subsection{Student Optimization Problem}

We define expected net-reward matrices $D^{\Pi,l,i}$ and $D^{T,l,i}$ corresponding to each student $i$ of type $l$ where each $D^{\Pi,l,i}_{a,b}$ entry denotes the expected steady state net-reward received by the student $i$ if she chooses pure strategy $\Lambda_b$ and instructor chooses the pure strategy $\Delta_a$. Similarly $D^{T,l,i}_{a,b}$ denotes the expected transient aggregate net-reward (over time T) received by student $i$ of $Type_l$ when she chooses pure strategy $\Lambda_b$ and instructor chooses the pure strategy $\Delta_a$.

Now, suppose the instructor has fixed a (pure/mixed) strategy $\phi$ and the student $i$  has fixed a (pure/mixed) strategy $\psi^i$.  Using the above definitions, the expected steady state utility $U_{\Pi}^{l,i}$ and the expected transient aggregate utility $U_{T}^{l,i}$ (over the course duration $T$) to a student~$i$ of $Type_l$ on the forum is defined by: 
\begin{align}
\label{Utility_Student_OEF}  U_{\Pi}^{l,i} = \sum\limits_{a=1}^{v}  \sum\limits_{b=1}^{w} & D^{\Pi,l,i}_{a,b} \phi_a \psi_b^i;\quad   U_{T}^{l,i} = \sum\limits_{a=1}^{v} \sum\limits_{b=1}^{w} D^{T,l,i}_{a,b} \phi_a \psi_b^i\\
\nonumber \text{ where, }   D^{\Pi,l,i}_{a,b} =& \text{ } {\overline{R}_{\overline{\Pi}}^{l,i}} \text{ over the CTMC }\overline{X}_i(t)\text{(See Theorem~\ref{Thorem_SteadyState_reward_student_and_instructor}~(a))},\\
\nonumber D^{T,l,i}_{a,b} =& \text{ } {\overline{R}_T^{l,i}} \text{ over the CTMC }\overline{X}_i(t) \text{(See Theorem~\ref{Thorem_Transient_reward_student_and_instructor}~(a))},\\
\nonumber \text{ and } {\overline{X}_i}(t) \text{ is } & \text{defined by } \mu = \Delta_a \text{ and } \lambda_i = \Lambda_b (\text{See Equation }\ref{lumped_generator_matrix}).
\end{align}

\begin{proposition}\label{optimization_utilities_student_equal} 
Students $i,j$ belonging to the same type $Type_l$ receive equal utility (steady state and transient aggregate) if they choose the same policy i.e. if $\psi^i = \psi^j (= \psi^l)$ for students $ i,j \in Type_l$ then,
\begin{align}
 \nonumber U_{\Pi}^{l,i} = U_{\Pi}^{l,j} \text{, }  \hspace{4mm}
\nonumber  U_{T}^{l,i} = U_{T}^{l,j}.
\end{align}
\end{proposition}
\begin{proof}
Please check the Appendix D section of the paper for the detailed proof of this result.
\end{proof}

Using the above definitions, we now formulate the (steady state) optimization problem for student~$i$ of $Type_l$ when the instructor has fixed a strategy $\phi$. 
\begin{align}
\label{Optimization_student_optimization_problem}
 \psi^{i*} =& \mathop {arg\max}_{\psi^i}   \sum\limits_{a=1}^{v} \sum\limits_{b=1}^{w} D^{\Pi,l,i}_{a,b} \phi_a \psi_b^i, \quad \text { s.t. }  \sum\limits_{b=1}^{w} \psi_b^i = 1, \hspace{2mm}   \psi_b^i \geq 0.
\end{align}
Note that $\psi^{i*}$ is the optimal rate of arrival of student~$i$ of $Type_l$ in response to instructor strategy $\phi$. 

Note that students belonging to the same type can have different optimal strategies in response to the same instructor strategy as the optimization problem given in Equation~\ref{Optimization_student_optimization_problem} can have multiple solutions. Due to symmetry among students belonging to the same type, we assume that students belonging to the same type will choose the same optimal strategy i.e. $\psi^{i*} = \psi^{j*} = \psi^{l*} \text{ } \forall i,j \in Type_l$.

Hence, we can infer that, instead of solving the student (steady state) optimization problem (Equation ~\ref{Optimization_student_optimization_problem}) for each student $i,j$ belonging to a particular type $Type_l$, we can just solve the student optimization problem for a single representative $Type_l$ student (now represented as $l$)  and her   optimal policy $\psi^{l*}$ would be followed by each student $i$ of $Type_l$. Thus the (steady state) optimization problem to be solved by each representative student $l$  $( 1 \leq l \leq L)$ will be :

\begin{align}
\label{Optimization_type_l_optimization_problem}
 \psi^{l*} =& \mathop {arg\max}_{\psi^l}   \sum\limits_{a=1}^{v} \sum\limits_{b=1}^{w} D^{\Pi,l}_{a,b} \phi_a \psi_b^l, \quad \text { s.t. } \sum\limits_{b=1}^{w} \psi_b^l = 1, \hspace{2mm}   \psi_b^l \geq 0\\
 \nonumber  \text{where, }&  D^{\Pi,l}_{a,b} =  D^{\Pi,l,i}_{a,b} \text{ for any } i \in Type_l.
\end{align} 

The optimization problem for the expected transient aggregate utilities will be similar to the steady-state counterpart discussed above. 

\subsection{Instructor Optimization Problem}

We now define similarly the steady state utility $U_{\Pi}^{I}$ and the aggregate transient utility $U_{T}^{I}$ (over time $T$) to the instructor when she has fixed her   strategy as $\phi$ and the $n$ students have fixed their policies as $\psi^i$ ($1 \leq i \leq n$). We define net-reward matrices $B^{\Pi,I,i}$ and $B^{T,I,i}$ for the instructor corresponding to each student $i$ where each $B^{\Pi,I,i}_{a,b}$ entry denotes the steady state utility received by the instructor and each $B^{T,I,i}_{a,b}$ entry denotes the aggregate utility (over time T) received by the instructor w.r.t. student $i$'s arrivals, if the student $i$ chooses pure strategy $\Lambda_b$ and the instructor chooses the pure strategy $\Delta_a$.

\begin{align} \label{Utility_Instructor_OEF}
U_{\Pi}^{I} =  \sum\limits_{a=1}^{v}  \sum\limits_{b=1}^{w} &  \sum\limits_{i=1}^{n} B^{\Pi,I,i}_{a,b} \phi_a \psi_b^i ; \quad U_{T}^{I} =\sum\limits_{a=1}^{v} \sum\limits_{b=1}^{w} \sum\limits_{i=1}^{n}  B^{T,I,i}_{a,b} \phi_a \psi_b^i\\
\nonumber \text{ where, }   B^{\Pi,I,i}_{a,b} &= \text{ } c_i {\overline{R}_{\overline{\Pi}}^{I,i}} \text{ over the CTMC}\overline{X}_i(t)\text{ (See Theorem~\ref{Thorem_SteadyState_reward_student_and_instructor}~(b))},\\
\nonumber B^{T,I,i}_{a,b} &= \text{ } c_i {\overline{R}_T^{I,i}} \text{ over the CTMC }\overline{X}_i(t)\text{ (See Theorem~\ref{Thorem_Transient_reward_student_and_instructor}~(b))},\\
\nonumber \text{ and }  {\overline{X}_i}(t) & \text{ is }  \text{defined by } \mu = \Delta_a \text{ and } \lambda_i = \Lambda_b.\text{ (See Equation ~\ref{lumped_generator_matrix})}.
\end{align}

\begin{proposition}\label{optimization_instructor_matrices_w.r.t_a_type_equal}
The net-reward matrices $B^{\Pi,I,i}, B^{\Pi,I,j}$ (steady state) and $B^{T,I,i},B^{T,I,j}$ (transient aggregate) for the instructor w.r.t. students $i,j$ belonging to $Type_l$ have the following property:
\begin{align}
\nonumber B^{\Pi,I,i} = B^{\Pi,I,j} (= B^{\Pi,I,l}), \hspace{4mm}
\nonumber B^{T,I,i} = B^{T,I,j} (= B^{T,I,l}).
\end{align}
\end{proposition}
\begin{proof}
Please check the Appendix D section of the paper for the detailed proof of this result.
\end{proof}

The instructor is the leader, so she chooses her   policy $\phi$ first and then each student (follower) observes the strategy chosen by the instructor and then decides the policy $\psi^{i}$ such that she can maximize her utility. The instructor will maximize her utility by choosing a solution $\phi^{*}$ to the following (steady state) optimization problem:
\begin{align}
\phi^{*} =& \mathop{ \text{arg}\max}_{\phi}  \sum\limits_{a=1}^{v} \sum\limits_{b=1}^{w} \sum\limits_{i=1}^{n} B^{\Pi,I,i}_{a,b} \phi_a [\psi^{i*}(\phi)]_b\\
\nonumber & s.t. \sum\limits_{a =1}^{v} \phi_a = 1, \hspace{2mm} \phi_a \in [0 \ldots 1]
\end{align}
$\psi^{i*}(\phi)$ is the optimal strategy (See Equation~\ref{Optimization_student_optimization_problem}) of  student $i$ when $\phi$ is the instructor policy. We know that there are $n_l$ students belonging to $Type_l$. As given in previous section, we know that  $\psi^{i*}(\phi) = \psi^{j*}(\phi) = \psi^{l*}(\phi)$ $\forall i,j \in Type_l$. Using, Proposition~\ref{optimization_instructor_matrices_w.r.t_a_type_equal}, we get:
\begin{align}
\nonumber \phi^{*} =& \mathop{ \text{arg}\max}_{\phi}  \sum\limits_{a=1}^{v} \sum\limits_{b=1}^{w} \sum\limits_{l=1}^{L} n_l B^{\Pi,I,l}_{a,b} \phi_a [\psi^{l*}(\phi)]_b\\
\nonumber & s.t. \sum\limits_{a =1}^{v} \phi_a = 1, \hspace{2mm} \phi_a \in [0 \ldots 1]
\end{align}
Dividing the objective function of this (steady state) optimization problem by $n$, we get
\begin{align}
\nonumber \phi^{*} =& \mathop{ \text{arg}\max}_{\phi} \sum\limits_{a=1}^{v} \sum\limits_{b=1}^{w} \sum\limits_{l=1}^{L} p^l B^{\Pi,I,l}_{a,b} \phi_a [\psi^{l*}(\phi)]_b\\
\nonumber & s.t. \sum\limits_{a =1}^{v} \phi_a = 1, \hspace{2mm} \phi_a \in [0 \ldots 1]
\end{align}
Here, $p^l = n_l / n$ i.e. the proportion of the class that belongs to $Type_l$. Using the student optimization problem in Equation~\ref{Optimization_type_l_optimization_problem}, we can rewrite the above problem as:
\begin{align}
\label{Optimization_final_optimization_instructor}
\phi^{*} =& \mathop{ \text{arg}\max}_{\phi} \sum\limits_{a=1}^{v} \sum\limits_{b=1}^{w} \sum\limits_{l=1}^{L} p^l B^{\Pi,I,l}_{a,b} \phi_a \psi^{l*}_b\\
 \nonumber  s.t., & \text{ } \psi^{l*} = \mathop {arg\max}_{\psi^l}   \sum\limits_{a=1}^{v} \sum\limits_{b=1}^{w} D^{\Pi,l}_{a,b} \phi_a \psi_b^l, \\
& \nonumber \sum\limits_{a =1}^{v} \phi_a = 1, \hspace{2mm} \phi_a \in [0 \ldots 1],\quad  \sum\limits_{b=1}^{w} \psi_b^l = 1, \hspace{2mm} \nonumber  \psi_b^l \geq 0.
\end{align}

Note that the size of the optimization problem reduces considerably with the introduction of the similarity among students belonging to the same type. Now we have to solve the optimization problem w.r.t to only $L+1$ agents instead of $n+1$ agents and in general $L<<n$.
The optimization problem using the expected transient aggregate utilities for the instructor will be similar to the steady-state counterpart discussed above. 

\subsection{MILP Formulation}

The (steady state) optimization problem given in Equation~\ref{Optimization_final_optimization_instructor} is essentially a mixed integer quadratic program (MIQP).  We convert this MIQP to a mixed integer linear program (MILP) by following the approach in \citep{PARACHURI08} where they solve a Bayesian Stackelberg game by converting a similar bi-level optimization problem into an MILP (See \textbf{Proposition~2} in \citep{PARACHURI08}). Thus, we get the following:

\begin{align}\label{MILP_Formulation}
\displaystyle \max_{\psi,\xi,\eta} & \sum\limits_{l=1}^{L} \sum\limits_{a=1}^{v} \sum\limits_{b=1}^{w} p^l B^{\Pi,I,l}_{a,b} {\xi}_{ab}^l\\
\nonumber s.t. & \sum\limits_{a =1}^{v} \sum\limits_{b = 1}^{w} \xi_{ab}^l = 1, \forall l  \hspace{1.5mm}; \hspace{1.5mm}
\sum\limits_{b =1}^{w} \xi_{ab}^l \leq 1, \forall l\\
\nonumber & \psi_b^l \leq \sum\limits_{a=1}^{v} \xi_{ab}^l \leq 1, \forall l, \forall b \hspace{1.5mm};\hspace{1.5mm}
\sum\limits_{b=1}^{w} \psi_b^l = 1, \forall l\\
\nonumber & 0 \leq (\eta^l - \sum\limits_{ a = 1}^{v} D^{\Pi,l}_{ab}(\sum\limits_{h =1}^{w} \xi^l_{ah})) \leq (1-\psi_b^l)N,\forall l, \forall b \hspace{1.5mm};\hspace{1.5mm} \eta \in \mathbb{R}\\
\nonumber & \sum\limits_{b=1}^{w} \xi_{ab}^l = \sum\limits_{b=1}^{w} \xi_{ab}^1,\forall l, \forall a \hspace{1.5mm};\hspace{1.5mm} \psi_b^l \in \{0,1\}\forall l \hspace{1.5mm};\hspace{1.5mm} \xi_{ab}^l \in [0 \ldots 1]\forall l
\end{align}
Here, $ {\xi}_{ab}^l = \phi_a \psi_b^l$, $\eta$ is derived through duality, and  $N$ is a large constant
 (The details can be seen in Parachuri~et.~al~\citep{PARACHURI08}).

Analogously, we can come up with MILP formulation considering the expected transient aggregate utilities and they are given below.

\begin{align}\label{MILP_Formulation2}
\displaystyle \max_{\psi,\xi,\eta} & \sum\limits_{l=1}^{L} \sum\limits_{a=1}^{v} \sum\limits_{b=1}^{w} p^l B^{T,I,l}_{a,b} {\xi}_{ab}^l\\
\nonumber s.t. & \sum\limits_{a =1}^{v} \sum\limits_{b = 1}^{w} \xi_{ab}^l = 1, \forall l  \hspace{1.5mm}; \hspace{1.5mm}
\sum\limits_{b =1}^{w} \xi_{ab}^l \leq 1, \forall l\\
\nonumber & \psi_b^l \leq \sum\limits_{a=1}^{v} \xi_{ab}^l \leq 1, \forall l, \forall b \hspace{1.5mm};\hspace{1.5mm}
\sum\limits_{b=1}^{w} \psi_b^l = 1, \forall l\\
\nonumber & 0 \leq (\eta^l - \sum\limits_{ a = 1}^{v} D^{T,l}_{ab}(\sum\limits_{h =1}^{w} \xi^l_{ah})) \leq (1-\psi_b^l)N,\forall l, \forall b \hspace{1.5mm};\hspace{1.5mm} \eta \in \mathbb{R}\\
\nonumber & \sum\limits_{b=1}^{w} \xi_{ab}^l = \sum\limits_{b=1}^{w} \xi_{ab}^1,\forall l, \forall a \hspace{1.5mm};\hspace{1.5mm} \psi_b^l \in \{0,1\}\forall l \hspace{1.5mm};\hspace{1.5mm} \xi_{ab}^l \in [0 \ldots 1]\forall l
\end{align}

\section{Simulations}
\label{Section_simulations}

\vspace{-0.05in}
The solution to the optimization problem~(\ref{MILP_Formulation}) (based on steady state utilities) and optimization problem~(\ref{MILP_Formulation2}) (based on transient utilities) provide us with the optimal strategies $\phi^*$ and $\psi^{i*}$'s for the instructor and the students respectively. We solve the MILP using ILOG-CPLEX~\citep{ilog-cplex} software and study the changing dynamics of the student-instructor interactions in an online classroom by varying the different parameters of the model. Table~\ref{Table_Parameters} lists the values and ranges for various parameters used in the simulations. 

\vspace{-0.08in}
\subsection{OEF-CTMC: Transient Vs Steady State}
\vspace{-0.05in}
We substantiate the importance of transient analysis \citep{narahari1994transient} of the proposed CTMC for online classroom setting by studying the variation of the normalized error between the student utility values obtained through the steady and transient analyses. We define
\begin{align}
\nonumber &\text{Normalized Error } =\left| \frac{U_{\Pi}^{l,i} - \frac{U_{T}^{l,i}}{T}}{U_{\Pi}^{l,i}} \right|  \text{, where } U_{\Pi}^{l,i} \text{ and } U_{T}^{l,i} \text{ are respectively}&  \\
\nonumber &\text{the steady state and transient utilities of a student as defined in Equation~\ref{Utility_Student_OEF}.}
\end{align} 

\begin{figure}[h]
\begin{minipage}{16cm}
 \begin{center}
  \includegraphics[scale = .35]{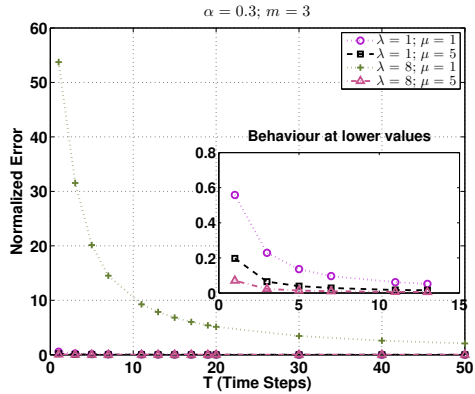}
 \end{center}
 \vspace{-0.2in}
 \caption{
\label{Simulations_Error_Vs_T}  Study normalized error in student utility using the steady state and transient analyses as time evolves.} 

\end{minipage} 
 \end{figure}

For illustration purposes, we assume that only $Type_1$ students are present in the classroom and fix the max-reward and cost per arrival for each of these students as: $m = 3$, $\alpha = 0.3$. We observe the variation in the normalized error with time (Figure~\ref{Simulations_Error_Vs_T}).

 \begin{table*}[ht]
\tablefont{2.6mm}
\setlength{\extrarowheight}{2pt}
\centering
\renewcommand{\arraystretch}{1}
\begin {tabular} {|c|l|l|l|}
\hline
{\textbf{Parameter}}  & \textbf{Definition} &  {\textbf{Range/Value} (For Model)} &  {\textbf{Range/Value} (Fixed for Simulations)}\\
\hline
	$n$ & Total no. of students & Any finite value & $1000$\\
	 & in the online classroom &&\\
\hline
	$M$ & Maximum number of answers & Any finite value & $100$\\
	& from a student (Last state of &&\\
		& the lumped CTMC) 	&	 &\\
\hline
	$L$ & Number of types of & Any finite value ($\leq n$) & 2  \\
	& students being considered &&\\
\hline
$Type_l$ & Set containing the & Finite set of students & $|Type_1|$ = $|Type_2|$ = $n/2$ \\
& students of Type $l$ &&\\
\hline
	$T$ & Duration of the & Any finite value & 11\\
	& online course (Units: Weeks) &&\\
\hline
   $\lambda$ & Rate of arrival of & Any finite set of values & $\{1,4,5,6,7,8,10\}$\\ 
   & \textbf{one} student (per Week)  &&\\
\hline
	$\mu$ & Rate of arrival of & Any finite set of values & $\{1,2, \ldots,10\}$\\ 
	& the instructor (per Week) &&\\
\hline
	$\alpha_l$  & Cost per arrival & $\left(0,1 \right)$  &  $\alpha_1 \in $ $ \{0.1,0.2,0.3 \}$  ($Type_1$ - High Caliber/Low Cost)\\
	$(1 \leq l \leq L)$	& to a $Type_l$ student && $\alpha_2 \in $  $\{0.6,0.8,0.9 \}$ ($Type_2$ - Low Caliber/High Cost)  \\
\hline
	$\beta$ &  Cost per arrival & $(0,1)$ & $0.5$\\ 	
	& to the instructor &&\\
\hline
	$\delta$ & Willingness of the & $(0,1)$ & $0.9$\\ 			
	& instructor to reward &&\\
\hline
	$m_l$ & Maximum number of & Any finite value & $\{ 2,6,10\}$\\
	& rewards per question &&\\
	      &  for each  student $i \in Type_l$  & & \\ 
\hline
	$c_i$ & Bias of the instructor & $c_i \in (0,1)$, $\sum_{i=1}^{n} c_i = 1 $  & Extreme Bias towards $Type_1$ $\left(\text{or, }Type_2\right)$ students: \\
	& towards a student $i$ &$c_i=c_j \text{ } \forall i,j \in Type_l$ & $c_i = \frac{0.99}{n_1} \left(\text{or, } \frac{0.01}{n_1}  \right) \forall i \in Type_1$,\\
\hline
\end {tabular}
\caption{{Parameters of the Model}}\label{Table_Parameters}
\end{table*}

 We observe from Figure~\ref{Simulations_Error_Vs_T} that the normalized error tends to zero as time progresses. But this error becomes negligible only after 10-15 weeks and is quite significant before that period. The typical duration of online classrooms is around 4-10 weeks \citep{MLR2013} and this necessitates the use of transient analysis over steady state analysis in such a setting for accurate results. Note that the normalized error can be any value greater than or equal to zero as the steady state and avg. aggregate utilities can have the same or opposite signs at a time $t$.  Also, note that we observe a similar behavior for other parameter configurations as well and hence we do not discuss them here. Also, due to the significance of the transient analysis, we shall provide the simulation results from only the transient analysis of our model. The results based on steady state analysis follow a similar trend. 

 \subsection{Heterogeneous Student Participation: Effect of Budget and Instructor Rate}

\begin{figure*}[h]
\centering
\begin{tabular}{|c|c|}
\hline
\begin{minipage}{7 cm}
\centering \vspace{0.2cm}
 \includegraphics[scale = .5]{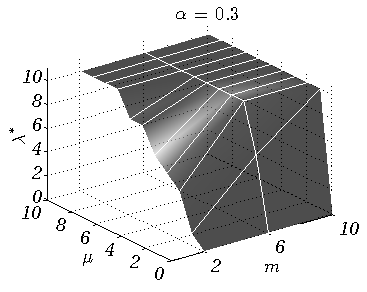}
\end{minipage}
&
\begin{minipage}{7 cm}
\centering
\includegraphics[scale = .5]{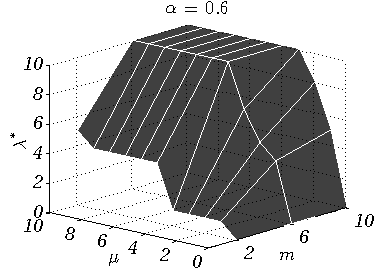}
\end{minipage}
\\
(a)&(b)\\
\hline
\begin{minipage}{7 cm}
\centering 
 \includegraphics[scale = .5]{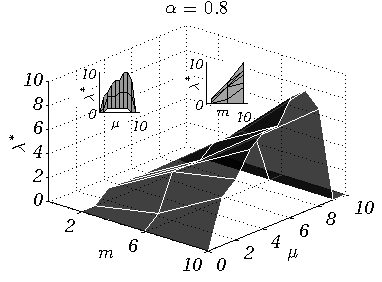}
\end{minipage}
&
\begin{minipage}{7 cm}
\vspace{1mm} 
\centering
\includegraphics[scale = .5]{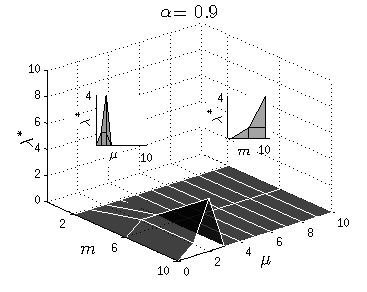}
\end{minipage}
\\
(c)&(d)\\
\hline
\end{tabular}
\caption{ Variation of the optimal participation rates ($\lambda^*$) of  each student belonging to the different types (characterized by four different $\alpha$'s) with change in the instructor's arrival rate ($\mu$) and the maximum number of rewards ($m$) given to that type. The $X-Z$ and $Y-Z$ projections of Figures~\ref{Simulations_3D_Graphs}c and ~\ref{Simulations_3D_Graphs}d  are given as insets to aid the understanding of the graphs.  
\label{Simulations_3D_Graphs}
}
\end{figure*}

We study the variation in the optimal student participation rate ($\lambda^*$) of the heterogeneous student population with $\mu$ and $m$ for different student types in Figure~\ref{Simulations_3D_Graphs}. We observe that, if we keep $\mu$ constant and increase the reward $m$, then $\lambda^*$ either increases or remains the same. Hence, giving higher rewards can increase student participation but only up to a certain extent.

If we fix $m$ and study the variation of $\lambda^*$ with changing $\mu$ in Figure~\ref{Simulations_3D_Graphs}(a) to Figure~\ref{Simulations_3D_Graphs}(d), we observe that as the cost per arrival ($\alpha$) of a student increases, $\lambda^*$ becomes  sensitive to $\mu$. This means that the students with lower $\alpha$ are self motivated and have only a limited impact of $\mu$ (See Figure~\ref{Simulations_3D_Graphs}(a)). They act as super-posters and aggressively post on the OEF being only slightly affected by the instructor participation rates. This phenomenon is also corroborated in our empirical study as shown in Figure~\ref{Figure_OEF2}(a).

However, we observe non-monotonic participation patterns for students with high cost per arrival ($\alpha=0.8, 0.9$) in Figures~\ref{Simulations_3D_Graphs}(c) and~\ref{Simulations_3D_Graphs}(d) as $\lambda^*$  initially increases with increasing $\mu$ and then, with any further increase in $\mu$, $\lambda^*$ starts falling. This trend has been noted in literature (for example: \citep{GHOSH13}). 
Our model is thus able to corroborate  real-world behavior of heterogeneous students in an OEF varying from the disinterested students to those who are highly active. We observe that the active students are self-motivated and will visit the forum even for low rewards and their level of participation generally increases as the instructor arrives more frequently on the OEF. However, the not-so-active students want high rewards and lower instructor participation i.e., they want lesser questions (and more time) with higher incentives.
 
\subsection{Instructor-Student Arrival Rates: Effect of Instructor Bias}

It is generally observed that instructors have varying notions of bias towards different types of students. Some instructors may value the participation from the weaker students higher than good students while others may focus on the brighter students. In our model, the parameter $c_i$ (See Table~\ref{Table_Parameters}) captures the inherent instructor bias towards a student $i$. This parameter is an input parameter to the model and is set based on how the instructor values an answer from the different types of students. 

\begin{figure}[h]
\vspace{-.3cm}
\hspace{-0.2in}
\begin{tabular}{ll}
\begin{minipage}{7.5cm}
\centering
\includegraphics[scale = .17]{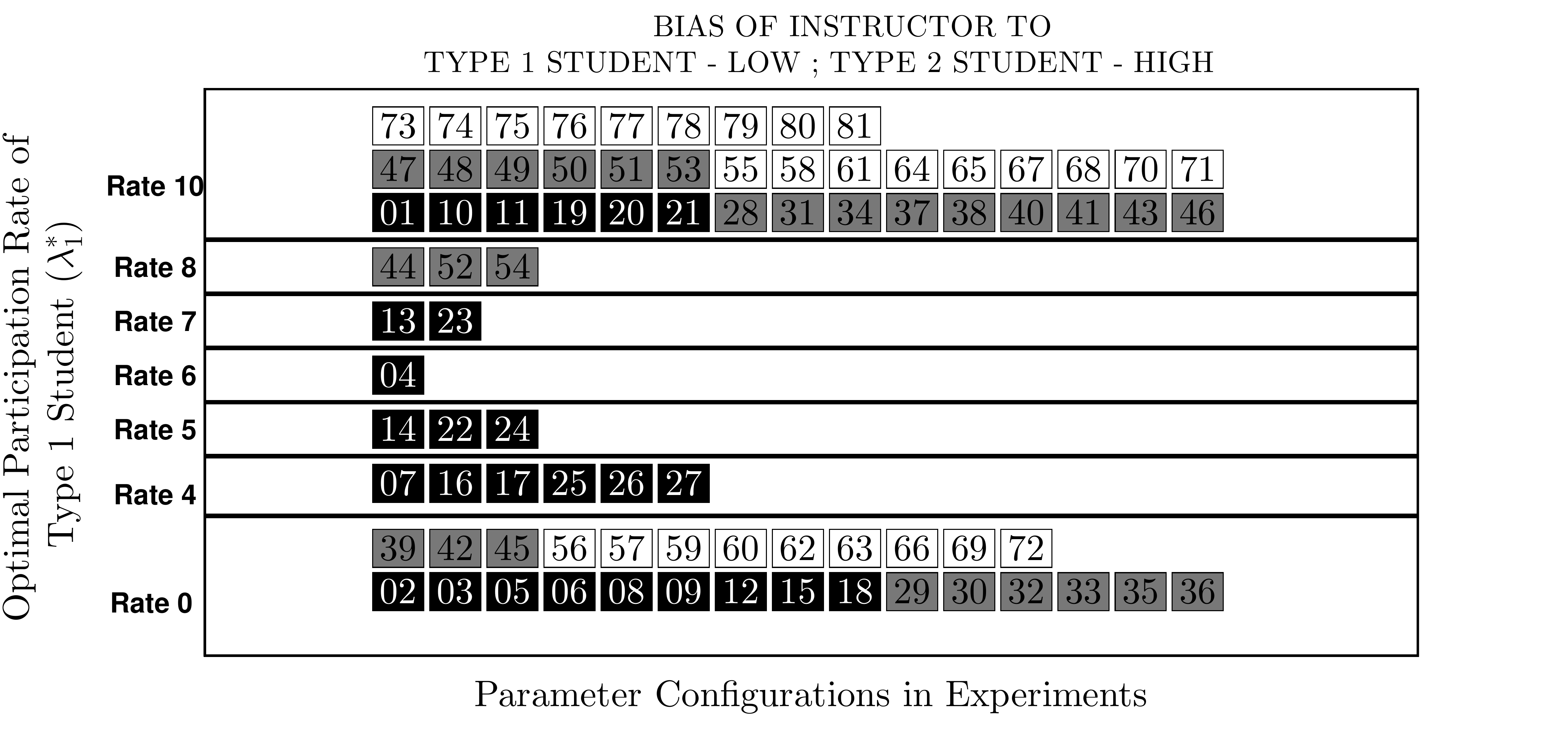}
\end{minipage}
&
\begin{minipage}{7.5cm}
\centering \includegraphics[scale = .17]{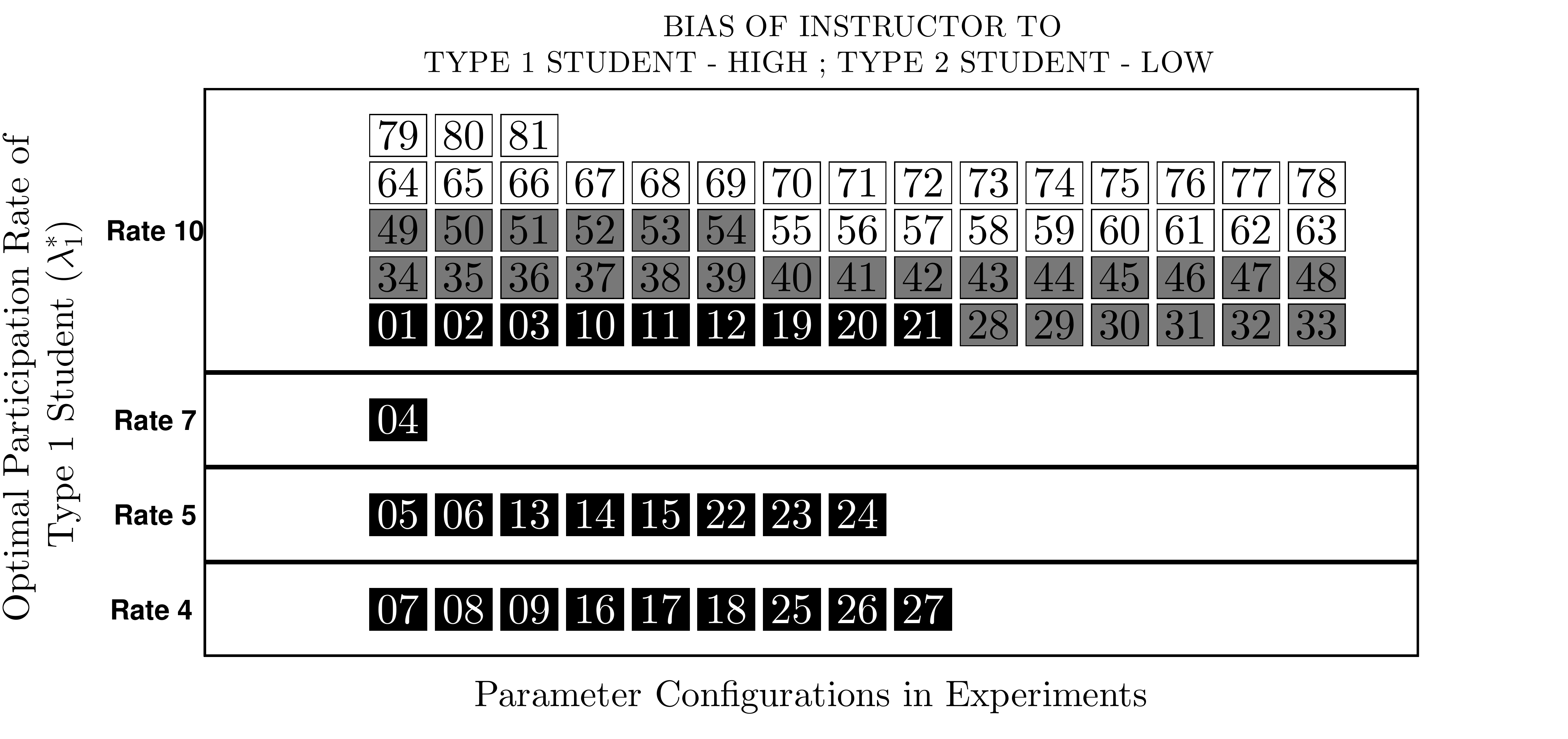}
\end{minipage}
\\
\begin{minipage}{7.5cm}
\centering (a)  
\end{minipage}
&
\begin{minipage}{7.5cm}
\centering (b)  
\end{minipage}
\\
\begin{minipage}{7.5cm}
\centering
\includegraphics[scale = .17]{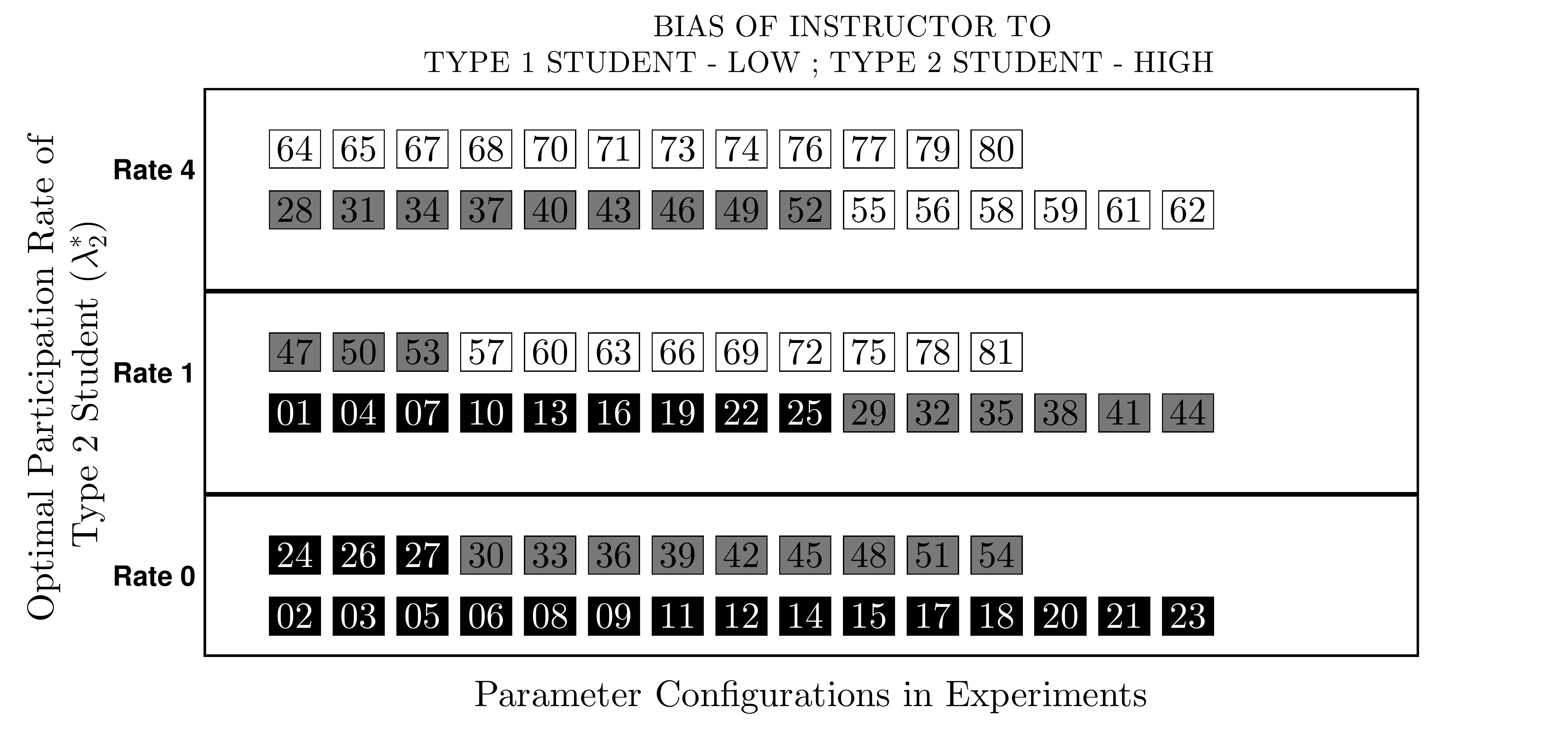}
\end{minipage}
&
\begin{minipage}{7.5cm}
\centering \includegraphics[scale = .17]{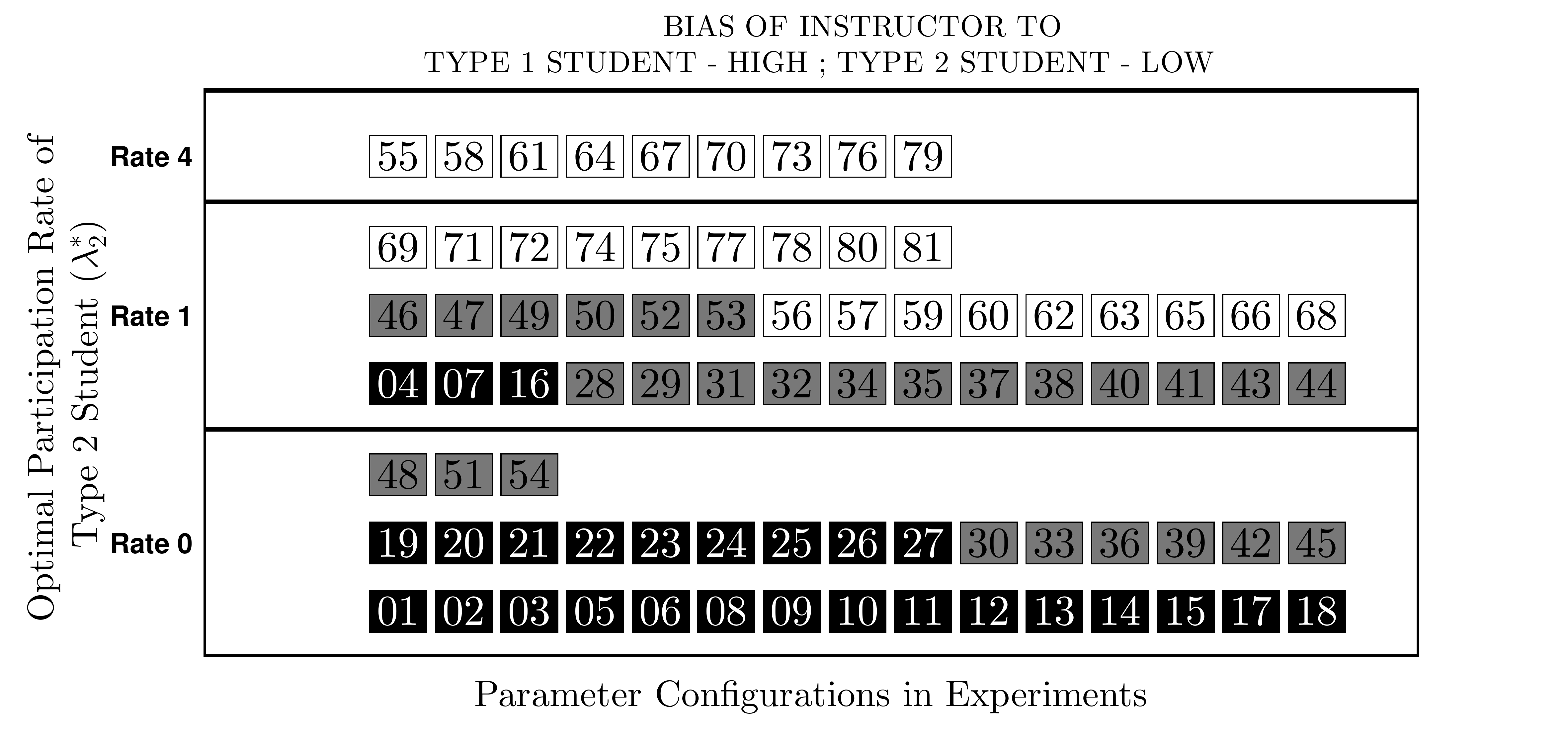}
\end{minipage}
\\
\begin{minipage}{7.5cm}
\centering (c)  
\end{minipage}
&
\begin{minipage}{7.5cm}
\centering (d)  
\end{minipage}
\\
\begin{minipage}{7.5cm}
\centering
\includegraphics[scale = .17]{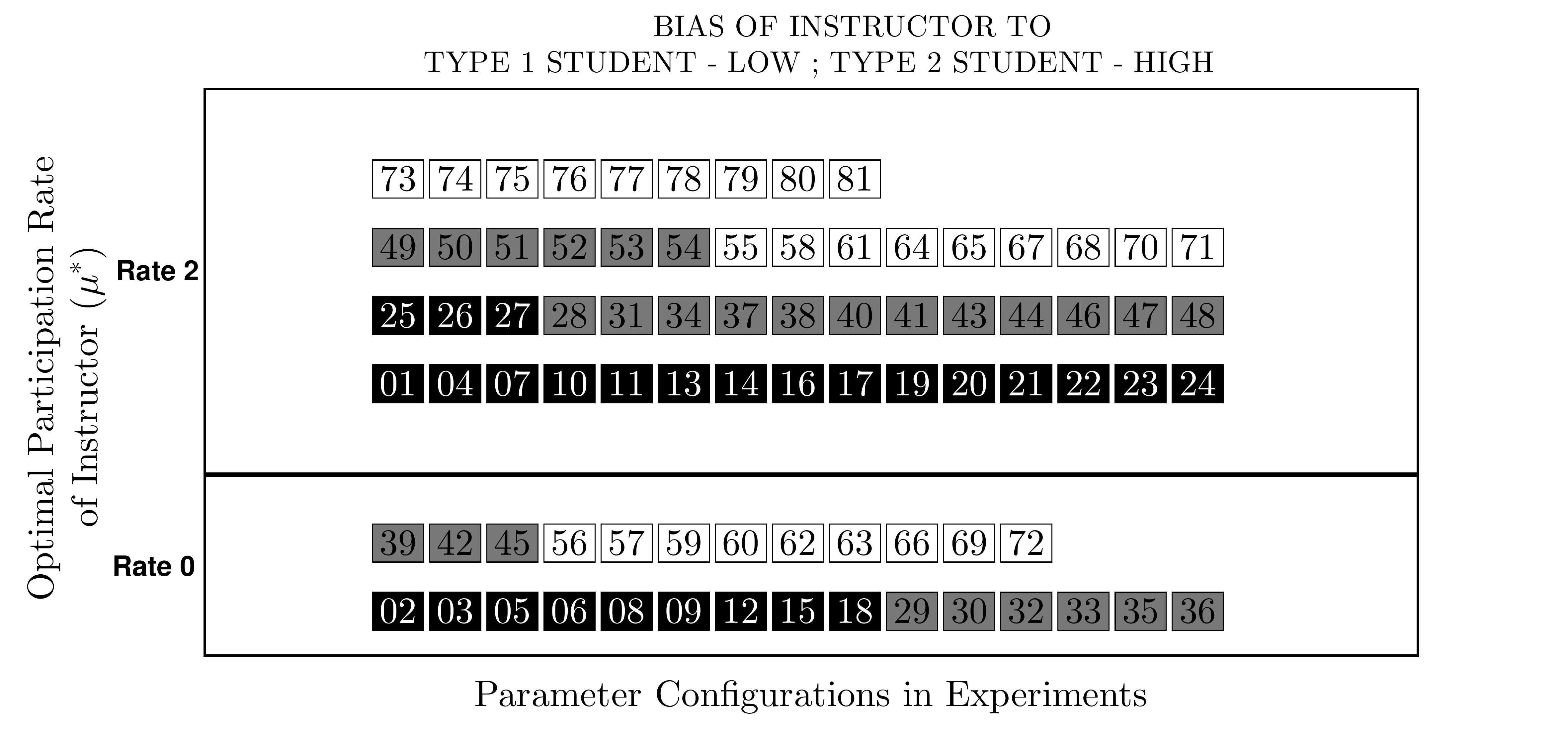}
\end{minipage}
&
\begin{minipage}{7.5cm}
\centering \includegraphics[scale = .17]{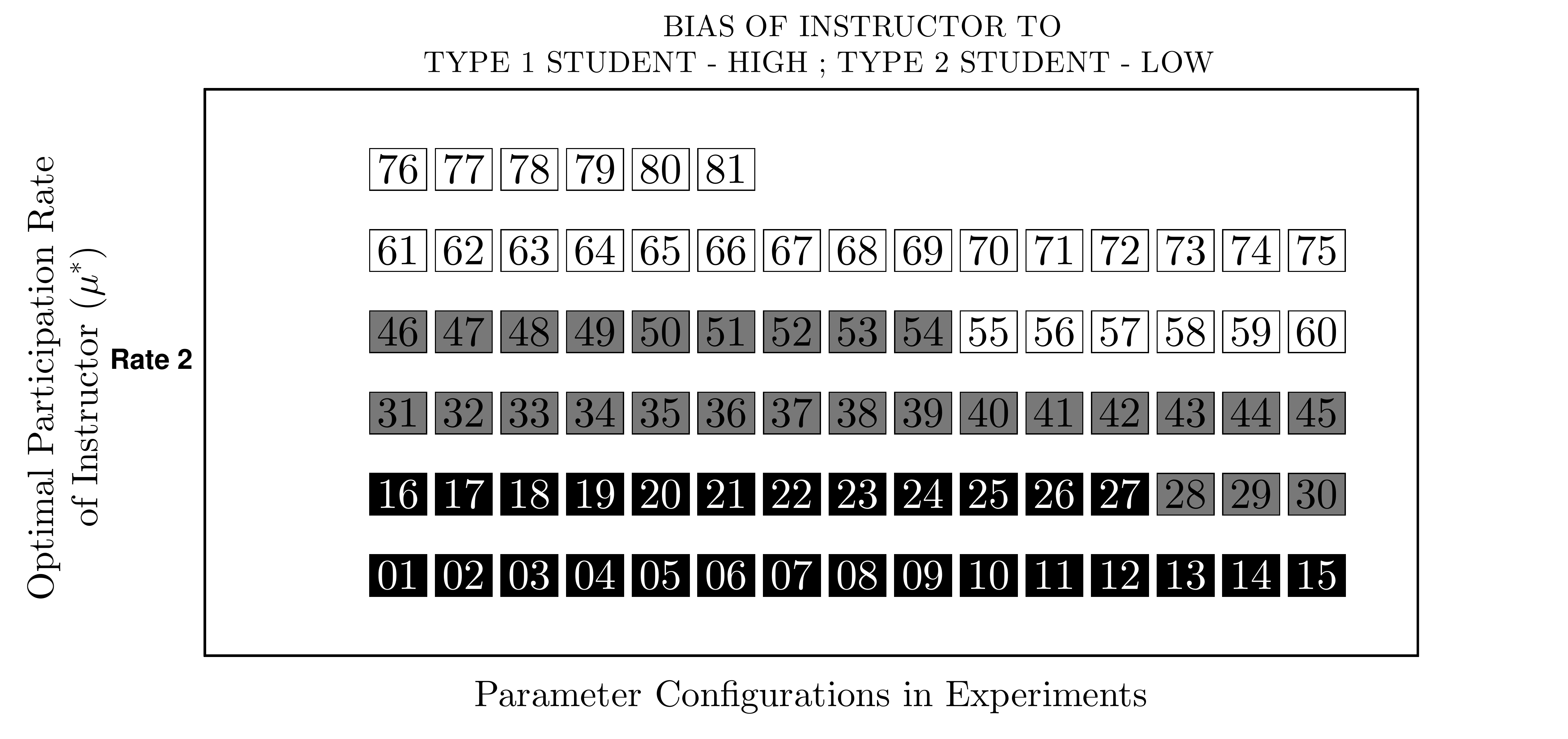}
\end{minipage}
\\
\begin{minipage}{7.5cm}
\centering (e)  
\end{minipage}
&
\begin{minipage}{7.5cm}
\centering (f)  
\end{minipage}
\vspace{-0.1in}
\end{tabular}
\caption{\footnotesize Effect of Instructor bias on the optimal participation rates of a Type~$1$ student. Each square represents the respective optimal participation rate of a Type~$1$ student depending on  parameter configuration in experiment. Black, Grey and White squares indicate configurations  where $m_1=2$(low rewards), $m_1=6$ (medium rewards) and $m_1=10$ (high rewards) respectively.\label{experimentsScatterPlots}}

\vspace{-0.1in}
\end{figure}

We consider a class with 1000 students categorized into two types (Table~\ref{Table_Parameters}). For illustration purposes, we consider 2 students from this class: Student~$1$ (belonging to $Type_1$ with low $\alpha_1$) who is very active on the forum  and Student~$2$ (belonging to $Type_2$ with high $\alpha_2$) who does not like to post frequently on the OEF. We fix the following parameters: $\alpha_1$, $\alpha_2$, $m_1$ (budget allocated per question to answers from Type 1 students), $m_2$ (budget for Type 2 students) where each parameter takes values: $\alpha_1 \in \{ 0.01, 0.1,0.2 \}, \alpha_2 \in \{0.8, 0.9,0.99 \}, m_1,m_2 \in \{ 2, 6, 10\}$ resulting in 81 (i.e., $3\times3\times3\times 3$) configurations. We first set the instructor bias as a low value ($c_1=(0.01/n_1)$)  for Student~1 and as a high value ($c_2=(0.99/n_2)$) for Student~2 where $n_1$ and $n_2$ are number of students of $Type_1$ and $Type_2$ respectively. We examine the optimal rates of participation for Student~1, Student~2, and instructor for each of the possible $81$ configurations in the presence of instructor bias towards certain type of students. Note that a configuration basically corresponds to a fixed value for the parameters: ($\alpha_1$, $\alpha_2$, $m_1$, $m_2$). 

We will initially consider an instructor with low bias towards Student~$1$ and very high bias to Student~$2$ i.e., we set $c_1=(0.01/n_1)$ and $c_2=(0.99/n_2)$ where $n_1$ and $n_2$ are the number of students of $Type_1$ and $Type_2$ respectively and $c_i$ is the bias of instructor towards Student~$i$. Note that these parameters affect the utility values of the instructor as described in Equation~\ref{Utility_Instructor_OEF}. Under this setting, we generate scatter plots depicting optimal participation rates where each point in the scatter plot corresponds to one of the $81$ parameter configurations. These 81 parameter configurations are numbered uniquely so that we can compare the results when the instructor reverses his bias values. The scatter plots are given in Figure~\ref{experimentsScatterPlots}(a), Figure~\ref{experimentsScatterPlots}(c), and Figure~\ref{experimentsScatterPlots}(e) and these correspond to the optimal rates of participation for Student~1, Student~2, and instructor respectively. Note that these plots correspond to the scenario when the instructor has low bias to Student~1 and high bias to Student~2. We change the instructor behaviour to have high bias towards Student~1 and very low bias towards Student~2 fixing $c_1=(0.99/n_1)$ and $c_2=(0.01/n_2)$ and run the experiments similarly for the $81$ parameter configurations as given above. The optimal participation rates of Student~1, Student~2, and instructor are given in Figure~\ref{experimentsScatterPlots}(b), Figure~\ref{experimentsScatterPlots}(d), and Figure~\ref{experimentsScatterPlots}(f) respectively. 

We observe that, if the instructor's bias towards Student~$1$ is low then there are configurations when Type~$1$ student will not participate even for medium and high rewards (for example: configs 45, 56 in Figure~\ref{experimentsScatterPlots}(a) have optimal rate as 0)  whereas if the instructor's bias towards Type~$1$ students is high, then these students start participating enthusiastically with high rates for medium and high rewards and increase their participation levels even for the lower rewards (for example: configs 45, 56 in Figure~\ref{experimentsScatterPlots}(a) have optimal rate 10 while config 09 improved from rate 0 in Figure~\ref{experimentsScatterPlots}(a) to rate 4 in Figure~\ref{experimentsScatterPlots}(b)). 
We also see that there is a dip in participation of Student~2 for many configurations when the instructor decreases his bias to Student~2 (See Figure~\ref{experimentsScatterPlots}(c) and Figure~\ref{experimentsScatterPlots}(d)). For example, see configs 01, 28 in Figure~\ref{experimentsScatterPlots}(c) and Figure~\ref{experimentsScatterPlots}(d). In Figure~\ref{experimentsScatterPlots}(c) for Config 01, Student~2 participates with Rate~1 whereas when the instructor decreases bias to Student~2, for this configuration, Student~2 optimal rate decreases to Rate~0 as shown in Figure~\ref{experimentsScatterPlots}(d). Similarly Student~2 decreases his optimal rate from Rate~4 to Rate~2 when the instructor decreases the bias for Student~2 for the configuration 28 as shown in Figure~\ref{experimentsScatterPlots}(c) and Figure~\ref{experimentsScatterPlots}(d). 

We also observe that the instructor needs to increase his/her participation when there is a higher bias to Student~1 than the scenario when there is a higher bias to Student~2. This is beacuse Student~1 is an enthusiastic student compared to Student~2 and hence, to keep the attention of Student~1 , the instructor needs to post more questions on the forum. In the scenario where the instructor has high bias to Student~2, the optimal rate of the instructor can be low as Student~2 dislikes answering too many questions on the forum as the cost of answering the questions overshadows the rewards obtained by Student~2. 

To summarize, these experiments shed interesting insights into the role of instructor bias parameter of our proposed OEF model and thus, makes our proposed OEF model more realistic. It has been observed that participation of students is indeed affected by instructor biases towards certain students and the instructor bias parameter enables us to capture this effect in our proposed model for OEFs.

\vspace{-0.1in}
\section{Conclusions}
\label{Section_conclusions}
\vspace{-0.05in}

In this work, we performed empirical analysis of online education forums (OEFs) of the Game Theory course offered in the Indian Institute of Science in 2012 and 2014. We identified several key parameters which dictate the activities of OEFs like student heterogeneity, effective incentive design, super-posters phenomenon, etc. Motivated by empirical observations, we proposed a continuous time Markov chain (CTMC) model to capture instructor -student interactions in an OEF. Using concepts from lumpability of CTMCs, we performed steady state and transient analysis evaluate expected steady state and transient aggregate net-rewards for the instructor and the students. We then formulated a mixed-integer linear program  which views the OEFs strategically as a single-leader-multiple-follower Stackelberg game. We undertook detailed simulations and develop new insights into the activities of OEF by studying the effects of parameters like participation costs, budget and instructor bias  on the student/instructor participation rates. Our model corroborated with some key empirical observations and could recommend an optimal plan to the instructor for maximizing student participation in OEFs. 

\section{Future Work}
\label{futurework}
The analysis used in the paper assumed certain parameters (for eg: $m_l$ parameter in Table~\ref{CTMCnotationtable1}) to be fixed. An additional preprocessing stage could be introduced to optimally solve for such parameters which could improve the results of the model. Further, students could collude on their strategies to improve utilities while reducing their participation levels. Devising instructor-driven collusion-resistant strategies could be an interesting direction. Our work does not consider effort modeling of the students. Information about the effort put by the students could guide the incentive design and may result in better quality of participation among the students in the OEFs.

\bibliographystyle{unsrtnat}
\bibliography{onlineedforums-ctmc-stackelberg}

\setcounter{proposition}{0}
\setcounter{theorem}{4}
\setcounter{lemma}{0}

\appendix

\section{Proofs : Lumpability}

\begin{theorem} 
 \textbf{(i)} $X(t) = (\mathcal{S},Q)$ is lumpable w.r.t. partition ${\overline{S}_i} = \{ {\overline{S}^a_i} | a \in \{ 0,1,\ldots,M \} \} $. 

\textbf{(ii)} The quotient (lumped) Markov chain ${\overline{X}_i}(t) = ({\overline{S}_i},{\overline{Q}_i})$ we get on lumping the CTMC $X(t)$ w.r.t. partition ${\overline{S}_i}$  ($ i \in \{ 1,2,\ldots,n \}$) is given as :

\begin{align}
\nonumber {\overline{Q}_i}({\overline{S}^a_i},{\overline{S}^b_i}) =&
			   \begin{cases} 
 	  				\lambda_i   & b=a+1, \\
					\mu   	    & b=0\neq a,\\
					\omega	    & b=a,\\
					0 	    & o/w.
			   \end{cases} \\
\nonumber  \text{where, } \omega = -\sum\limits_{c \in D \setminus \{ a \} } & {\overline{Q}_i}({\overline{S}^a_i},{\overline{S}^c_i}),D = \{0,1,\ldots,M \}.
\end{align}

\end{theorem}

\begin{proof}
As given in Definition ~\ref{define_lumping}, $X(t)$ is lumpable w.r.t ${\overline{S}_i}$ if for any two blocks ${\overline{S}^a_i}, {\overline{S}^b_i} \in {\overline{S}_i}$ and for every $v,y \in {\overline{S}^a_i}$ we have $q(v,{\overline{S}^b_i}) = q(y,{\overline{S}^b_i})$ i.e. the rate of transition from each state in block ${\overline{S}^a_i}$ to block ${\overline{S}^b_i}$ should be equal. Fig. ~\ref{Diagram_Lumpability_Proof} shows the two blocks ${\overline{S}^a_i}$ and ${\overline{S}^b_i}$.

 \begin{figure}[!hbtp]
 \begin{center}
  \includegraphics[scale = .4]{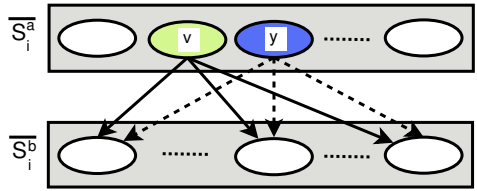}
 \end{center}
 \caption[Proof Sketch]{Two blocks  $S_i^a$ and $S_i^b$ of the partition $S_i$ are shown. The solid lines depict transition from state $v$ in the $S_i^a$ block to states in the $S_i^b$ block. And the dotted lines depict transition from state $y$ in the  $S_i^a$ block to states in the $S_i^b$ block. We need to prove that the sum of rates of solid transition lines is equal to the sum of rates of dotted transition lines. 
\label{Diagram_Lumpability_Proof}}
 \end{figure}

 By definition,  ${\overline{S}^a_i} = \{(x_i, x_{-i})\in\mathcal{S} | x_i = a \} $  and ${\overline{S}^b_i} = \{(x_i, x_{-i}) \in\mathcal{S} | x_i = b \} $. Now states $v,y \in {\overline{S}^a_i}$, hence these maybe represented as: $v = (a, v_{-i})$ and $y = (a, y_{-i})$ respectively. Also we represent a state $z\in S^b_i$ as: $z= (b, z_{-i})$. Let $D = \{ 0,1,\ldots,M\}$ and $\mathcal{S}_{-i} = \{ (x_{-i})| (x_i,x_{-i}) \in  \mathcal{S} \}$. Now we need to prove that  $\forall {\overline{S}^a_i},{\overline{S}^b_i} \in {\overline{S}_i}$ and for any   $v,y \in {\overline{S}^a_i}$,
\begin{align}
     \nonumber & q(v,{\overline{S}^b_i}) = q(y,{\overline{S}^b_i}) \quad \text{or, } \quad \sum\limits_{z \in {\overline{S}^b_i}} Q(v,z) = \sum\limits_{z \in {\overline{S}^b_i}} 
      Q(y,z) \\
\nonumber  \text{or, }  & \sum\limits_{z_{-i} \in \mathcal{S}_{-i}} Q((a,v_{-i}),( b, z_{-i})) =   
    \sum\limits_{z_{-i} \in \mathcal{S}_{-i}} Q((a,y_{-i}),( b, z_{-i}))
\end{align}

We consider the following exhaustive cases. 
\begin{enumerate}
	\item $\underline{ b \neq a}$
		\begin{itemize} 

			\item $\underline{ b = a + 1}$\\
				\begin{flalign}
					\nonumber \textbf{LHS }  =& \sum\limits_{z_{-i} \in \mathcal{S}_{-i}} Q((a,v_{-i}),( b, z_{-i})) = \sum\limits_{z_{-i} \in \mathcal{S}_{-i}} Q((a,v_{-i}),( a+1, z_{-i}))  &\\
					 \nonumber   =& Q((a,v_{-i}),( a+1, v_{-i})) +   \sum\limits_{z_{-i} \in  \mathcal{S}_{-i} \setminus \{ v_{-i} \} } Q((a,v_{-i}),( a+1, z_{-i})) = \lambda_i + 0 = \lambda_i  &\\
					 \nonumber    & (\text{As the first }  Q \text{ expression satisfies Case}~(i)  \text{ and all the other $Q$ }  \text{expressions in the}& \\
					\nonumber     &  \text{summation satisfy Case}~(iv)  \text{ of  Equation~\ref{generator_matrix}}.)  &
					\end{flalign}
					
				\begin{flalign}
				\nonumber \textbf{RHS }  =& \sum\limits_{z_{-i} \in \mathcal{S}_{-i}} Q((a,y_{-i}),( b, z_{-i})) = \sum\limits_{z_{-i} \in \mathcal{S}_{-i}} Q((a,y_{-i}),( a+1, z_{-i}))  & \\
				\nonumber =& Q((a,y_{-i}),( a+1, y_{-i})) +   \sum\limits_{z_{-i} \in  \mathcal{S}_{-i} \setminus \{ y_{-i} \} } Q((a,y_{-i}),( a+1, z_{-i})) = \lambda_i + 0  =  \lambda_i & \\
					 \nonumber    & (\text{As the first }  Q \text{ expression satisfies Case}~(i)  \text{ and all the other $Q$ expressions in the}  & \\
					\nonumber     & \text{summation} \text{ satisfy Case}~(iv)  \text{ of  Equation~\ref{generator_matrix}}.)  &
					\end{flalign}
				$\therefore$ $LHS = RHS$ in this case.

			\item $\underline{ b > a + 1}$\\
				  Similar analysis as above yields $LHS=RHS=0$ as Case~$(iv)$ of Equation~\ref{generator_matrix}  is satisfied for the whole 
				  summation. 

			\item $\underline{ b < a , b \neq 0}$\\
				  Similar analysis as above yields $LHS=RHS=0$ as Case~$(iv)$ of Equation~\ref{generator_matrix}  is satisfied for the whole 
				  summation. 

			\item $\underline{ b = 0 , b \neq a}$
				\begin{flalign}
				\nonumber \textbf{LHS } =& \sum\limits_{z_{-i} \in \mathcal{S}_{-i}} Q((a,v_{-i}),( b, z_{-i})) = \sum\limits_{z_{-i} \in \mathcal{S}_{-i}} Q((a,v_{-i}),( 0, z_{-i}))  \\
				\nonumber =& Q((a,v_{-i}),( 0,0,\ldots,0)) +  \sum\limits_{z_{-i} \in  \mathcal{S}_{-i} \setminus \{ (0,\ldots,0) \} } Q((a,v_{-i}),( 0, z_{-i})) = \mu + 0 = \mu& \\
				\nonumber & (\text{As the first }  Q \text{ expression satisfies Case}~(ii) \text{ of  Equation~\ref{generator_matrix}  and all the other $Q$} & \\
				\nonumber &  \text{ expressions in the summation satisfy Case}~(iv) \text{ of Equation~\ref{generator_matrix}.}  ) &
					\end{flalign}
				\begin{flalign}
				\nonumber \textbf{RHS } =& \sum\limits_{z_{-i} \in \mathcal{S}_{-i}} Q((a,y_{-i}),( b, z_{-i})) = \sum\limits_{z_{-i} \in \mathcal{S}_{-i}} Q((a,y_{-i}),( 0, z_{-i}))& \\
				\nonumber =& Q((a,y_{-i}),( 0,0,\ldots,0)) +  \sum\limits_{z_{-i} \in  \mathcal{S}_{-i} \setminus \{ (0,\ldots,0) \} } Q((a,y_{-i}),( 0, z_{-i})) =  \mu + 0  =  \mu& \\
				\nonumber & (\text{As the first }  Q \text{ expression satisfies Case}~(ii) \text{ of  Equation~\ref{generator_matrix}  and all the other $Q$} & \\
				\nonumber &  \text{ expressions in the summation satisfy Case}~(iv) \text{ of  Equation~\ref{generator_matrix}.}  )  &
					\end{flalign}
				$\therefore$ $LHS = RHS$ in this case.	
		\end{itemize}

	\item $\underline{ b = a}$
				\begin{flalign}
				\nonumber \textbf{LHS } =& \sum\limits_{z_{-i} \in \mathcal{S}_{-i}} Q((a,v_{-i}),( b, z_{-i})) = \sum\limits_{z_{-i} \in \mathcal{S}_{-i}} Q((a,v_{-i}),( a, z_{-i})) & \\
				\nonumber =& \sum\limits_{z_{-i} \in \mathcal{S}_{-i} \setminus \{v_{-i}\} } Q((a,v_{-i}),( a, z_{-i})) + Q((a,v_{-i}),( a, v_{-i}))& \\
				\nonumber & ( \text{ Last } Q \text{ expression satisfy Case}~(iii) \text{ of Equation~\ref{generator_matrix}.}) & \\
				\nonumber =& \sum\limits_{z_{-i} \in \mathcal{S}_{-i} \setminus \{ v_{-i}  \} } Q((a,v_{-i}),( a, z_{-i})) -  \sum\limits_{(z_i,z_{-i}) \in \mathcal{S} \setminus \{(a,v_{-i}) \}}  Q((a,v_{-i}),(z_i , z_{-i}))& \\
				\nonumber & \text{( Expanding the second summand) } & \\
				\nonumber =& \sum\limits_{(z_{-i} \in \mathcal{S}_{-i} \setminus \{ v_{-i}  \} )}
					    	 Q((a,v_{-i}),( a, z_{-i})) -  \sum\limits_{(z_{-i} \in \mathcal{S}_{-i} \setminus \{v_{-i}\})}
					     	 Q((a,v_{-i}),(a , z_{-i}))-& \\
				\nonumber & \quad \hspace{1mm} \sum\limits_{(z_i \in D \setminus \{ a \})} \sum\limits_{(z_{-i} \in \mathcal{S}_{-i})}
					     	 Q((a,v_{-i}),(z_i , z_{-i}))& \\
				\nonumber =& -\sum\limits_{(z_i \in D \setminus \{ a \})} \sum\limits_{(z_{-i} \in \mathcal{S}_{-i})}
					     	 Q((a,v_{-i}),(z_i , z_{-i}))&
\end{flalign}			
			
			Similarly on solving for $RHS$ we get, \begin{flalign}
		\nonumber 	\textbf{RHS } =& -\sum\limits_{(z_i \in D \setminus \{ a \})}
			 \sum\limits_{(z_{-i} \in \mathcal{S}_{-i})} Q((a,y_{-i}),(z_i , z_{-i}))&
\end{flalign}			
			 Now, from \textbf{ Case~$1$ } we see that,
				\begin{align}
			    	\forall c \neq a,
			 	\nonumber &\sum\limits_{z_{-i} \in \mathcal{S}_{-i}}  Q((a,v_{-i}), 	( c, z_{-i}))  = \sum\limits_{z_{-i}
			 		\in \mathcal{S}_{-i}} Q((a,y_{-i}),( c, z_{-i})),  \\
				\nonumber \Rightarrow &
			  		\sum\limits_{(z_i \in D \setminus \{ a \})}  \sum\limits_{z_{-i} \in \mathcal{S}_{-i}} Q((a,v_{-i}),
			 		(z_i, z_{-i})) = \sum\limits_{(z_i \in D \setminus \{ a \})} \sum\limits_{z_{-i} \in 
			 		\mathcal{S}_{-i}}  Q((a,y_{-i}),( z_i, z_{-i})) \\
				\nonumber \Rightarrow &
			  		-\sum\limits_{(z_i \in D \setminus \{ a \})}  \sum\limits_{z_{-i} \in \mathcal{S}_{-i}} Q((a,v_{-i}),
			 		( z_i, z_{-i})) = -\sum\limits_{(z_i \in D \setminus \{ a \})} \sum\limits_{z_{-i} \in 
			 		\mathcal{S}_{-i}} Q((a,y_{-i}),( z_i, z_{-i}))\\
			 	\nonumber \therefore \text{ } LHS =& RHS.
				\end{align}		
\end{enumerate}

Thus we prove that $\forall {\overline{S}^a_i},{\overline{S}^b_i} \in {\overline{S}_i}$ and for any   $v,y \in {\overline{S}^a_i}$, $ q(v,{\overline{S}^b_i}) = q(y,{\overline{S}^b_i}) $, thus the CTMC $X(t)$ is lumpable w.r.t partition ${\overline{S}_i}$.
\end{proof}

\begin{proof}(ii)
Also, as stated in Definition~\ref{define_lumping}$(ii)$, the generator matrix for the  quotient Markov chain ${\overline{X}_i}(t)=({\overline{S}_i},{\overline{Q}_i})$ will be determined by ${\overline{Q}_i}({\overline{S}_i^a},{\overline{S}_i^b})=q(x,{\overline{S}_i^b})$  for any ${\overline{S}_i^a},{\overline{S}_i^b} \in {\overline{S}_i}$, where $x$ is any state in ${\overline{S}_i^a}$. And these quantities have already been computed in the Cases~$1$ and $2$ above. Hence Proved.
\end{proof}
   
\begin{lemma}
${\overline{\pi}^t_i}(\overline{ x}) =  \sum\limits_{x \in {\overline{S}_i^{\overline{x}}}} \pi^t(x)$
\end{lemma}
\begin{proof}
The CTMC $X(t)$ being considered has a finite state space $S$ and the arrival rates of the students and the instructor on the CTMC are already known. Also because of the nature of the problem, the arrival rate $\lambda_i$ for a student $i$ ($i \in \{ 1,2,\ldots,n \}$)  and the arrival of the instructor $\mu$ will be some finite values. Thus each value in the generator matrix $Q$ of $X(t)$ defined in Equation~\ref{generator_matrix} will be a finite value. Following a similar argument we can say that each value in the generator matrix ${\overline{Q}_i}$ of ${\overline{X}_i}(t)$ (for any $i \in \{ 1,2,\ldots,n \}$) defined in Equation~\ref{generator_matrix} will be a finite value. Thus there exists a finite number $\widehat{q} < \infty$ which bounds the rate entries in the rate matrices $Q$ and ${\overline{Q}_i}$ ($\forall i \in \{ 1,2,\ldots,n \}$).

Thus using Theorem ~\ref{dtmc_underlying_and_lumping_other_paper_theorem} for CTMCs $X(t)$ and ${\overline{X}_i}(t)$ we can say that:
\begin{enumerate}
\item
\begin{enumerate}
\item there exists an underlying DTMC $Y_k$ for the CTMC $X(t)$ with transition matrix $P = Q/\widehat{q} + I$.
\item there exists an underlying DTMC ${\overline{Y}^i_k}$ for the CTMC ${\overline{X}_i}(t)$ with transition matrix $\overline{P_i} = {\overline{Q}_i}/\widehat{q} + I$.
 \end{enumerate}
\item 
\begin{enumerate}
\item the distribution vector $\pi^t$ of the CTMC $X(t)$ at time $t$ starting with $\pi^0$ at time $0$ is given by $\pi^t = \sum\limits_{k=0}^{\infty} \exp(-\widehat{q} t)\left[ (\widehat{q} t)^k / k! \right] \pi^k$, where $\pi^k$ is the distribution vector of $Y_k$ after $k$ jumps.
\item the distribution vector ${\overline{\pi}_i^t}$ of the CTMC ${\overline{X}_i}(t)$ at time $t$ starting with ${\overline{\pi}^0_i}$ at time $0$ is given by ${\overline{\pi}_i^t} = \sum\limits_{k=0}^{\infty} \exp(-\widehat{q} t)\left[ (\widehat{q} t)^k / k! \right] {\overline{\pi}^k_i}$, where ${\overline{\pi}^k_i}$ is the distribution vector of ${\overline{Y}^i_k}$ after $k$ jumps.
\end{enumerate}

 \item  $X(t)$ is lumpable on $Q$ w.r.t. the partition ${\overline{S}_i}$ (Theorem~\ref{lumpable_theorem}) thus $Y_k$ is also lumpable on $P$ w.r.t the partition ${\overline{S}_i}$.
\end{enumerate}

Also, the initial state distributions $\pi^0$ and ${\overline{\pi}^0_i}$ for the CTMCs $X(t)$ and ${\overline{X}_i}(t)$ respectively are defined  as $\pi^0(0,0,\ldots,0)=1$, $\pi^0(y) = 0 \forall y \in \mathcal{S} \setminus \{(0,0,\ldots,0) \} $, ${\overline{\pi}^0_i}(0) =1$, and  ${\overline{\pi}^0_i}(\overline{x}) =0 \forall \overline{x} \in {\overline{S}_i} \setminus \{ 0\}$. Note that ${\overline{\pi}^0_i}(\overline{x}) = \sum\limits_{y \in \overline{x}} \pi^0(y) \forall \overline{x} \in {\overline{S}_i}$ holds in this case.
Now:
\begin{align*}
\overline{\pi}_{i}^{t} (\overline{x})
	=& \sum\limits_{k=0}^{\infty} e^{-\widehat{q}t} \frac{ {(\widehat{q}t)}^k}{ k!} {\overline{\pi}_{i}^{k}} (\overline{x}) \hspace{2mm} \text{ (by statement 2(b) above)}\\
  	=& \sum\limits_{k=0}^{\infty} e^{-\widehat{q}t} \frac{ {(\widehat{q}t)}^k}{ k!} \sum\limits_{y \in \overline{S}^{\overline{x}}_i} \pi^{k} (y) \hspace{2mm} \textrm{ (using Theorem ~\ref{transient_state_other_paper_theorem})}\\
 	=&  \sum\limits_{y \in \overline{S}_i^{\overline{x}}} \sum\limits_{k=0}^{\infty} e^{-\widehat{q}t} \frac{ {(\widehat{q}t)}^k}{ k!}  \pi^{k} (y) (\text{using Tonelli's theorem \citep{TAO11}})\\
 	=& \sum\limits_{y \in \overline{S}_i^{\overline{x}}} \pi^t(y) \hspace{2mm} \text{(by statement 2(a) above)}.\\
\end{align*}
\end{proof}
    
\begin{lemma}
$ R_t^{l,i} = {\overline{ R}_t^{l,i} }$ 
\end{lemma}
\begin{proof}
\begin{align}
&  R_t^{l,i}
 	\nonumber =\sum \limits_{x \in \mathcal{S}} (R^{l,i}(x))\pi^t(x)
 	=\sum\limits_{ \overline{z} \in {\overline{ S}_i}}  \sum\limits_{x \in  {\overline{S}_i^{\overline{z}}}} (R^{l,i}(x))\pi^t(x)\\
 	\nonumber       & (\text{Note: } R^{l,i}(x) = R^{l,i}(y) = {\overline{R}^{l,i}}(\overline{z}) \forall x,y \in {\overline{S}_i^{\overline{z}}} \text{ using Equations~\ref{net_reward_mainCTMC},~\ref{net_reward_lumpedCTMC}}&\\ \nonumber &\text{ and the lumping criteria: }x_i = y_i = \overline{z}  \forall x,y \in {\overline{S}_i^{\overline{z}}})\\
 	 \nonumber =&  \sum\limits_{ \overline{z} \in {\overline{ S}_i}}  \sum\limits_{x \in  {\overline{S}_i^{\overline{z}}}} {\overline{R}^{l,i}}(\overline{z}) \pi^t(x)
 	 \nonumber =  \sum\limits_{ \overline{z} \in {\overline{ S}_i}}   {\overline{R}^{l,i}}(\overline{z}) \sum\limits_{x \in  {\overline{S}_i^{\overline{z}}}} \pi^t(x)\\
 	 \nonumber =&  \sum\limits_{ \overline{z} \in {\overline{ S}_i}}    {\overline{R}^{l,i}}(\overline{z}) {\overline{\pi}_i^t} (\overline{z}) \quad (\text{Using Lemma~\ref{lemma_state_probabilities_equivalence}})\\
 	 \nonumber =& {\overline{ R}_t^{l,i} }
\end{align}
\end{proof}

\begin{lemma} 
$ R_t^{I} = \sum\limits_{ 1 \leq i \leq n}    c_i  {\overline{ R}_t^{I,i} } $ 
\end{lemma}
\begin{proof}
\begin{align}
\nonumber 	 R_t^{I} =& \sum \limits_{x \in \mathcal{S}} (R^{I}(x))\pi^t(x)
			  =\sum \limits_{x \in \mathcal{S}} \sum\limits_{ 1 \leq i \leq n} c_i ( x_i \delta^{\log\mu} - \beta ) \pi^t(x)\\
 \nonumber 	 	=& \sum\limits_{ 1 \leq i \leq n}  \sum \limits_{x \in \mathcal{S}} c_i ( x_i \delta^{\log\mu} - \beta ) \pi^t(x)
=\sum\limits_{ 1 \leq i \leq n} \sum\limits_{ \overline{z} \in {\overline{ S}_i}}  \sum\limits_{x \in  {\overline{S}_i^{\overline{z}}}}   c_i ( x_i \delta^{\log\mu} - \beta ) \pi^t(x)\\
 \nonumber 	 	=& \sum\limits_{ 1 \leq i \leq n}    c_i  \sum\limits_{ \overline{z} \in {\overline{ S}_i}}  \sum\limits_{x \in  {\overline{S}_i^{\overline{z}}}}( x_i \delta^{\log\mu} - \beta ) \pi^t(x)\\
\nonumber  	 	 & ( \text{Note: }  \forall x,y \in {\overline{S}_i^{\overline{z}}}   \text{, } ( x_i \delta^{\log\mu} - \beta )= (y_i \delta^{\log\mu} - \beta) =  {\overline{ R}^{I,i}}(\overline{z}) \\ 
\nonumber              &\text{ due to lumping critera } x_i = y_i = \overline{z}  \forall x,y \in {\overline{S}_i^{\overline{z}}}) \\
 \nonumber 	 	=& \sum\limits_{ 1 \leq i \leq n}    c_i  \sum\limits_{ \overline{z} \in {\overline{ S}_i}}  \sum\limits_{x \in  {\overline{S}_i^{\overline{z}}}}{\overline{ R}^{I,i}}(\overline{z}) \pi^t(x)
 =\sum\limits_{ 1 \leq i \leq n}    c_i  \sum\limits_{ \overline{z} \in {\overline{ S}_i}}  {\overline{ R}^{I,i}}(\overline{z}) \sum\limits_{x \in  {\overline{S}_i^{\overline{z}}}}  \pi^t(x)\\	 	
 \nonumber 	 	=& \sum\limits_{ 1 \leq i \leq n}    c_i  \sum\limits_{ \overline{z} \in {\overline{ S}_i}}  {\overline{ R}^{I,i}}(\overline{z}) {\overline{\pi}^t_i}( \overline{z}) \quad (\text{Using Lemma~\ref{lemma_state_probabilities_equivalence}})\\
 \nonumber 	 	=& \sum\limits_{ 1 \leq i \leq n}    c_i  {\overline{ R}_t^{I,i} }
\end{align}
\end{proof}

\begin{lemma}
The CTMC $X(t)=(\mathcal{S},Q)$ is irreducible and positive recurrent and a steady state vector $\Pi$ exists for this CTMC.
\end{lemma}
\begin{proof}
A steady state vector $\Pi$ exists for a CTMC if it is irreducible and positive recurrent. And a CTMC $X(t)$ is irreducible and positive recurrent if the underlying DTMC $Y_k$ is irreducible and positive recurrent. Also, a finite-DTMC is irreducible and positive recurrent if it has a single communicating class \citep{VISHWANADHAM92}.

As argued in the proof of Lemma~\ref{lemma_state_probabilities_equivalence} an underlying DTMC $Y_k= (\mathcal{S},P)$ exists for our CTMC $X(t) = (\mathcal{S},Q)$ and the transition probability matrix $P$ is given by: $P= Q/\widehat{q} + I$, where $\widehat{q}$ is an upper-bound on the entries in the $Q$ matrix. Using Equation~\ref{generator_matrix} we thus define $P$ as:  
\begin{align}
\label{ProbabilityTransitionMatrix}
P(x,y) = \begin{cases} 			
 	  \lambda_j/ \widehat{q}  & \text{if } \sum\limits_{i} |x_i - y_i| =  1  \text{ \& } \exists j: y_j-x_j = 1 \text{ (Case~$i$)}\\
 	  \mu  / \widehat{q}     & \text{if } y = (0,\ldots,0) \neq x \text{ (Case~$ii$)}\\
 	     1 +\sum \limits_{y' \in \mathcal{S} \setminus \{x \}} \frac{-Q(x ,y')}{\widehat{q}}  & \text{if } x = y \text{ (Case~$iii$)}\\
	  0 &o/w \text{ (Case~$iv$)}.   \\
      \end{cases} 
\end{align} 

We shall now prove that the DTMC $Y_k$ has a single communicating class by arguing that the state $x : (0,0,\ldots,0)$ is reachable from each state $y \in \mathcal{S}$ and also each of these states $y$ is reachable from the state $x$.
This in turn implies that any two states $y$ and $z$ are accessible from each other.
\begin{itemize}

\item The one step transition probability $p_{yx}(1)$  from a state $y \in \mathcal{S} \setminus \{x\}$ to the state $x:(0,0,\ldots,0)$ is non-zero (i.e. Case~$(ii)$ in Equation~\ref{ProbabilityTransitionMatrix}). Also a state is always accessible from itself in $0$ steps. Thus the state $x:(0,0,\ldots,0)$ is accessible from all states $y \in \mathcal{S}$.

\item Consider a state $y = (y_1,y_2,\ldots,y_n) \in \mathcal{S}$. Now, if the DTMC $Y_k$ is in state $x$ and student 1 starts arriving on the forum one by one $y_1$ times then the following state sequence will be generated: $(0,0,\ldots,0)$, $(1,0,\ldots,0)$, $(2,0,\ldots,0)$, $\ldots$, $(y_1,0,\ldots,0)$ and then student 2 starts arriving one by one $y_2$ times, then the following state sequence is incremented to the existing state sequence: $(y_1,1,\ldots,0)$, $(y_1,2,\ldots,0)$, $\ldots$, $(y_1,y_2,\ldots,0)$. Similarly now if the student 3 arrives one by one $y_3$ times (then the DTMC $Y_k$ reaches state $(y_1,y_2,y_3,0,\ldots,0)$ ) and so on till the student n also arrives one by one $y_n$ times, we finally reach state $y:(y_1,y_2,\ldots,y_n)$.

Now each of the transition between two consecutive states in this state sequence satisfies the Case~$(i)$ of Equation~\ref{ProbabilityTransitionMatrix}, thus all transitions in this sequence have some positive transition probability. Thus a $\sum\limits_{i = 1}^{n} y_i$ (number of transitions/steps) step walk exists from state $x$ to state $y$ and thus state $y$ is accessible from state $x$ \citep{GALLAGER09}. 

\end{itemize}

Thus all states in $\mathcal{S}$ are accessible from each other and so  the DTMC $Y_k$ has a single communicating class. Now as $Y_k$ is finite and has a single communicating class, thus the underlying DTMC $Y_k$ is irreducible and positive recurrent. Therefore the CTMC  $X(t)$ is also irreducible and positive recurrent. Thus the steady state vector (or limiting probabilities) $\Pi$ exist \citep{VISHWANADHAM92} for the CTMC $X(t)$.
\end{proof}

\begin{lemma}
The lumped-CTMC ${\overline{X}_i}(t)=({\overline{S}_i},{\overline{Q}_i})$ is irreducible and positive recurrent and a steady state vector ${\overline{\Pi}_i}$ exists for this CTMC. Also each component of ${\overline{\Pi}_i}$ will be obtained as: ${\overline{\Pi}_i}(\overline{x}) = \sum\limits_{x \in {\overline{S}^{\overline{x}}_i} } \Pi(x). $
\end{lemma}
\begin{proof}
The CTMC $X(t)=(\mathcal{S},Q)$ is irreducible with stationary distribution $\Pi$ (From Lemma~\ref{Lemma_CTMC_irreducible_and_steady_state_exists}). Also ${\overline{X}_i}(t) = ({\overline{S}_i},{\overline{Q}_i})$ is the lumped chain w.r.t. the partition ${\overline{S}_i}$ of the state space $\mathcal{S}$ of $X(t)$. Thus using Theorem $~\ref{steady_state_other_paper_theorem}$ we can say that the lumped chain ${\overline{X}_i}(t)$ also has a stationary distribution ${\overline{\Pi}_i}$ and whose components will be obtained as: ${\overline{\Pi}_i}(\overline{x}) = \sum\limits_{x \in {\overline{S}^{\overline{x}}_i} } \Pi(x). $
\end{proof}

\begin{theorem}
The expected transient aggregate net-rewards over time $T$ and the expected steady-state net-rewards received by the students and the instructor  is the same when calculated using the original CTMC $X(t)$ or the $n$ lumped CTMCs ${\overline{X}_i}(t)$ $i \in \{1,2,\ldots,n\}$. 
\begin{itemize} 
\item Expected transient aggregate net-rewards over time $T$.\\
a) $  R_T^{l,i}  = {\overline{ R}_T^{l,i} }$ \hspace{3mm} 
b) $  R_T^{I} =  {\overline{ R}_T^{I} } $.

\item Expected steady-state net-rewards.	\\
a) $  R_{\Pi}^{l,i} = {\overline{ R}_{\Pi}^{l,i} }$ \hspace{3mm}
b) $ R_{\Pi}^{I} = {\overline{ R}_{\Pi}^{I} } $.	
\end{itemize}
\end{theorem}

\begin{proof}
\begin{enumerate}
\item Expected transient aggregate net-rewards over time $T$ \\
(a) To Student $i$.
		\begin{align}
		\nonumber		R_T^{l,i}     
				&= \int\limits_{t=0}^{T} R_t^{l,i} dt
				= \int\limits_{t=0}^{T} {\overline{R}_t^{l,i}} dt
				\hspace{2mm} \text{(Using Lemma~\ref{Lemma_Student_Utilities_equal_in_original_And_Lumped})}
				= {\overline{ R}_T^{l,i} } 
		\end{align}
(b) To Instructor.
		\begin{multline}
		\nonumber		R_T^{I} 
				= \int\limits_{t=0}^{T} R_t^{I} \text{ dt}
				= \int\limits_{t=0}^{T} \sum\limits_{ 1 \leq i \leq n}    c_i  {\overline{ R}_t^{I,i} } \text{ dt} \hspace{2mm}(\text{Using Lemma~\ref{Lemma_Instructor_Utility_can_be_calculated_in_original_And_Lumped}}) \\ = \sum\limits_{ 1 \leq i \leq n}  c_i  \int\limits_{t=0}^{T}   {\overline{ R}_t^{I,i} } \text{ dt}
				=  \sum\limits_{ 1 \leq i \leq n}  c_i  {\overline{R}_T^{I,i}}  \text{ dt} 
				= {\overline{R}_T^{I}}
		\end{multline}
\item Expected steady-state net-rewards\\
(a) To Student $i$.
		\begin{align}
		\nonumber	R_{\Pi}^{l,i}     
		=& \sum \limits_{x \in \mathcal{S}} (R^{l,i}(x))\Pi(x) 
		=\sum\limits_{ \overline{z} \in {\overline{ S}_i}}  \sum\limits_{x \in  {\overline{S}_i^{\overline{z}}}} (R^{l,i}(x))\Pi(x)\\
 	\nonumber       & (\text{Note: } R^{l,i}(x) = R^{l,i}(y) = {\overline{R}^{l,i}}(\overline{z}) \forall x,y \in {\overline{S}_i^{\overline{z}}} \text{ using Equations~\ref{net_reward_mainCTMC},~\ref{net_reward_lumpedCTMC}}&\\ \nonumber &\text{ and the lumping criteria: }x_i = y_i = \overline{z}  \forall x,y \in {\overline{S}_i^{\overline{z}}})\\
 	 \nonumber      =&  \sum\limits_{ \overline{z} \in {\overline{ S}_i}}  \sum\limits_{x \in  {\overline{S}_i^{\overline{z}}}} {\overline{R}^{l,i}}(\overline{z}) \Pi(x)
 	 =  \sum\limits_{ \overline{z} \in {\overline{ S}_i}}   {\overline{R}^{l,i}}(\overline{z}) \sum\limits_{x \in  {\overline{S}_i^{\overline{z}}}} \Pi(x)\\
 	\nonumber       =&  \sum\limits_{ \overline{z} \in {\overline{ S}_i}}   {\overline{R}^{l,i}}(\overline{z}) {\overline{\Pi}_i} (\overline{z}) \quad  (\text{Using Lemma~\ref{Lemma_CTMC_lumped_steady_state_exists}})\\
 	\nonumber       =& {\overline{ R}_{\Pi}^{l,i} }
		\end{align}
	(b) To Instructor.
		\begin{align}
		\nonumber      		 R_{\Pi}^{I} =& \sum \limits_{x \in \mathcal{S}} R^{I} (x)\Pi(x)
		= \sum \limits_{x \in \mathcal{S}} \sum\limits_{ 1 \leq i \leq n} c_i ( x_i \delta^{\log\mu} - \beta ) \Pi(x)
		=\sum\limits_{ 1 \leq i \leq n}  \sum \limits_{x \in \mathcal{S}} c_i ( x_i \delta^{\log\mu} - \beta ) \Pi(x)\\
 	 	\nonumber      =& \sum\limits_{ 1 \leq i \leq n} \sum\limits_{ \overline{z} \in {\overline{ S}_i}}  \sum\limits_{x \in  {\overline{S}_i^{\overline{z}}}}   c_i ( x_i \delta^{\log\mu} - \beta ) \Pi(x)
 	 	= \sum\limits_{ 1 \leq i \leq n}    c_i  \sum\limits_{ \overline{z} \in {\overline{ S}_i}}  \sum\limits_{x \in  {\overline{S}_i^{\overline{z}}}}( x_i \delta^{\log\mu} - \beta ) \Pi(x)\\
 	 	\nonumber       & ( \text{Note: }  \forall x,y \in {\overline{S}_i^{\overline{z}}}   \text{, } ( x_i \delta^{\log\mu} - \beta )= (y_i \delta^{\log\mu} - \beta)=  {\overline{ R}^{I,i}}(\overline{z})) &\\
 	 	\nonumber      =& \sum\limits_{ 1 \leq i \leq n}    c_i  \sum\limits_{ \overline{z} \in {\overline{ S}_i}}  \sum\limits_{x \in  {\overline{S}_i^{\overline{z}}}}{\overline{ R}^{I,i}}(\overline{z}) \Pi(x)
 	 	=\sum\limits_{ 1 \leq i \leq n}    c_i  \sum\limits_{ \overline{z} \in {\overline{ S}_i}}  {\overline{ R}^{I,i}}(\overline{z}) \sum\limits_{x \in  {\overline{S}_i^{\overline{z}}}}  \Pi(x)\\
 	 	\nonumber      =& \sum\limits_{ 1 \leq i \leq n}    c_i  \sum\limits_{ \overline{z} \in {\overline{ S}_i}}  {\overline{ R}^{I,i}}(\overline{z}) {\overline{\Pi}_i}( \overline{z})       \quad (\text{Using Lemma~\ref{Lemma_CTMC_lumped_steady_state_exists} } )\\
 	 	\nonumber      =& \sum\limits_{ 1 \leq i \leq n}    c_i  {\overline{ R}_{\Pi}^{I,i} }
 	 	= {\overline{R}_{\Pi}^{I}}
		\end{align}			
\end{enumerate}
\end{proof}

\section{ Proofs : Steady State Analysis}

\begin{lemma}
The solution to equations ~\ref{eqnSteady1}, ~\ref{eqnSteady2}, and ~\ref{eqnSteady3} is:  
\begin{align} \nonumber {\overline{\Pi}_i}( \overline{x}) =& \frac{\rho^{ \overline{x}}_i}{(1+\rho_i)^{ \overline{x}+1}} \text{, } \forall \overline{x}: 0 \leq  \overline{x} < M \text{ and}
\quad  {\overline{\Pi}_i}( M) = \frac{\rho_i^{M}}{(1+\rho_i)^{M}} .\text{ where, } \rho_i = \frac{\lambda_i}{\mu}.
\end{align}
\end{lemma}

\begin{proof}
\begin{align}
\nonumber      & \textbf{Inflow in state 0} = \textbf{Outflow from state 0}\\
\nonumber      & \sum\limits_{\overline{y} \geq 1} { \mu  {\overline{\Pi}_i}( \overline{y}) } =  \lambda_i {\overline{\Pi}_i}( 0) \\ 
\nonumber      \Rightarrow & \mu  (1-{\overline{\Pi}_i}(0)) = \lambda_i {\overline{\Pi}_i}(0) 
\nonumber       \Rightarrow   {\overline{\Pi}_i}(0) (\mu + \lambda_i) = \mu \\
\nonumber      \Rightarrow &  {\overline{\Pi}_i}(0) = \frac{\mu}{\mu + \lambda_i} = \frac{1}{1 + \rho_i}  \text{ }(\text{Taking } \rho_i = \frac{\lambda_i}{\mu} )
\end{align}

\begin{align}
\nonumber      & \textbf{Similarly for any state }  \overline{x} \in \{ 1,\ldots,M-1\}\\
\nonumber      & \text{Inflow in state }\overline{x} = \text{Outflow from state } \overline{x}\\
\nonumber      &  \lambda_i  {\overline{\Pi}_i}(\overline{x}-1)  =  \mu {\overline{\Pi}_i}(\overline{x}) + \lambda_i {\overline{\Pi}_i}(\overline{x})\\
\nonumber      \Rightarrow & {\overline{\Pi}_i}(\overline{x}) = \frac{\lambda_i}{\mu + \lambda_i} {\overline{\Pi}_i}(\overline{x}-1)  = \frac{{\rho_i}}{1 + {\rho_i}} {\overline{\Pi}_i}(\overline{x}-1) \\
\nonumber      & \text{Expanding RHS till }{\overline{\Pi}_i}( 0) \text{, we get}\\
\nonumber      \Rightarrow & {\overline{\Pi}_i}(\overline{x}) = \left(\frac{{\rho_i}}{1 + {\rho_i}} \right)^{\overline{x}} {\overline{\Pi}_i}(0) = \frac{{\rho_i}^{\overline{x}}}{(1 + {\rho_i})^{\overline{x}+1}}   \\ \nonumber
\\
\nonumber      & \textbf{Inflow in state M} = \textbf{Outflow from state M}\\
\nonumber      &  \lambda_i  {\overline{\Pi}_i}( M-1)  =  \mu {\overline{\Pi}_i}(M) \\
\nonumber      \Rightarrow & {\overline{\Pi}_i}(M) = \frac{\lambda_i}{\mu } {\overline{\Pi}_i}(M-1) \\
\nonumber      \Rightarrow & {\overline{\Pi}_i}(M) = {\rho_i}  \frac{{\rho_i}^{M-1}}{(1 + {\rho_i})^{M}}= \frac{{\rho_i}^{M}}{(1 + {\rho_i})^{M}} 
\end{align}
\end{proof}

\begin{claim}
\label{Claim_summation_a_into_f_power_a}
\begin{align} \nonumber \sum \limits_{\overline{ x} = 0}^{m_l} \overline{ x} {\overline{ \Pi}_i}(\overline{ x}) = {\rho_i} \left(  1 - (m_l + 1) \left(  \frac{{\rho_i}}{1 + {\rho_i}}  \right)^{m_l}  + m_l \left(  \frac{{\rho_i}}{1 + {\rho_i}}  \right)^{m_l + 1}  \right)
\end{align}

\end{claim}
\begin{proof}
\begin{align}
&\nonumber \sum \limits_{\overline{ x} = 0}^{m_l} \overline{ x} {\overline{ \Pi}_i}(\overline{ x}) 				= \sum \limits_{\overline{ x} = 0}^{m_l} \overline{ x} \frac{{\rho_i} ^ {\overline{ x}}}{{(1 + {\rho_i})}^{\overline{ x}+1}}
=\frac{1}{(1 + {\rho_i})} \sum \limits_{\overline{ x} = 0}^{m_l} \overline{ x} \left( \frac{{\rho_i}}{1 + {\rho_i}} \right)^ {\overline{ x}}\\
\nonumber 	 				=& \frac{1 - (m_l + 1) \left(  \frac{{\rho_i}}{1 + {\rho_i}}  \right)^{m_l}  + m_l \left(  \frac{{\rho_i}}{1 + {\rho_i}}  \right)^{m_l + 1}   }{\left(  \frac{1}{1 + {\rho_i}}  \right)^{2}} \times \frac{1}{(1 + {\rho_i})} \frac{{\rho_i}}{(1 + {\rho_i})}\\
\nonumber 	 				=& {\rho_i} \left(  1 - (m_l + 1) \left(  \frac{{\rho_i}}{1 + {\rho_i}}  \right)^{m_l}  + m_l \left(  \frac{{\rho_i}}{1 + {\rho_i}}  \right)^{m_l + 1}  \right)
\end{align}
\end{proof}

\begin{claim}
\label{Claim_summation_f_power_a}
$\sum \limits_{\overline{ x} = 0}^{m_l}  {\overline{ \Pi}_i}(\overline{ x}) 								= 1 - \left( \frac{{\rho_i}}{1+ {\rho_i}} \right)^{m_l +1}$.

\end{claim}
\begin{proof}

\begin{multline}
\nonumber \sum \limits_{\overline{ x} = 0}^{m_l}  {\overline{ \Pi}_i}(\overline{ x}) 								= \sum \limits_{\overline{ x} = 0}^{m_l} \frac{{\rho_i} ^ {\overline{ x}}}{{(1 + {\rho_i})}^{\overline{ x}+1}}
\nonumber	 				= \frac{1}{(1 + {\rho_i})} \sum \limits_{\overline{ x} = 0}^{m_l} \left( \frac{{\rho_i}}{1 + {\rho_i}} \right)^ {\overline{ x}}
=\frac{1}{(1 + {\rho_i})} \frac{ 1 - \left( \frac{{\rho_i}}{1+ {\rho_i}} \right)^{m_l +1} }{ 1 - \frac{{\rho_i}}{1+ {\rho_i}} }
\nonumber	 			\\	= 1 - \left( \frac{{\rho_i}}{1+ {\rho_i}} \right)^{m_l +1}
\end{multline}
\end{proof}

\begin{claim}
\label{Claim_summation_a_into_f_power_a_for_M}
$\sum \limits_{\overline{ x} = 0}^{M}  \overline{x} {\overline{ \Pi}_i}(\overline{ x})					={\rho_i} \left(  1  - \left(  \frac{{\rho_i}}{1 + {\rho_i}}  \right)^{M}  \right)	.
$
\end{claim}
\begin{proof}

\begin{flalign}
&\nonumber \sum \limits_{\overline{ x} = 0}^{M}  \overline{x} {\overline{ \Pi}_i}(\overline{ x}) =  \sum \limits_{\overline{ x} = 0}^{M-1}  \overline{x} {\overline{ \Pi}_i}(\overline{ x}) +M {\overline{ \Pi}_i}(M)&	 \\	
\nonumber 					 & \text{(Using Claim ~\ref{Claim_summation_a_into_f_power_a})}&\\
\nonumber 					=& 	 {\rho_i} \left(  1 - M \left(  \frac{{\rho_i}}{1 + {\rho_i}}  \right)^{M-1}  + (M-1) \left(  \frac{{\rho_i}}{1 + {\rho_i}}  \right)^{M}  \right)	+ M \left(  \frac{{\rho_i}}{1 + {\rho_i}}  \right)^{M}	 \\								
\nonumber 					=&   {\rho_i} \left(  1 - M \left(  \frac{{\rho_i}}{1 + {\rho_i}}  \right)^{M-1}  - \left(  \frac{{\rho_i}}{1 + {\rho_i}}  \right)^{M}  \right) +		   M \left( 1 + {\rho_i} \right) \left(  \frac{{\rho_i}}{1 + {\rho_i}}  \right)^{M} \\
\nonumber 					= &   {\rho_i} \left(  1 - M \left(  \frac{{\rho_i}}{1 + {\rho_i}}  \right)^{M-1}  - \left(  \frac{{\rho_i}}{1 + {\rho_i}}  \right)^{M}  \right)	+ M \left(  \frac{{\rho_i}}{1 + {\rho_i}}  \right)^{M-1}  \\
\nonumber 					=& {\rho_i} \left(  1  - \left(  \frac{{\rho_i}}{1 + {\rho_i}}  \right)^{M}  \right)					 
\end{flalign}
\end{proof}

\begin{theorem}

 The  expected steady state net-rewards of the student $i$ and the instructor in ${\overline{X}_i}(t)$ are given by:					
\begin{align}
 \nonumber \textbf(a)\text{ } {\overline{ R}_\Pi^{l,i} } =&  \delta^{\log\mu} {\rho_i}  \left(  1 - \left(  \frac{{\rho_i}}{1 + {\rho_i}}  \right)^{m_l}   \right) -   \alpha_l   {\rho_i} \left(  1  - \left(  \frac{{\rho_i}}{1 + {\rho_i}}  \right)^{M}  \right)	\\
\nonumber \textbf(b)\text{ } {\overline{ R}_{\Pi}^{I,i} }	=&  \delta^{\log\mu}  {\rho_i} \left(  1  - \left(  \frac{{\rho_i}}{1 + \rho_i}  \right)^{M}  \right) - \beta  
\end{align}
\end{theorem}

\begin{proof}\textbf(a)

\begin{flalign}
& \nonumber {\overline{ R}_\Pi^{l,i} } = \sum \limits_{\overline{ x} \in {\overline{ S}_i}}  {\overline{R}^{l,i}}(\overline{ x}) {\overline{ \Pi}_i}(\overline{x})
    =\sum \limits_{\overline{ x} = 0}^{M}  {\overline{R}^{l,i}}(\overline{ x}) {\overline{ \Pi}_i}(\overline{ x}) = \sum \limits_{\overline{ x} = 0}^{M} \left(  {\overline{r}^{l,i} }(\overline{x}) - \alpha_l \overline{x} \right)  {\overline{ \Pi}_i}(\overline{ x})&
\end{flalign}
\begin{flalign}
\nonumber 					=&  \sum \limits_{\overline{ x} = 0}^{m_l} \left( \overline{x} \delta^{\log\mu} - \alpha_l \overline{x}   \right) {\overline{ \Pi}_i}(\overline{ x}) + \sum \limits_{\overline{ x} = m_l +1}^{M} \left( m_l \delta^{\log\mu} - \alpha_l \overline{x}   \right) {\overline{ \Pi}_i}(\overline{ x}) &\\
\nonumber 					=& \sum \limits_{\overline{ x} = 0}^{m_l} \left( \overline{x} \delta^{\log\mu}   \right) {\overline{ \Pi}_i}(\overline{ x}) + \sum \limits_{\overline{ x} = m_l +1}^{M} \left( m_l \delta^{\log\mu}  \right) {\overline{ \Pi}_i}(\overline{ x})   - \sum \limits_{\overline{ x} = 0}^{M} \alpha_l \overline{x} {\overline{ \Pi}_i}(\overline{ x})  &
\end{flalign}
\begin{flalign}
\nonumber 					=& \sum \limits_{\overline{ x} = 0}^{m_l} \left( \overline{x} \delta^{\log\mu}   \right) {\overline{ \Pi}_i}(\overline{ x})    +  \left( m_l \delta^{\log\mu}  \right) \sum \limits_{\overline{ x} = m_l +1}^{M}  {\overline{ \Pi}_i}(\overline{ x})  - \sum \limits_{\overline{ x} = 0}^{M} \alpha_l \overline{x} {\overline{ \Pi}_i}(\overline{ x}) &\\
\nonumber 					=& \delta^{\log\mu}  \sum \limits_{\overline{ x} = 0}^{m_l}  \overline{x}  {\overline{ \Pi}_i}(\overline{ x})  +  \left( m_l \delta^{\log\mu}  \right) \left( 1 - \sum \limits_{\overline{ x} = 0}^{m_l}  {\overline{ \Pi}_i}(\overline{ x}) \right) - \sum \limits_{\overline{ x} = 0}^{M} \alpha_l \overline{x} {\overline{ \Pi}_i}(\overline{ x}) &\\
\nonumber 					& \text{(Using Claims 
					 ~\ref{Claim_summation_a_into_f_power_a} and 
					 ~\ref{Claim_summation_f_power_a} )}  &\\
\nonumber 					=&  \delta^{\log\mu} {\rho_i}  \left(  1 - (m_l + 1) \left(  \frac{{\rho_i}}{1 + {\rho_i}}  \right)^{m_l}  + m_l \left(  \frac{{\rho_i}}{1 + {\rho_i}}  \right)^{m_l + 1}  \right)    - \sum \limits_{\overline{ x} = 0}^{M} \alpha_l \overline{x} {\overline{ \Pi}_i}(\overline{ x}) +  \left( m_l \delta^{\log\mu}  \right) \left( \frac{{\rho_i}}{1+ {\rho_i}} \right)^{m_l +1}  &\\
\nonumber 					=&  \delta^{\log\mu} {\rho_i}  \left(  1 - (m_l + 1) \left(  \frac{{\rho_i}}{1 + {\rho_i}}  \right)^{m_l}   \right) +   \left( m_l \delta^{\log\mu}  \right) \left( \frac{{\rho_i}}{1+ {\rho_i}} \right)^{m_l +1} \left(1+ {\rho_i} \right) -   \sum \limits_{\overline{ x} = 0}^{M} \alpha_l \overline{x} {\overline{ \Pi}_i}(\overline{ x})&\\
\nonumber 					=&  \delta^{\log\mu} {\rho_i}  \left(  1 - (m_l + 1) \left(  \frac{{\rho_i}}{1 + {\rho_i}}  \right)^{m_l}   \right) + \left( m_l \delta^{\log\mu}  \right) \frac{{\rho_i}^{m_l +1}  }{ \left( 1+ {\rho_i} \right)^{m_l}  }  - 
					     \sum \limits_{\overline{ x} = 0}^{M} \alpha_l \overline{x} {\overline{ \Pi}_i}(\overline{ x})& \\
\nonumber 					=&  \delta^{\log\mu} {\rho_i}  \left(  1 - \left(  \frac{{\rho_i}}{1 + {\rho_i}}  \right)^{m_l}   \right) - \sum \limits_{\overline{ x} = 0}^{M} \alpha_l \overline{x} {\overline{ \Pi}_i}(\overline{ x})&\\	
\nonumber 					=&  \delta^{\log\mu} {\rho_i}  \left(  1 - \left(  \frac{{\rho_i}}{1 + {\rho_i}}  \right)^{m_l}   \right) - \alpha_l \sum \limits_{\overline{ x} = 0}^{M}  \overline{x} {\overline{ \Pi}_i}(\overline{ x})	&\\
\nonumber 					 & \text{(Using Claim ~\ref{Claim_summation_a_into_f_power_a_for_M})}&\\
\nonumber 					=&  \delta^{\log\mu} {\rho_i}  \left(  1 - \left(  \frac{{\rho_i}}{1 + {\rho_i}}  \right)^{m_l}   \right) -   \alpha_l   {\rho_i} \left(  1  - \left(  \frac{{\rho_i}}{1 + {\rho_i}}  \right)^{M}  \right)		&			 
\end{flalign}

\end{proof}

\begin{proof} \textbf(b)

\begin{flalign}
\nonumber {\overline{ R}_{\Pi}^{I,i} }	=& \sum \limits_{\overline{ x} = 0}^{M} \left(  {\overline{R}^{I,i}}(\overline{ x})\right) {\overline{ \Pi}_i}(\overline{ x})	= \sum \limits_{\overline{ x} = 0}^{M} \left(  {\overline{ r}^{I,i}}(\overline{x}) - \beta  \right) {\overline{ \Pi}_i}(\overline{ x}) = \sum \limits_{\overline{ x} = 0}^{M} \left(  \overline{x} \delta^{\log\mu} - \beta  \right) {\overline{ \Pi}_i}(\overline{ x})&\\
\nonumber 					=& \sum \limits_{\overline{ x} = 0}^{M} \overline{x} \delta^{\log\mu}   {\overline{ \Pi}_i}(\overline{ x}) - \sum \limits_{\overline{ x} = 0}^{M} \beta {\overline{ \Pi}_i}(\overline{ x})
=\delta^{\log\mu}  \sum \limits_{\overline{ x} = 0}^{M} \overline{x}   {\overline{ \Pi}_i}(\overline{ x}) - \beta \\
\nonumber 					=&  \delta^{\log\mu}  {\rho_i} \left(  1  - \left(  \frac{{\rho_i}}{1 + {\rho_i}}  \right)^{M}  \right) - \beta  \quad	\text{(Using Claim ~\ref{Claim_summation_a_into_f_power_a_for_M})}				 
\end{flalign}
\end{proof}

\section{ Proofs : Transient Analysis}

\begin{lemma}
Given the initial distribution ${\overline{\pi}_{i}^0}$ for $\overline{X}_i(t)$ as ${\overline{\pi}_{i}^0}(0) =1$ and ${\overline{\pi}_{i}^0}(\overline{x}) =0$ $\forall \overline{x} \in \{1,2,\ldots,M \}$.
The solution to the differential equations~\ref{eq1_transient}, ~\ref{eq2_transient}, and ~\ref{eq3_transient} is given by:  
\begin{align}
\nonumber {\overline{\pi}_i^t}(0) =& \frac{\mu}{{K}_i} + \frac{\lambda_i}{{K}_i} e^{-{K}_it}\\
\nonumber {\overline{\pi}_i^t}(\overline{x}) =& \left( -  \left(\frac{\lambda_i}{{K}_i}\right)^{\overline{x}} \frac{\mu}{{K}_i}    - \sum\limits_{\overline{y}=1}^{\overline{x}-1} \frac{\lambda_i^ {\overline{x}} t^{\overline{y}} \mu}{\overline{y}! {K}_i^{\overline{x}-\overline{y}+1}} + \frac{{\lambda_i}^{\overline{x}+1} t^{\overline{x}}}{\overline{x}! {K}_i} \right)e^{-{K}_{i} t} 
+ \left(\frac{\lambda_i}{{K}_i}\right)^{\overline{x}} \frac{\mu}{{K}_i}  \hspace{2mm} \forall \overline{x}: 0 < \overline{x} < M\\
\nonumber {\overline{\pi}_{i}^t}(M) =& 1 - \sum\limits_{\overline{x}=0}^{M-1} {\overline{\pi}_i^t}(\overline{x}) \text {, where } {K}_i = (\lambda_i + \mu)
\end{align}

\end{lemma}					

\begin{proof}

\begin{itemize}

\item \underline{\textbf{For state $0$}}
\begin{align}
\nonumber &\frac{d {\overline{\pi}_i^t}(0)}{dt} = -\lambda_i {\overline{\pi}_i^t}(0) + \sum\limits_{\overline{y}=1}^{M} \mu  {\overline{\pi}_i^t}(\overline{y})&\\
\nonumber  \Rightarrow & \frac{d {\overline{\pi}_i^t}(0)}{dt} + \lambda_i {\overline{\pi}_i^t}(0) =   \mu   (1 -  {\overline{\pi}_i^t}(0))& \\
\nonumber  \Rightarrow & \frac{d {\overline{\pi}_i^t}(0)}{dt} + (\lambda_i + \mu) {\overline{\pi}_i^t}(0) =   \mu & \\
 \nonumber \Rightarrow & {\overline{\pi}_i^t}(0) = \frac{\mu}{\mu + \lambda_i} + c e^{-(\mu + \lambda_i)t}&\\
\nonumber &\text{ (Now using the initial distributions }{\overline{\pi}_i^0}(0) =1\text{)}&\\
\nonumber &{\overline{\pi}_i^0}(0) = \frac{\mu}{\mu + \lambda_i} + c e^{-(\mu + \lambda_i)0}=1  \Rightarrow   c = 1 - \frac{\mu}{\mu + \lambda_i} =  \frac{\lambda_i}{\mu + \lambda_i}&\\
 \nonumber &  \text{Thus we get, } {\overline{\pi}_i^t}(0) = \frac{\mu}{K_i} + \frac{\lambda_i}{K_i} e^{-K_it}, \text{where }K_i = \mu + \lambda_i.
\end{align}

\item \underline{\textbf{For state $1$}}

\begin{align}
\nonumber &\frac{d {\overline{\pi}_i^t}(1)}{dt} = -(\lambda_i+\mu) {\overline{\pi}_i^t}(1) + \lambda_i {\overline{\pi}_i^t}(0)&\\
\nonumber \Rightarrow &\frac{d {\overline{\pi}_i^t}(1)}{dt} = -(\lambda_i+\mu) {\overline{\pi}_i^t}(1) + \lambda_i \left( \frac{\mu}{K_i} + \frac{\lambda_i}{K_i} e^{-K_i t} \right)&\\
\nonumber \Rightarrow &\frac{d {\overline{\pi}_i^t}(1)}{dt} + K_i {\overline{\pi}_i^t}(1) = W&\\
\nonumber &\text{ where } K_i = (\mu + \lambda_i) \text{ and } W = \left( \frac{\lambda_i}{K_i} \left( \mu + \lambda_i e^{-K_i t} \right) \right)&
\end{align}
\begin{align}
\nonumber \Rightarrow & {\overline{\pi}_i^t}(1) e^{\int K_i dt} = c + \int W e^{\int K_i dt} dt&\\
\nonumber \Rightarrow & {\overline{\pi}_i^t}(1) e^{ K_i t} = c + \int \left( \frac{\lambda_i}{K_i} \left( \mu + \lambda_i e^{-K_i t} \right) \right) e^{ K_i t} dt&\\
\nonumber \Rightarrow & {\overline{\pi}_i^t}(1) e^{ K_i t} = c + \frac{\lambda_i \mu}{K_i} \int e^{ K_i t} dt + \frac{\lambda_i^2}{K_i} \int 1 dt&\\
\nonumber \Rightarrow & {\overline{\pi}_i^t}(1) e^{ K_i t} = c + \frac{\lambda_i \mu}{K_i} \frac{e^{ K_i t}}{K_i} + \frac{\lambda_i^2}{K_i} t&\\
\nonumber \Rightarrow & {\overline{\pi}_i^t}(1)  = \frac{\lambda_i \mu}{K_i^2}  + \left( c  + \frac{\lambda_i^2}{K_i} t \right) e^{ - K_i t}&\\
\nonumber &\text{ (Now using the initial distribution }{\overline{\pi}_i^0}(1) =0\text{)}&\\
\nonumber &{\overline{\pi}_i^0}(1) = \frac{\lambda_i \mu}{K_i^2}  + \left( c \right) e^{0} = 0 \Rightarrow c = - \frac{\lambda_i \mu}{K_i^2} &\\
 \nonumber &  \text{Thus we get, } {\overline{\pi}_i^t}(1)  = \frac{\lambda_i \mu}{K_i^2}  + \left( - \frac{\lambda_i \mu}{K_i^2}  + \frac{\lambda_i^2 t}{K_i}  \right) e^{ - K_i t}&
\end{align}

\item				
Similarly on solving for \textbf{States 2 and 3} we get:
\begin{align}
\nonumber &{\overline{\pi}_i^t}(2)  = \frac{\lambda_i^2 \mu}{K_i^3}  + \left( - \frac{\lambda_i^2 \mu}{K_i^3}   + \frac{\lambda_i^3  t^2}{2K_i} - \frac{\lambda_i^2 \mu t}{K_i^2} \right) e^{ - K_i t}\\
\nonumber &{\overline{\pi}_i^t}(3)  = \frac{\lambda_i^3 \mu}{K_i^4}  + \left( - \frac{\lambda_i^3 \mu}{K_i^4} - \frac{\lambda_i^3 \mu t}{K_i^3}   + \frac{\lambda_i^4  t^3}{3! K_i} - \frac{\lambda_i^3 \mu t^2}{2 K_i^2} \right) e^{ - K_i t}\\
\nonumber & \text{ where } K_i = (\mu + \lambda_i)
\end{align}

\item
Observing from ${\overline{\pi}_i^t}(1)$, ${\overline{\pi}_i^t}(2)$, ${\overline{\pi}_i^t}(3)$, we claim and prove by induction that $\forall \overline{x} \in \{1,\ldots,M-1\}$:
\begin{align}
\label{induction_steady_state_eqn}
{\overline{\pi}_i^t}(\overline{x}) =& \left( -  \left(\frac{\lambda_i}{{K}_i}\right)^{\overline{x}} \frac{\mu}{{K}_i}    - \sum\limits_{\overline{y}=1}^{\overline{x}-1} \frac{\lambda_i^ {\overline{x}} t^{\overline{y}} \mu}{\overline{y}! {K}_i^{\overline{x}-\overline{y}+1}} + \frac{{\lambda_i}^{\overline{x}+1} t^{\overline{x}}}{\overline{x}! {K}_i} \right)e^{-{K}_{i} t} + \left(\frac{\lambda_i}{{K}_i}\right)^{\overline{x}} \frac{\mu}{{K}_i} \hspace{4mm} (\text{ where } K_i = (\mu + \lambda_i))
\end{align}

Now let us assume that this holds for some state $\overline{z}$ s.t. $ 1 \leq \overline{z} < M-1$. Thus we have the value of ${\overline{\pi}_i^t}( \overline{z})$ by substituting in place of $\overline{x}$ in the equation above. Now we shall prove that the Equation~\ref{induction_steady_state_eqn} holds for state $\overline{z} +1$ also.

\underline{\textbf{For state $ \overline{z} + 1$}}
\begin{align}
\nonumber &\frac{d {\overline{\pi}_i^t}( \overline{z} + 1)}{dt} = -(\lambda_i+\mu) {\overline{\pi}_i^t}( \overline{z} + 1) + \lambda_i {\overline{\pi}_i^t}( \overline{z})& \\
\nonumber \Rightarrow &\frac{d {\overline{\pi}_i^t}( \overline{z} + 1)}{dt}+ (\lambda_i+\mu) {\overline{\pi}_i^t}( \overline{z} + 1) = \lambda_i {\overline{\pi}_i^t}( \overline{z})&
\end{align}
\begin{align}
\nonumber \Rightarrow & {\overline{\pi}_i^t}( \overline{z} + 1) e^{K_i t} = c + \int W e^{K_i t} dt \hspace{2mm} &\\
\nonumber &(\text{where } K_i = (\lambda_i + \mu) \text{ and }  W = \lambda_i {\overline{\pi}_i^t}( \overline{z}))&\\
\nonumber \Rightarrow & {\overline{\pi}_i^t}( \overline{z} + 1) e^{K_i t} = c + \int \left( \frac{\lambda_i}{K_i} \right)^{\overline{z}} \frac{\lambda_i \mu}{K_i} e^{K_i t} dt - \int  \left(\frac {\lambda_i}{K_i}\right)^{\overline{z}}  \frac{\lambda_i \mu}{K_i} dt - \sum\limits_{\overline{y} = 1}^{\overline{z}-1} \int \frac{\lambda^{\overline{z}+1} \mu  t^{\overline{y}}} { \overline{y}! K^{\overline{z} - \overline{y} + 1}_i} dt + \int \frac{\lambda_i^{\overline{z} + 2}  t^{\overline{z}}}{ \overline{z}! K_i} dt&\\
\nonumber \Rightarrow & {\overline{\pi}_i^t}( \overline{z} + 1) e^{K_i t} = c +  \left( \frac{\lambda_i}{K_i} \right)^{\overline{z}+1} \frac{\mu}{K_i} e^{K_i t} - \left(\frac {\lambda_i}{K_i}\right)^{\overline{z} + 1}  \mu t - \sum\limits_{\overline{y} = 1}^{\overline{z}-1}  \frac{\lambda^{\overline{z}+1} \mu t^{\overline{y} + 1}} { (\overline{y} +1)! K^{\overline{z} - \overline{y} + 1}_i}  + \frac{\lambda_i^{\overline{z} + 2} t^{\overline{z } + 1}}{ (\overline{z} + 1 )! K_i} &\\
\nonumber & \text{Now, set } \overline{u}= \overline{y} + 1 \text{ in the summation.}&\\
\nonumber \Rightarrow & {\overline{\pi}_i^t}( \overline{z} + 1) e^{K_i t} = c +  \left( \frac{\lambda_i}{K_i} \right)^{\overline{z}+1} \frac{\mu}{K_i} e^{K_i t} - \left(\frac {\lambda_i}{K_i}\right)^{\overline{z} + 1}  \mu t - \sum\limits_{\overline{u} = 2}^{\overline{z}}  \frac{\lambda^{\overline{z}+1} \mu  t^{\overline{u} } } { (\overline{u})! K^{\overline{z} - \overline{u} + 2}_i}+ \frac{\lambda_i^{\overline{z} + 2} t^{\overline{z } + 1}}{ (\overline{z} + 1 )! K_i} &\\
\nonumber &\text{ Now using the initial distribution }{\overline{\pi}_i^0}(\overline{z}+1) =0  \text{ we get } c = -\left( \frac{\lambda_i}{K_i} \right)^{\overline{z}+1} \frac{\mu}{K_i} \text{ and putting this}\\ 
\nonumber &  \text{ value above we get,}&\\
\nonumber \Rightarrow & {\overline{\pi}_i^t}( \overline{z} + 1) =  \left( \frac{\lambda_i}{K_i} \right)^{\overline{z}+1} \frac{\mu}{K_i} + \left(  - \left(\frac {\lambda_i}{K_i}\right)^{\overline{z} + 1}  \mu t - \sum\limits_{\overline{u} = 2}^{\overline{z}}  \frac{\lambda^{\overline{z}+1} \mu t^{\overline{u} }} { (\overline{u})! K^{\overline{z} - \overline{u} + 2}_i}  + \frac{\lambda_i^{\overline{z} + 2} t^{\overline{z } + 1}}{ (\overline{z} + 1 )! K_i}  \right) e^{-K_i t} &\\    \nonumber &   \quad\quad\quad\quad \quad -\left( \left( \frac{\lambda_i}{K_i} \right)^{\overline{z}+1} \frac{\mu}{K_i} \right) e^{-K_i t}  &\\
\nonumber \Rightarrow & {\overline{\pi}_i^t}( \overline{z} + 1) =  \left( \frac{\lambda_i}{K_i} \right)^{\overline{z}+1} \frac{\mu}{K_i}  + \left( - \left( \frac{\lambda_i}{K_i} \right)^{\overline{z}+1} \frac{\mu}{K_i}  - \sum\limits_{\overline{u} = 1}^{\overline{z}}  \frac{\lambda^{\overline{z}+1} \mu t^{\overline{u} }} { (\overline{u})! K^{\overline{z} - \overline{u} + 2}_i}  + \frac{\lambda_i^{\overline{z} + 2} t^{\overline{z } + 1}}{ (\overline{z} + 1 )! K_i}  \right) e^{-K_i t} &\\
\nonumber & \text{ which is what we will get if we replace }\overline{x} \text{ by } \text{ } (\overline{z} + 1) \text{ in Equation~\ref{induction_steady_state_eqn}}. 
\end{align}
Hence proved by induction that we may calculate the transient distributions of states between $1$ and $M-1$ using Equation ~\ref{induction_steady_state_eqn}.

\item				
Now as the total probability of being in any of the states at any time $t$ is equal to $1$, we calculate the transient probability ${\overline{\pi}_i^t}( M)$ of being in state $M$ at time $t$ as:
\begin{align}
\nonumber {\overline{\pi}_{i}^t}(M) =& 1 - \sum\limits_{\overline{x}=0}^{M-1} {\overline{\pi}_i^t}(\overline{x})
\end{align}				

\end{itemize}
				

\end{proof}		

\newpage	
\begin{theorem}
 The  expected  transient aggregate net-rewards of the student $i$ and the instructor in ${\overline{X}_i}(t)$ are given by:	
 
\begin{align}
\nonumber \text{(a) } {\overline{ R}_T^{l,i}} =&  \delta^{\log\mu} \sum \limits_{\overline{ x} = 0}^{m_l}  ( \overline{x} - m_l)     \int\limits_{0}^{T}  {\overline{ \pi}_i^t}(\overline{ x}) dt  + m_l \delta^{\log\mu} T-  \alpha_l \sum \limits_{\overline{ x} = 0}^{M}   \overline{x}   \int\limits_{0}^{T} {\overline{ \pi}_i^t}(\overline{ x}) dt \\ \nonumber 		
\text{(b) } {\overline{ R}_{T}^{I,i} } =& \delta^{\log\mu}    \sum \limits_{\overline{ x} = 0}^{M} \overline{x}  \int\limits_{0}^{T}  {\overline{ \pi}_i^t}(\overline{ x}) dt -  \beta T.
\end{align}
 \end{theorem}
 			
 \begin{proof}(i)
 
\begin{flalign}
 \nonumber {\overline{ R}_T^{l,i}}=& \int\limits_{0}^{T} \sum \limits_{\overline{ x} \in {\overline{ S}_i}}  {\overline{R}^{l,i}}(\overline{ x}) {\overline{ \pi}_i^t}(\overline{ x}) dt = \int\limits_{0}^{T} \sum \limits_{\overline{ x} = 0}^{M}  {\overline{R}^{l,i}}(\overline{ x}) {\overline{ \pi}_i^t}(\overline{ x}) dt& \\
\nonumber =& \int\limits_{0}^{T} \sum \limits_{\overline{ x} = 0}^{M} \left(  {\overline{r}^{l,i} }(\overline{x}) - \alpha_l \overline{x} \right)  {\overline{ \pi}_i^t}(\overline{ x}) dt&\\					
\nonumber =& \int\limits_{0}^{T} \sum \limits_{\overline{ x} = 0}^{m_l} \left( \overline{x} \delta^{\log\mu} - \alpha_l \overline{x}   \right) {\overline{ \pi}_i^t}(\overline{ x}) dt + \int\limits_{0}^{T} \sum \limits_{\overline{ x} = m_l +1}^{M} \left( m_l \delta^{\log\mu} - \alpha_l \overline{x}   \right) {\overline{ \pi}_i^t}(\overline{ x}) dt& 
\end{flalign}
\begin{flalign}
\nonumber 					=& \int\limits_{0}^{T} \sum \limits_{\overline{ x} = 0}^{m_l}  \overline{x} \delta^{\log\mu}  {\overline{ \pi}_i^t}(\overline{ x}) dt - \int\limits_{0}^{T} \sum \limits_{\overline{ x} = 0}^{M}  \alpha_l \overline{x} {\overline{ \pi}_i^t}(\overline{ x}) dt + \int\limits_{0}^{T} \sum \limits_{\overline{x} = m_l +1}^{M}  m_l \delta^{\log\mu}   {\overline{ \pi}_i^t}(\overline{ x})   dt  &\\
\nonumber=&  \delta^{\log\mu}  \int\limits_{0}^{T} \sum \limits_{\overline{ x} = 0}^{m_l}  \overline{x}   {\overline{ \pi}_i^t}(\overline{ x}) dt - \alpha_l \int\limits_{0}^{T} \sum \limits_{\overline{ x} = 0}^{M}   \overline{x} {\overline{ \pi}_i^t}(\overline{ x}) dt + m_l \delta^{\log\mu} \int\limits_{0}^{T} \sum \limits_{\overline{x} = m_l +1}^{M}    {\overline{ \pi}_i^t}(\overline{ x})   dt &
\end{flalign}
\begin{flalign}
\nonumber =&  \delta^{\log\mu}  \int\limits_{0}^{T} \sum \limits_{\overline{ x} = 0}^{m_l}  \overline{x}   {\overline{ \pi}_i^t}(\overline{ x}) dt - \alpha_l \int\limits_{0}^{T} \sum \limits_{\overline{ x} = 0}^{M}   \overline{x} {\overline{ \pi}_i^t}(\overline{ x}) dt   + m_l \delta^{\log\mu} \int\limits_{0}^{T} \left( 1 - \sum \limits_{\overline{x} = 0}^{m_l}    {\overline{ \pi}_i^t}(\overline{ x}) \right)  dt &\\		 		
\nonumber=&  \delta^{\log\mu}  \int\limits_{0}^{T} \sum \limits_{\overline{ x} = 0}^{m_l}  \overline{x}   {\overline{ \pi}_i^t}(\overline{ x}) dt - \alpha_l \int\limits_{0}^{T} \sum \limits_{\overline{ x} = 0}^{M}   \overline{x} {\overline{ \pi}_i^t}(\overline{ x}) dt +  m_l \delta^{\log\mu} T - m_l \delta^{\log\mu} \int\limits_{0}^{T} \sum \limits_{\overline{x} = 0}^{m_l}    {\overline{ \pi}_i^t}(\overline{ x})  dt 
\end{flalign}
\begin{flalign}
\nonumber=&   m_l \delta^{\log\mu} T  + \delta^{\log\mu}  \int\limits_{0}^{T} \sum \limits_{\overline{ x} = 0}^{m_l}  ( \overline{x} - m_l)   {\overline{ \pi}_i^t}(\overline{ x}) dt -  \alpha_l \int\limits_{0}^{T} \sum \limits_{\overline{ x} = 0}^{M}   \overline{x} {\overline{ \pi}_i^t}(\overline{ x}) dt&
\end{flalign}
\begin{flalign}
\nonumber & \text{ Interchanging the integral and the summation}\text{ we get: }&\\
\nonumber 					=& m_l \delta^{\log\mu} T + \delta^{\log\mu} \sum \limits_{\overline{ x} = 0}^{m_l}   \int\limits_{0}^{T}  ( \overline{x} - m_l)   {\overline{ \pi}_i^t}(\overline{ x}) dt -  \alpha_l \sum \limits_{\overline{ x} = 0}^{M}   \int\limits_{0}^{T}  \overline{x} {\overline{ \pi}_i^t}(\overline{ x}) dt&\\
\nonumber 					=& m_l \delta^{\log\mu} T + \delta^{\log\mu} \sum \limits_{\overline{ x} = 0}^{m_l}  ( \overline{x} - m_l)     \int\limits_{0}^{T}  {\overline{ \pi}_i^t}(\overline{ x}) dt -  \alpha_l \sum \limits_{\overline{ x} = 0}^{M}   \overline{x}   \int\limits_{0}^{T} {\overline{ \pi}_i^t}(\overline{ x}) dt&
 \end{flalign}
 
 \end{proof}

\begin{proof}  (ii)

\begin{align}
 \nonumber {\overline{ R}_{T}^{I,i} }
\nonumber 					=& \int\limits_{0}^{T}  \sum \limits_{\overline{ x} = 0}^{M} \left(  {\overline{R}^{I,i}}(\overline{ x})\right) {\overline{ \pi}_i^t}(\overline{ x}) dt&\\
\nonumber 					=&\int\limits_{0}^{T}  \sum \limits_{\overline{ x} = 0}^{M} \left(  {\overline{ r}^{I,i}}(\overline{x}) - \beta  \right) {\overline{ \pi}_i^t}(\overline{ x}) dt
= \int\limits_{0}^{T}  \sum \limits_{\overline{ x} = 0}^{M} \left(  \overline{x} \delta^{\log\mu} - \beta  \right) {\overline{ \pi}_i^t}(\overline{ x}) dt&\\
\nonumber 					=& \int\limits_{0}^{T}  \sum \limits_{\overline{ x} = 0}^{M} \overline{x} \delta^{\log\mu}   {\overline{ \pi}_i^t}(\overline{ x}) dt - \int\limits_{0}^{T}  \sum \limits_{\overline{ x} = 0}^{M} \beta {\overline{ \pi}_i^t}(\overline{ x})dt &\\
\nonumber 					=& \delta^{\log\mu}  \int\limits_{0}^{T}   \sum \limits_{\overline{ x} = 0}^{M} \overline{x}   {\overline{ \pi}_i^t}(\overline{ x}) dt - \int\limits_{0}^{T}  \beta dt
= \delta^{\log\mu}  \int\limits_{0}^{T}   \sum \limits_{\overline{ x} = 0}^{M} \overline{x}   {\overline{ \pi}_i^t}(\overline{ x}) dt -  \beta T&\\
\nonumber 					 & \text{ Interchanging the integral and the summation }\text{ we get: }&\\
\nonumber 					=& \delta^{\log\mu}    \sum \limits_{\overline{ x} = 0}^{M} \overline{x}  \int\limits_{0}^{T}  {\overline{ \pi}_i^t}(\overline{ x}) dt -  \beta T	&						
\end{align}

\end{proof}

 \begin{claim}
 \label{Claim_Integral_pi_x_t_less_between_0andM}
 \begin{align}
 \nonumber  \int\limits_{0}^{T} {\overline{ \pi}_i^t}(\overline{x}) \text{ } dt    &= \begin{cases} 
  \nonumber \frac{\mu T}{{K}_i} + \frac{\lambda_i( -e^{-{K}_i T} + 1)}{{K}_i^2}   & \text{ if } {\overline{x}}=0 \\
  \nonumber  =\left(\frac{\lambda_i}{{K}_i}\right)^{\overline{x}} \frac{\mu}{{K}_i} T +\frac {\lambda_i^{\overline{x}}\mu}{{K}^{\overline{x}+2}_l} \left( e^{-{K}_i T} - {\overline{x}} \right)	\\
 \nonumber +\frac {\lambda_i^{\overline{x}}\mu}{{K}^{\overline{x}+2}_l} \sum \limits_{{\overline{y}}=1}^{\overline{x}-1}
  \left(\frac {\Gamma({\overline{y}}+1,{K}_{i} T)} {{\overline{y}}!} \right) + \frac{\lambda_i^{\overline{x}+1}}{{K}_i^{\overline{x}+2}} \left( 1 - \frac {\Gamma(\overline{x}+1,{K}_i T)}{\overline{x}!}  \right) & \text{ if } 0< { \overline{x} }<M \\
  \nonumber     = T - \sum\limits_{\overline{x} = 0}^{ M-1} \int\limits_{0}^{T} {\overline{ \pi}_{i}^t}(\overline{x}) dt    & \text{ if } {\overline{x}} = M.
       \end{cases} . \\
 \nonumber   & where\text{ }{K}_i= \lambda_i + \mu.
 \end{align} 
 \end{claim}
\begin{proof}\textbf{State $0$}
 \begin{align}
\nonumber &	\int\limits_{0}^{T} {\overline{ \pi}_i^t}(\overline{0}) \text{ } dt
	=\int\limits_{0}^{T} \left(  \frac{\mu}{{K}_i} + \frac{\lambda_i}{{K}_i} e^{-{K}_it} \right)  \text{ } dt\\
\nonumber & =  \frac{\mu}{{K}_i} T + \left[ \frac{\lambda_i}{{K}_i} \frac{e^{-{K}_it}}{-{K}_i}  \right]_0^T
= \frac{\mu}{{K}_i} T + \frac{\lambda_i}{({K}_i)^2} \left( 1 - e^{- {K}_i T} \right)
\end{align}

\textbf{State $x$} $\forall \overline{x}: 1 \leq \overline{x} \leq M-1$
 \begin{align}
\nonumber & \int\limits_{0}^{T}  {\overline{\pi}_i^t}(\overline{x})  \text{ } dt\\
\nonumber =&  \int\limits_{0}^{T}  \left( -  \left(\frac{\lambda_i}{{K}_i}\right)^{\overline{x}} \frac{\mu}{{K}_i}    - \sum\limits_{\overline{y}=1}^{\overline{x}-1} \frac{\lambda_i^ {\overline{x}} t^{\overline{y}} \mu}{\overline{y}! {K}_i^{\overline{x}-\overline{y}+1}} + \frac{{\lambda_i}^{\overline{x}+1} t^{\overline{x}}}{\overline{x}! {K}_i} \right)e^{-{K}_{i} t} dt  + \int\limits_{0}^{T}   \left(\frac{\lambda_i}{{K}_i}\right)^{\overline{x}} \frac{\mu}{{K}_i}  \text{ } dt\\
\nonumber =&  \int\limits_{0}^{T}  \left(\frac{\lambda_i}{{K}_i}\right)^{\overline{x}} \frac{\mu}{{K}_i}  \text{ } dt - \int\limits_{0}^{T} \left(\frac{\lambda_i}{{K}_i}\right)^{\overline{x}} \frac{\mu}{{K}_i} e^{-{K}_{i} t} dt  - \int\limits_{0}^{T} \sum\limits_{\overline{y}=1}^{\overline{x}-1} \frac{\lambda_i^ {\overline{x}} t^{\overline{y}} \mu}{\overline{y}! {K}_i^{\overline{x}-\overline{y}+1}} e^{-{K}_{i} t}  \text{ } dt +  \int\limits_{0}^{T}  \frac{{\lambda_i}^{\overline{x}+1} t^{\overline{x}}}{\overline{x}! {K}_i}   e^{-{K}_{i} t} \text{ } dt\\
\nonumber =& \left(\frac{\lambda_i}{{K}_i}\right)^{\overline{x}} \frac{\mu}{{K}_i} T  -  \left[ \left(\frac{\lambda_i}{{K}_i}\right)^{\overline{x}} \frac{\mu}{{K}_i} \frac{e^{-{K}_{i} t}}{-K_i} \right]^T_0 -  \sum\limits_{\overline{y}=1}^{\overline{x}-1} \frac{\lambda_i^ {\overline{x}}  \mu}{\overline{y}! {K}_i^{\overline{x}-\overline{y}+1}} \int\limits_{0}^{T}   t^{\overline{y}}  e^{-{K}_{i} t}  \text{ } dt +    \frac{{\lambda_i}^{\overline{x}+1} }{\overline{x}! {K}_i}    \int\limits_{0}^{T} t^{\overline{x}}  e^{-{K}_{i} t} \text{ } dt\\
\nonumber &\text{ (Using Claim ~\ref{Claim_integral_to_gamma_function})}\\
\nonumber =&  \left(\frac{\lambda_i}{{K}_i}\right)^{\overline{x}} \frac{\mu}{{K}_i} T  +  \left(\frac{\lambda_i}{{K}_i}\right)^{\overline{x}} \frac{\mu}{\left({K}_i\right)^2} \left(e^{-{K}_{i} T} - 1\right)  -  \sum\limits_{\overline{y}=1}^{\overline{x}-1} \frac{\lambda_i^ {\overline{x}}  \mu}{\overline{y}! {K}_i^{\overline{x}-\overline{y}+1}} \frac{1}{({{K_i})^{\overline{y} + 1}}} \left(      \overline{y}!     -        \Gamma(\overline{y} + 1 , K_i T)       \right) &\\
\nonumber & +    \frac{{\lambda_i}^{\overline{x}+1} }{\overline{x}! {K}_i}    \frac{1}{({{K_i})^{\overline{x} + 1}}} \left(      \overline{x}!     -        \Gamma(\overline{x} + 1 , K_i T)       \right)&\\
\nonumber =&  \left(\frac{\lambda_i}{{K}_i}\right)^{\overline{x}} \frac{\mu}{{K}_i} T 
+  \left(\frac{\lambda_i}{{K}_i}\right)^{\overline{x}} \frac{\mu}{\left({K}_i\right)^2} \left(e^{-{K}_{i} T} - 1\right) -  \sum\limits_{\overline{y}=1}^{\overline{x}-1} \frac{\lambda_i^ {\overline{x}}  \mu}{ {K}_i^{\overline{x}+2}} + \sum\limits_{\overline{y}=1}^{\overline{x}-1} \frac{\lambda_i^ {\overline{x}}  \mu}{ {K}_i^{\overline{x}+2}} \left(  \frac{\Gamma(\overline{y} + 1, K_i T) }{\overline{y}!} \right) &\\
\nonumber & +    \frac{{\lambda_i}^{\overline{x}+1} }{ {K}_i ^{\overline{x}+2}}
- \frac{{\lambda_i}^{\overline{x}+1} }{ {K}_i ^{\overline{x}+2}} \left(  \frac{\Gamma(\overline{x} + 1, K_i T) }{\overline{x}!} \right)&\\
\nonumber =&  \left(\frac{\lambda_i}{{K}_i}\right)^{\overline{x}} \frac{\mu}{{K}_i} T 
+  \left(\frac{\lambda_i}{{K}_i}\right)^{\overline{x}} \frac{\mu}{\left({K}_i\right)^2} \left(e^{-{K}_{i} T} - 1\right) -  \left( {\overline{x}-1} \right)  \frac{\lambda_i^ {\overline{x}}  \mu}{ {K}_i^{\overline{x}+2}} 
+ \frac{\lambda_i^ {\overline{x}}  \mu}{ {K}_i^{\overline{x}+2}}  \sum\limits_{\overline{y}=1}^{\overline{x}-1}   \left(  \frac{\Gamma(\overline{y} + 1, K_i T) }{\overline{y}!} \right) &\\
\nonumber & +    \frac{{\lambda_i}^{\overline{x}+1} }{ {K}_i ^{\overline{x}+2}}
- \frac{{\lambda_i}^{\overline{x}+1} }{ {K}_i ^{\overline{x}+2}} \left(  \frac{\Gamma(\overline{x} + 1, K_i T) }{\overline{x}!} \right)&\\
\nonumber =&  \left(\frac{\lambda_i}{{K}_i}\right)^{\overline{x}} \frac{\mu}{{K}_i} T 
+  \left(\frac{\lambda_i}{{K}_i}\right)^{\overline{x}} \frac{\mu}{\left({K}_i\right)^2} \left(e^{-{K}_{i} T} - \overline{x} \right) + \frac{\lambda_i^ {\overline{x}}  \mu}{ {K}_i^{\overline{x}+2}}  \sum\limits_{\overline{y}=1}^{\overline{x}-1}   \left(  \frac{\Gamma(\overline{y} + 1, K_i T) }{\overline{y}!} \right) +  \frac{{\lambda_i}^{\overline{x}+1} }{ {K}_i ^{\overline{x}+2}} \left( 1-  \frac{\Gamma(\overline{x} + 1, K_i T) }{\overline{x}!} \right)&
\end{align}

\underline{\textbf{State $M$}}
 \begin{flalign}
\nonumber & \int\limits_0^T {\overline{\pi}_{i}^t}(M) \text{ } dt = \int\limits_0^T    \left( 1 - \sum\limits_{\overline{x}=0}^{M-1} {\overline{\pi}_i^t}(\overline{x}) \right)  \text{ } dt
= T - \int\limits_0^T \sum\limits_{\overline{x}=0}^{M-1} {\overline{\pi}_i^t}(\overline{x}) \text{ } dt
= T - \sum\limits_{\overline{x}=0}^{M-1}   \int\limits_0^T {\overline{\pi}_i^t}(\overline{x}) \text{ } dt&
\end{flalign}

\end{proof}

\begin{claim}
\label{Claim_integral_to_gamma_function}
\begin{align}
 \nonumber \int\limits_{0}^{T} t^{\overline{x}} e^{-{K_i} t} \text{ } dt 
=  \frac{1}{({{K_i})^{\overline{x} + 1}}} \left(      \overline{x}!     -        \Gamma(\overline{x} + 1 , K_i T)       \right)
\end{align}
\end{claim}
\begin{proof}

\begin{align}
\nonumber & \int\limits_{0}^{T} t^{\overline{x}} e^{-{K_i} t} \text{ } dt 
= \int\limits_{0}^{K_i T} {\left( \frac{t'}{K_i}\right)}^{\overline{x}} e^{-t'} \text{ } \frac{dt'}{K_i}  \text{ (Taking $t' = K_i t$)}\\
\nonumber = & \frac{1}{({{K_i})^{\overline{x} + 1}}} \int\limits_{0}^{K_i T} \left(  t' \right)^{\overline{x}}   e^{-t'} \text{ } dt'
=\frac{1}{({{K_i})^{\overline{x} + 1}}} \left(      \int\limits_{0}^{\infty} \left(  t' \right)^{\overline{x}}   e^{-t'} \text{ } dt'       -       \int\limits_{K_i T}^{\infty} \left(  t' \right)^{\overline{x}}   e^{-t'} \text{ } dt'       \right)\\
\nonumber =& \frac{1}{({{K_i})^{\overline{x} + 1}}} \left(      \Gamma(\overline{x} + 1 )     -        \Gamma(\overline{x} + 1 , K_i T)       \right) 
= \frac{1}{({{K_i})^{\overline{x} + 1}}} \left(      \overline{x}!     -        \Gamma(\overline{x} + 1 , K_i T)       \right)
\end{align}
\end{proof}

\section{ OEF  as a Stackelberg Game}

\begin{proposition}
Students $i,j$ belonging to the same type $Type_l$ receive equal utility (steady state and aggregate) if they choose the same policy i.e. if $\psi^i = \psi^j (= \psi^l)$ for students $ i,j \in Type_l$ then,
\begin{align}
 \nonumber U_{\Pi}^{l,i} = U_{\Pi}^{l,j} \text{, }  \hspace{4mm}
 \nonumber U_{T}^{l,i} = U_{T}^{l,j} 
\end{align}
\end{proposition}
\begin{proof}
Given: $\psi^i = \psi^j \Rightarrow \psi_b^i = \psi_b^j \forall b: 1\leq b \leq w$.

Also, $ {\overline{R}_{\overline{\Pi}}^{l,i}} = {\overline{R}_{\overline{\Pi}}^{l,j}}$ if $\lambda_i = \lambda_j= \Delta_b$ (See Claim ~\ref{Claim_optimization_expected_rewards_equal}).
\begin{align}
\nonumber \therefore & {\overline{R}_{\overline{\Pi}}^{l,i}} \phi_a \psi^i_b =  {\overline{R}_{\overline{\Pi}}^{l,j}} \phi_a \psi^j_b\\
\nonumber \Rightarrow & D_{a,b}^{\Pi,l,i} \phi_a \psi^i_b = D_{a,b}^{\Pi,l,j} \phi_a \psi^j_b  \qquad \forall a \;\forall b\\
\nonumber \Rightarrow & \sum\limits_{a=1}^{v} \sum\limits_{b=1}^{w}  D_{a,b}^{\Pi,l,i} \phi_a \psi^i_b =  \sum\limits_{a=1}^{v} \sum\limits_{b=1}^{w} D_{a,b}^{\Pi,l,j} \phi_a \psi^j_b  \text{ } \\
\nonumber \Rightarrow &  U_{\Pi}^{l,i} = U_{\Pi}^{l,j} .
\end{align}
Following a similar line of proof we can also prove that $U_{T}^{l,i} = U_{T}^{l,j} $.
\end{proof}

\begin{claim}
\label{Claim_optimization_expected_rewards_equal} The rewards (steady state and aggregate) received by the students $i,j$ belonging to a $Type_l$ is same if $ \lambda^i = \lambda^j $. Also the rewards (steady state and aggregate) the instructor receives w.r.t. the students $i,j$ belonging to a $Type_l$ is same if $ \lambda^i = \lambda^j$. Thus $\forall i,j \in Type_l$, if $ \lambda^i = \lambda^j$ then:
\begin{align}
\nonumber {\overline{R}_{\overline{\Pi}}^{l,i}} = {\overline{R}_{\overline{\Pi}}^{l,j}}
\text{, } \hspace{2mm}
{\overline{R}_{\overline{\Pi}}^{I,i}} = {\overline{R}_{\overline{\Pi}}^{I,j}}
\text{, } \hspace{2mm}
{\overline{R}_{T}^{l,i}} = {\overline{R}_{T}^{l,j}}
\text{, } \hspace{2mm}
{\overline{R}_{T}^{I,i}} = {\overline{R}_{T}^{I,j}}
\end{align}

\end{claim}
\begin{proof}
This is trivially true as these quantities are equal if we put $\lambda_i = \lambda_j$ for students $i,j$ belonging to $Type_l$ in the expressions given in Theorems ~\ref{Thorem_SteadyState_reward_student_and_instructor}  and ~\ref{Thorem_Transient_reward_student_and_instructor}.
\end{proof}

\begin{proposition}
The net-reward matrices $B^{\Pi,I,i}, B^{\Pi,I,j}$ (steady state) and $B^{T,I,i},B^{T,I,j}$ (aggregate) for the instructor w.r.t. students $i,j$ belonging to $Type_l$ have the following property:
\begin{align}
\nonumber B^{\Pi,I,i} = B^{\Pi,I,j} (= B^{\Pi,I,l}), \hspace{4mm}
\nonumber B^{T,I,i} = B^{T,I,j} (= B^{T,I,l}) .
\end{align}
\end{proposition}

\begin{proof}

Given students $i,j \in Type_l$ and considering the matrices $B^{\Pi,I,i}$ and $B^{\Pi,I,j}$:
\begin{itemize}
\item For entries $B^{\Pi,I,i}_{a,b}$ and $B^{\Pi,I,j}_{a,b}$, $\lambda_i = \lambda_j = \Lambda_b$ thus ${\overline{R}_{\overline{\Pi}}^{I,i}} = {\overline{R}_{\overline{\Pi}}^{I,j}}$ (Claim ~\ref{Claim_optimization_expected_rewards_equal}). 
Also $c_i = c_j$ $\forall i,j \in Type_l$ (as discussed in section ~\ref{section_rewards}).

\item  Thus, 
$
B^{\Pi,I,i}_{a,b} = c_i {\overline{R}_{\overline{\Pi}}^{I,i}} = c_j {\overline{R}_{\overline{\Pi}}^{I,j}}  = B^{\Pi,I,j}_{a,b}.
$
\end{itemize} 
Thus $B^{\Pi,I,i} = B^{\Pi,I,j} (= B^{\Pi,I,l} say)$ $\forall i,j \in Type_l$. Using similar arguments we can also prove that $B^{T,I,i}_{a,b} = B^{T,I,j}_{a,b} (= B^{T,I,l} say) \text{ } \forall i,j \in Type_l$.

\end{proof}

\endproof

\end{document}